\documentclass[11pt]{article}
\usepackage{amsmath,amssymb}
\usepackage{amsthm}
\usepackage{mathtools}
\usepackage[noend]{algorithmic}
\usepackage[ruled,vlined]{algorithm2e}
\usepackage{setspace}
\usepackage{url}
\usepackage{makeidx}
\usepackage{enumerate}
\usepackage[top=1in, bottom=1in, left=1in, right=1in]{geometry}
\usepackage{graphicx,float,psfrag,epsfig,caption}
\usepackage[usenames,dvipsnames,svgnames,table]{xcolor}
\definecolor{darkgreen}{rgb}{0.0,0,0.9}
\usepackage[pagebackref,colorlinks=true,pdfpagemode=UseNone,citecolor=OliveGreen,linkcolor=BrickRed,urlcolor=BrickRed,pdfstartview=FitH]{hyperref}
\usepackage{epstopdf}
\usepackage{color}
\usepackage{xr}
\usepackage{subfig}
\usepackage{caption}
\usepackage{graphicx}
\usepackage[utf8]{inputenc}
\usepackage{cancel}
\usepackage{algorithm2e}

\usepackage{scalerel,stackengine}

\stackMath
\newcommand\reallywidehat[1]{%
\savestack{\tmpbox}{\stretchto{%
  \scaleto{%
    \scalerel*[\widthof{\ensuremath{#1}}]{\kern.1pt\mathchar"0362\kern.1pt}%
    {\rule{0ex}{\textheight}}
  }{\textheight}%
}{2.4ex}}%
\stackon[-6.9pt]{#1}{\tmpbox}%
}
\usepackage[mathscr]{euscript}

\DeclareSymbolFont{rsfs}{U}{rsfs}{m}{n}
\DeclareSymbolFontAlphabet{\mathscrsfs}{rsfs}

\numberwithin{equation}{section}

\newtheoremstyle{myexample} 
    {\topsep}                    
    {\topsep}                    
    {\rm }                   
    {}                           
    {\bf }                   
    {.}                          
    {.5em}                       
    {}  

\newtheoremstyle{myremark} 
    {\topsep}                    
    {\topsep}                    
    {\rm}                        
    {}                           
    {\bf}                        
    {.}                          
    {.5em}                       
    {}  

\theoremstyle{myremark}

\theoremstyle{myremark}

\theoremstyle{myexample}

\definecolor{darkgreen}{rgb}{0.0, 0.5, 0.0}

\newcommand{\bea}{\begin{eqnarray}}
\newcommand{\eea}{\end{eqnarray}}
\newcommand{\<}{\langle}
\renewcommand{\>}{\rangle}
\newcommand{\E}{{\mathbb E}}

\def\fr{\frac}
\def\fr12{\frac{1}{2}}

\def \EE{\mathcal{E}}

\def \mtt{\mathtt{m}}
\def \Ctt{\mathtt{C}}

\def \HH{{\cal H}}

\def\Unif{{\sf Unif}}

\def\eps{{\varepsilon}}

\def\btheta{{\boldsymbol{\theta}}}

\def\bx{{\boldsymbol{x}}}

\def\bb{{\boldsymbol{b}}}

\def\bJ{\boldsymbol{J}}

\def\de{{\rm d}}

\def\E{{\mathbb E}}

\def\<{\langle}
\def\>{\rangle}

\def\by{{\boldsymbol{y}}}

\def\b0{{\boldsymbol{0}}}

\def\br{{\boldsymbol r}}

\SetKw{KwInitialize}{Initialize}
\SetKw{KwDraw}{Draw}

\allowdisplaybreaks
\title{High-dimensional random dynamical systems: co-existence of chaotic attractors, 
phase transitions, and response to periodic drive}
\author{Samantha J. Fournier$^*$ \and Pierfrancesco Urbani\thanks{Université Paris-Saclay, CNRS, CEA, Institut de physique théorique, 91191, Gif-sur-Yvette, France}}

\makeindex 
\begin{document}

\hypersetup{linkcolor=RoyalPurple}

\maketitle
\begin{abstract}
We study the dynamical properties of a broad class of high-dimensional random dynamical systems exhibiting chaotic as well as fixed point and periodic attractors. We consider cases in which attractors can co-exists in some regions of the phase diagrams and we characterize their nature by computing the maximal Lyapunov exponent. For a specific choice of the dynamical system we show that this quantity can be computed explicitly in the whole chaotic phase due to an underlying integrability of a properly defined Schr\"odinger problem.
Furthermore, we consider the response of this dynamical systems to periodic perturbations. 
We show that these dynamical systems act as filters in the frequency-amplitude spectrum of the periodic forcing: only in some regions of the frequency-amplitude plane the periodic forcing leads to a synchronization of the dynamics.
All in all, the results that we present mirror closely the ones observed in the past forty years in the study of standard models of random recurrent neural networks. However, the dynamical systems that we consider are easier to study and we believe that this may be an advantage if one wants to go beyond random dynamical systems and consider specific training strategies.
\end{abstract}

\maketitle

\tableofcontents

\section{Introduction}
High-dimensional dynamical systems are the building blocks of several complex systems. To cite a few, they appear as (i) systems describing the evolution of parameters in optimization/machine learning models \cite{montanari2025dynamical}, (ii) instances of recurrent neural networks thought as models of microcircuits in the brain, playing a central role in computational neuroscience \cite{sompolinsky1988chaos, parisi1986asymmetric} and (iii) models of interacting species in well mixed ecological/economic/social networks \cite{rieger1989solvable, rieger1993dynamical}.

In many cases, the dynamics of the degrees of freedom can be described as the gradient descent (or stochastic versions of it) on an high-dimensional non-convex energy  landscape \cite{cugliandolo2002dynamics,cugliandolo2023recent}. This would suggest that, for such types of systems, some understanding on the properties of the dynamical system can be gained by looking at the structure of the energy landscape itself. This is for example the case of energy landscapes that are strongly convex with just one global minimizer attracting the descent dynamics. Surprisingly, it is also the case of specific models whose energy landscape is non-convex and rough with exponentially many local minima potentially attracting the dynamics. This situation emerges in simple mean field spin glasses \cite{cugliandolo1993analytical}. However, generally, even when the dynamics can be ascribed to a descent trajectory in an energy landscape, as soon as the corresponding loss function has a continuous manifold of ground states, understanding which of them is the final attractor requires to fully solve the corresponding dynamical system. This is for example the case of overparametrized loss landscapes minimized via gradient-based algorithms \cite{kamali2023dynamical} which is very relevant for artificial neural network applications \cite{montanari2025dynamical}.

In this work, we focus on dynamical systems for which there is no meaningful notion of an energy landscape. 
This is the case of recurrent neural networks or interacting agents in economic networks. In many of these cases, the heterogeneity of the interactions and the high-dimensional nature of the underlying dynamics conspire to generate chaotic dynamics.
Understanding chaos in high dimensions is an important problem because, to some extent, many of these dynamical systems can be trained or deformed according to some external perturbation/adaptation rule that can lead to some desired functionality or behavior. The way in which all this happens depends on the properties of the chaotic attractors where the dynamics lands. For example, random recurrent neural networks can be trained to perform some task such as  (i) motor control \cite{sussillo2009generating} or (ii) generation of desired configurations \cite{fournier2025generative}. In these cases, the chaotic dynamics has to be controlled because, on one hand, the level of chaos is essential to keep an endogenous dynamics but, at the same time, it cannot be so high that the collective motion  results in wild uncontrolled fluctuations.

Therefore, understanding chaos in high dimension is an important topic that is becoming central in many contexts. 
The main theoretical challenges that this poses can be summarized as follows:
\begin{enumerate}
    \item {\bf Properties of the chaotic dynamics}. Understand the properties of chaotic attractors, their geometry, size, dimension. When the underlying dynamical system is high-dimensional, it is typically unclear how to leverage standard tools of dynamical systems to characterize the chaotic phases of the models.
    \item {\bf Response of chaotic attractors to deformation/adaptation of the dynamical system}. Understand how to control the chaotic nature of the dynamics with external inputs and/or to perform a task. This could be some non-trivial motor control/cognitive task in neural networks or the adaptation to some change in the environment in economic networks.
\end{enumerate}
In this work, we focus on the first point as it is an essential step to understand the second one. Note that some preliminary work on the second point has been already presented in \cite{fournier2023statistical}.

In many contexts, chaos is induced by non-reciprocal interactions between the degrees of freedom. This is clear when it comes to neural networks and/or ecological communities: both neurons and species interact in a directed, non-reciprocal way. Non-reciprocity in the interactions is the main source of endogenous motion/activity in these systems. 

This work considers simple models of non-reciprocal interactions in abstract high-dimensional dynamical systems. The models that we focus on are not microscopically grounded (or at least this is not the way we introduce them). However, the main result that we will get is that
\begin{itemize}
    \item they can be analyzed in a rather detailed way. This allows us to study: (i) transient time dynamics and (ii) the properties of the chaotic attractors, their co-existence and their response to periodic drive.
    \item despite being abstract, these models show a phenomenology that mirrors directly the one found in specific models of recurrent neural networks.
\end{itemize}
Taken together, these two aspects suggest that the class of models that we consider are good models of recurrent neural networks and therefore can be used as general models. Their simplicity allows to use them in order to study problems usually almost intractable in standard models. This is for example the case for training. The models that we study in this work were used to investigate precisely this point in \cite{fournier2023statistical}.

Therefore, the main purpose of this work is to develop a set of theoretical tools to study simple high-dimensional chaotic dynamical systems and to show that they give rise to a phenomenology that is very close to the one found in more standard models. 
The main toolbox that we employ to derive our results is Dynamical Mean Field Theory (DMFT) which is a standard tool to understand the behavior of observables whose dynamics concentrates in the high-dimensional limit.

\paragraph*{Related Literature --}
The study of dynamical systems is a building block of statistical mechanics, mathematical physics and computer science \cite{katok1995introduction}. In low dimensions, one can use simple dynamical systems to study the emergence of chaotic attractors. This is a rich subject with a large number of applications to physics and engineering. 

In the past forty years there has been growing interest in dynamical systems where the number of degrees of freedom is large. This high-dimensional regime is rather different from the one that is typically considered in dynamical system theory and it is relevant in the study of a large class of complex systems. 
The study of dynamical systems in high dimension was started in the foundational work by Sompolinsky, Crisanti and Sommers \cite{sompolinsky1988chaos, crisanti2018path}. This work considers simple models of random recurrent neural networks and shows that depending on the strength of the interactions between the neurons in the system, the asymptotic state of the dynamics can be either a fixed point attractor, where the degrees of freedom converge to a particular random point in phase space, or a chaotic attractor where endogenous motion is kept steady.
Most importantly, this work shows how to use field theory techniques to study the properties of the chaotic attractor, in particular the Maximal Lyapunov Exponent (MLE). Field theory techniques to study neural networks are also the subject of recent reviews, see \cite{helias2020statistical, ringel2025applications}.
The class of models studied in \cite{sompolinsky1988chaos} became the building blocks of several subsequent studies of recurrent neural networks, including recent works \cite{mastrogiuseppe2018linking, pereira2023forgetting, clark2023dimension, engelken2023lyapunov, martorell2024dynamically, clark2024theory, huang2025freezing, hakim2025theory}.

Dynamical systems evolving without any relation to a Lyapunov function are harder to understand via static methods, defined as probabilistic approaches on the exploration of phase space configurations. For example, there is no meaningful notion of local minima that can be used as educated guess to understand the long-time dynamics. In recent years however, several works have tried to propose a \emph{landscape-dynamics correspondence}, also in this case by suggesting that the properties of the (unstable) equilibria of the dynamical systems may be used to understand the ones of the chaotic attractors, see \cite{wainrib2013topological, qiu2024optimization, stubenrauch2025fixed}.
This set of landscape-dynamics conjectures has been recently tested using the class of dynamical systems that we focus on in this work, see \cite{fournier2025non}. The main conclusion of this work is that any simple measure on the landscape of equilibria (equiprobability on unstable equilibria) is not enough to characterize the ones that play the most prominent role in chaotic dynamics and typical equilibria do not seem indicative of the properties of dynamics.
Therefore, how to twist the calculations of the statistics of equilibria in high-dimensional dynamical systems in order to capture the corresponding chaotic dynamics remains, to a large extent, a very interesting open problem.
Since typical equilibria are not helpful to understand training dynamics, it is important to develop powerful techniques and simple enough models to push the dynamical analysis as far as possible.

\paragraph*{Our contribution --}
The standard models of recurrent neural networks considered in \cite{sompolinsky1988chaos} can be studied via DMFT. This technique allows to control the steady state dynamics of the corresponding dynamical system in a rather accurate way. However, as soon as one is interested in studying either the transient behavior of the dynamics or what happens when the dynamical system is not fully random but trained in some way \cite{sussillo2009generating, fournier2023statistical}, the standard model in \cite{sompolinsky1988chaos} are hard to analyze.

The purpose of this work is to consider simplified models of high-dimensional dynamical systems and to characterize them in full detail. 
Our main contributions are the following:
\begin{itemize}
    \item We consider a set of abstract high-dimensional dynamical systems that can be studied exactly via DMFT techniques.
    \item We derive the DMFT equations describing the dynamics of the models via simple field theory techniques. The DMFT equations that we derive have been presented partially in \cite{fournier2023statistical, fournier2025non}. Some aspects of the corresponding phase diagram was presented in these references. We report them here for completeness.
    \item We study the phase diagram of these models. In particular, we consider cases in which the phase diagram develops multiple co-existing attractors. We study the corresponding phase transitions. 
    \item We develop a formalism to study the chaotic attractors of the sytems. Differently from \cite{sompolinsky1988chaos}, we do not rely extensively on field theory techniques that are rather complex, especially in more general situations in which several order parameters need to be taken into account. Instead, we develop a simplified formalism based on the formal definition of the maximal Lyapunov exponent (MLE), i.e. that it controls the sensitivity of the dynamics to initial conditions. However, at variance with \cite{sompolinsky1988chaos}, the dynamical systems that we consider are exactly soluble in the sense that we can compute explicitly the MLE in the whole chaotic phase, in some instances of the models. Correspondingly, we can give a much more accurate predictions on the chaotic dynamics. For example, additionally to the MLE we can also compute the prefactors of the exponential divergence of the distance between two closely initialized dynamical trajectories.
\end{itemize}

\paragraph*{Structure of the manuscript --}
This manuscript is organized as follows. 

In Sec.~\ref{sec_models} we define the class of models that we analyze. We emphasize that for such a class of models, the dynamics can be tracked also when the corresponding dynamical systems is discrete in time and we detail what are the main control parameters that we are interested in studying.

In Sec. \ref{sec_DMFT} we derive the DMFT equations that describe the dynamics of the dynamical systems in the infinite dimensional limit. We detail a derivation based on simple field theory techniques and also sketch how the same result can be obtained via the dynamical cavity method \cite{mezard1987spin}. We show how the DMFT equations can be extended to discrete time dynamical systems and we discuss how to integrate them numerically and efficiently.

In Sec. \ref{sec_steady_State} we study the steady state dynamics of the class of dynamical systems that we consider. The models we focus on exhibit a rich behavior with several order parameters and corresponding dynamical phases. We describe their phenomenology and the corresponding phase diagrams. 
We show that, in some cases, one can have dynamical systems with co-existing multiple chaotic attractors that are localized on different regions of phase space. 

In Sec. \ref{sec periodic drive} we discuss what happens to the steady state dynamics when the dynamical systems are subjected to incoherent periodic drive. We show that the collective response of these systems act as filters/resonators: when the amplitude and frequency of the perturbations are in some precise interval, the resulting steady state dynamics is a limit cycle.

In Sec. \ref{Sec_MLE} we develop the theoretical apparatus to compute the maximal Lyapunov exponent. We show that in several cases, this quantity can be computed exactly in the whole chaotic phase and we discuss how it behaves across phase transitions.

Through the presentation we validate our results via numerical simulations and show how the infinite dimensional results are approached at finite size. 
We conclude the manuscript with a discussion on what are the most promising applications of the results that we present.

\section{Models}\label{sec_models}
The main goal of this work is to study several properties of high-dimensional chaotic systems in a particular class of models that are particularly simple but retain all the essential features of more standard models, such as the ones studied in random recurrent neural networks.

\subsection{A large class of models}
The class of dynamical systems we focus on are described by a set of $N$ degrees of freedom $\bx=\{x_1,\ldots, x_N\}\in \mathbb{R}^N$ evolving according to the dynamical equations
\begin{equation}
    \partial_t x_i(t) = -\mu(t) x_i(t) + g r_i(\bx(t)) +  \frac{J_0}{N} \sum_j x_j(t) + h \equiv \EE_i(\bx(t)) \:.
    \label{generic_eq}
\end{equation}
While we mostly focus on continuous time dynamical systems, we may also consider the Euler discretized version of Eq.~\eqref{generic_eq}
\begin{equation}
    x_i(t+\de t) = x_i(t)+\de t\left[ -\mu(t) x_i(t) + g r_i(\bx(t)) +  \frac{J_0}{N} \sum_j x_j(t) + h\right]=x_i(t)+\de t \EE_i(\bx(t))\:.
    \label{discrete_time}
\end{equation}
where $\de t$ is just the timestep of the time discretization. 
We assume that the initial condition of Eq.~\eqref{generic_eq} at $t=0$ is extracted at random from a factorized probability distribution
\begin{equation}
    P_0(\bx) = \prod_{i=1}^N p_0(x_i)\:.
\end{equation}
The forcing fields $r_i$ are the generators of the chaotic behavior and are assumed to be Gaussian processes with covariance structure given by
\begin{equation}
\begin{split}
    \mathbb{E}[r_i(\bx)]&=0 \ \ \ \ \ \forall i=1,\ldots, N\\
    \mathbb{E}[r_i(\bx)r_j(\by)]&=\delta_{ij} f(\bx\cdot \by/N) \ \ \ \ \ \forall i,j=1,\ldots, N
\end{split}
\label{cov_r_i}
\end{equation}
In Eq.~\eqref{cov_r_i}, we have assumed that $f$ admits a Taylor expansion with positive semidefinite Taylor coefficients.
As soon as $f$ is non-linear, the corresponding dynamical system is non-linear.
We call \emph{pure} models the ones for which $f(z)$ is a pure monomial. Conversely, we address the corresponding models as \emph{mixed}. This terminology is directly derived from spin glasses \cite{nieuwenhuizen1995exactly,crisanti2004spherical,  chen2018tap,dembo2020dynamics} where the random fields $\br$ are actually the gradient of a spin glass Hamiltonian and Eq.~\eqref{generic_eq} describes the gradient flow dynamics on the corresponding energy landscape.
The control parameter $g$ sets the strength of the non-linear forcing of the dynamics.
We assume, without losing generality, that $f$ does not contain a constant term. Indeed this case would be equivalent with the model with an additional external field.
Furthermore, we assume that $f$ is not a linear function. This case corresponds to having a dynamical system that is integrable, see \cite{crisanti1987dynamics, stariolo2025zero}. 
We are interested in non-linear cases that cannot be understood in terms of an extensive number of uncoupled/weakly coupled degrees
of freedom.

The term proportional to $\mu(t)$ in Eq.~\eqref{generic_eq} is added to enforce that the forcing fields $\underline r$ do not lead to a divergence of the dynamics and therefore we call it the confining term.
The form of $\mu(t)$ can be rather arbitrary apart from two constraints:
\begin{itemize}
    \item We impose that $\mu(t)$ is a function of $C(t,t)=|\bx(t)|^2/N$ only. We call this function $\hat \mu(z)$ so that $\mu(t)=\hat \mu(C(t,t))$.
    \item The function $\hat \mu(z)$ can be non-convex. However, we assume that $\hat \mu(z)$ diverges sufficiently fast for $z\to \infty$ in such a way that the corresponding dynamical system is confined. This degree of divergence has to be sufficiently large to compensate the wild fluctuations of the forcing fields $\br$. In other words, if the maximal degree of $f$ is $d_f$, we need to have that $\hat\mu(z)$ diverges to infinity as $z^\nu$ with $\nu\geq \frac 12 (d_f-1)$, \cite{fournier2023statistical}. 
\end{itemize}
The last term in Eq.~\eqref{generic_eq} is a ferromagnetic term of strength proportional to $J_0$. This term can induce ferromagnetic phases on the model defined as regions of the phase diagrams where the magnetization
\begin{equation}
m(t)=\frac 1N \sum_ix_i(t)    
\end{equation}
is non zero on average. Correspondingly, we have also introduced an external field $h$, through which we can probe hysteresis loops in the phase diagram of the model.
The external field $h$ is assumed to be constant and can model input currents in a recurrent neural network. 
The extension to time varying external fields is straightforward and it is considered in Sec.~\ref{sec periodic drive}.

\subsubsection{Initial condition of the dynamical systems}\label{sec_init}
We assume that $p_0(z)$ is a Gaussian distribution such that
\begin{equation}
\begin{split}
    \int_{-\infty}^\infty \de z p_0(z)z^2&=\Ctt\\
    \int_{-\infty}^\infty \de z p_0(z)z&=\mtt\:.
\end{split}
\end{equation}
This implies that 
\begin{equation}
    \begin{split}
        \lim_{N\to \infty}\frac 1N \sum_{i=1}^Nx_i(0)^2 &= \Ctt = C(0,0)\\
        \lim_{N\to \infty}\frac 1N \sum_{i=1}^Nx_i(0) &= \mtt = m(0)\:.
    \end{split}
\end{equation}

\subsubsection{The simplest model}\label{sec_simplest}
A particularly simple choice for the dynamical system in Eq.~\eqref{generic_eq} is given by
\begin{equation}
  \partial_t x_i(t) = -\mu(t) x_i(t) +g\left[ \frac{g_1}{\sqrt N} \sum_{j=1}^N J^{(1)j}_i x_j(t) +\frac{g_2}{N} \sum_{j,k=1}^N J^{(2)jk}_i x_j(t) x_k(t)\right] + \frac{J_0}{N} \sum_j x_j(t) + h
  \label{def_simplest}
\end{equation}
where we have chosen
\begin{equation}
    \begin{split}
        r_i(t) &=  \frac{g_1}{\sqrt N} \sum_{j=1}^N J^{(1)j}_i x_j(t) +\frac{g_2}{N} \sum_{j,k=1}^N J^{(2)jk}_i x_j(t) x_k(t)
    \end{split}
\end{equation}
The vectors $\bJ^{(1)}_i$ and matrices $\bJ_i^{(2)}$ are respectively \emph{i.i.d.} random vectors and symmetric random matrices. Their entries have Gaussian statistics, namely
\begin{equation}
\begin{split}
    &\E [J_i^{(1)j}]=0\;\;\;\;\; \E[J_i^{(1)j}J_i^{(1)k}]=\delta_{jk}\\
    &J_i^{(2)jk} = J_i^{(2)kj} \;\;\;\;\; \E[{J_i^{(2)jk}}]=0 \;\;\;\;\; \E[{(J_i^{(2)jk})^2}]=1+\delta_{jk}\:.
\end{split}
\end{equation}
This implies that for this particular case we have $f(z) = g_1^2z+2 g_2^2 z^2$.
The form of the confining term is rather generic as far as it is a function of the norm of the vector $\bx$ only. 
Note that $g$ controls the strength of the linear drive represented by the Gaussian process $\br$ while $g_1$ and $g_2$ control the relative intensity of the linear and non-linear part in the drive $\br$. One of these three coupling constants is redundant but we keep all of them for clarity.
This model has been recently used in \cite{fournier2023statistical, fournier2025non} to investigate both the landscape-dynamics correspondence in chaotic high-dimensional systems as well as how to track specific training algorithms for recurrent neural networks.

\paragraph{The $g_2=0$ case --} If $g_2=0$ or, equivalently, when $f(z)$ is a linear function, the dynamical system is still non-linear in general, due to the non-linearities that may define the confining term in the dynamics. However, the corresponding models are integrable because one can construct $N$ explicit integrals of motion by going to the right and left eigenvectors of $\bJ^{(1)}$. This is similar to the case of the spherical 2-spin glass model \cite{cugliandolo1995full} where the corresponding Langevin dynamics can be integrated efficiently using tools in random matrix theory. Therefore we dub the corresponding model the \emph{integrable model}. This model is interesting also from a mathematical point of view given that the underlying integrability could be use to characterize the finite size corrections to the dynamics, as for example was done for the 2-spin spherical spin glass \cite{fyodorov2015large}. However, in the following, we want to focus on dynamical systems for which the interactions between the degrees of freedom is non-linear and therefore we will always set $g_2>0$.

In the following, we consider the model in Eq.~\eqref{def_simplest} and investigate the phase diagrams as a function of the control parameters.


\section{Dynamical mean field theory}\label{sec_DMFT}
In this section, we analyze the dynamical system in Eq.~\eqref{generic_eq} in the $N\to \infty$ limit. 
In this case a dimensional reduction allows to go from a high-dimensional set of ordinary differential equations (or discrete time equations) to either a self-consistent stochastic process for one degree of freedom or a set of integro-differential non-linear PDEs. 
We generically refer to the resulting equations as DMFT equations.

\subsection{Derivation of the DMFT equations}
Although there are many ways to derive the DMFT equations, here we follow a route typical in statistical physics of out of equilibrium systems and transform the dynamical equations in a path integral \cite{martin1973statistical,de1978field, cugliandolo2002dynamics, zinn2021quantum}. In the large $N$ limit, this path integral can be evaluated via a saddle point method. The corresponding saddle point equations are the DMFT equations.

\subsubsection{Path integral formalism}\label{path_integral_derivation}
We introduce a dynamical partition function obtained from an integral over possible trajectories of the system
\begin{equation}
\begin{split}
  Z[\hat{\bb}, \bb] &= \int \de \bx_0 P(\bx_0)\int_{\bx(0)=\bx_0} \mathcal{D}\bx \mathcal{D}\hat{\bx}\, e^{-S[{\bx}, {\hat{\bx}}] + i \sum_{k=1}^N\int_{t\geq 0}\de t\left[ \hat{x}_k(t) b_k(t) + \hat{b}_k(t) x_k(t)\right]}
\end{split}\label{dyn_Z}
\end{equation}
where the function $S$ is given by
\begin{equation}
  S[\bx, \hat\bx] = i\sum_{k=1}^N \int_{t\geq 0}\de t \hat{x}_k(t) \left( \partial_t x_k(t) -\EE_k(\bx(t)) \right)
\end{equation}
and plays the role of a dynamical action.
The dynamical partition function in Eq.~\eqref{dyn_Z} has the remarkable property that $Z[0,\bb]=1$. This follows from the causality of the dynamics. The external fields $\bb$ and $\hat \bb$ are introduced with the purpose of being external sources which can be used to compute dynamical correlation functions.

The dynamical partition function in Eq.~\eqref{dyn_Z} is a random quantity that depends on the specific realization of the random fields $\br$ and therefore we compute its average. The only term entering in $Z$ that has to be considered in this regard is
\begin{align}
\E\exp\left[i \int_t \hat{x}_i(t)r_i(\bx(t))\right] = \exp\left[-\frac{g^2}2 \int_{t,t'} \hat{x}_i(t) f(C(t,t')) \hat{x}_i(t')\right]\:.
\label{av_disor}
\end{align}
In Eq.~\eqref{av_disor} we have introduced the dynamical correlation function $C(t,t')$ defined as
\begin{equation}
  C(t,t') = \frac1N \sum_{j=1}^N x_j(t)x_j(t')\:.
\label{def_C}
\end{equation}
After averaging over the Gaussian random fields, 
the degrees of freedom are decoupled except for the the ferromagnetic term. 
To take this into account we introduce the magnetization defined as
\begin{equation}
m(t) = \frac 1N \sum_{i=1}^N x_i(t)\:.
\label{def_m}
\end{equation}
The definitions in Eq.~\eqref{def_C} and \eqref{def_m} can be enforced in the path integral by introducing auxiliary variables, $\hat C$ and $\hat m$ so that the dynamical partition function can be rewritten as
\begin{equation}
  \E{Z}[\hat{\bb}, \bb] = \int_{C(0,0)=\Ctt; m(0)=\mtt} \mathcal{D}C \mathcal{D}\hat{C} \mathcal{D}m \mathcal{D}\hat{m} \, e^{-N \mathcal{L}[{\bb}, {\hat{\bb}}, C, \hat{C}, m, \hat{m}]}
  \label{Z_order_param}
\end{equation}
where
\begin{equation}
\begin{split}
  \mathcal{L}[{\bb}, {\hat{\bb}}, C, \hat{C}, m, \hat{m}] &=i \int_{t,t'\geq 0}\de t\de t'\hat{C}(t,t')C(t,t') + i\int_{t\geq 0}\de t \hat{m}(t)m(t) - W[{\bb}, {\hat{\bb}}, C, \hat{C}, m, \hat m]\\
  W[{\bb}, {\hat{\bb}}, C, \hat{C}, m, \hat{m}] &= \frac1N \sum_k \ln \int \mathcal{D}x_k \mathcal{D}\hat{x}_k \, e^{-S_k[x_k, \hat{x}_k, C, \hat{C}] + \int_{t\geq 0}\de t\left[ i\hat{x}_k(t) b_k(t) + i\hat{b}_k(t) x_k(t)\right]} \\
  &\equiv \frac1N \sum_{k=1}^N \ln \mathcal{Z}_k[{b_k}, {\hat{b}_k}, C, \hat{C}, m, \hat{m}]\\
    S_k[x_k, \hat{x}_k, C, \hat{C}, m, \hat{m}] &= \int_{t\geq 0} \de t\,  \left(i\hat{x}_k(t) \left[ \partial_t x_k(t) + \hat \mu(C(t,t)) x_k(t) - J_0m(t) - h \right] -  i\hat{m}(t) x_k(t)\right)\\
    &- \int_{t,t'\geq 0}\de t \de t' \left( -\frac{g^2}2 \hat{x}_k(t) f(C(t,t')) i\hat{x}_k(t') + x_k(t) \hat{C}(t,t') x_k(t') \right)\:.
\end{split}
\end{equation}
Note that in Eq.~\eqref{Z_order_param} the initial condition for the dynamics is explicitly solved by imposing a constraint on the correlation function and the magnetization at initialization.

In the $N\to\infty$ limit, $\E{Z}[0, \bb]$ can be evaluated through the saddle point method. At the saddle point,  ${\cal L}=0$ because the dynamical partition function equals 1.
The corresponding saddle point equations are
\begin{equation}
    \begin{split}
    \frac{\delta \mathcal{L}}{\delta C(t,t')} &= i\hat{C}(t,t') +\frac{\hat \mu'(C(t,t))}N \sum_{k=1}^N \langle i\hat{x}_k(t) x_k(t) \rangle_{\mathcal{Z}_k^{(0)}}\\ 
    &-\frac{g^2}{2N} \sum_{k=1}^N f'(C(t,t')) \langle i\hat{x}_k(t) i\hat{x}_k(t') \rangle_{\mathcal{Z}_k^{(0)}} = 0
    \end{split}
\end{equation}
\begin{equation}
\frac{\delta \mathcal{L}}{\delta i\hat{C}(t,t')} = C(t,t') - \frac 1N \sum_{k=1}^N\langle x_k(t) x_k(t') \rangle_{\mathcal{Z}_k^{(0)}} = 0\label{SP_C}
\end{equation}
\begin{equation}
    \frac{\delta \mathcal{L}}{\delta m(t)} = i\hat{m}(t) -\frac{J_0}{N}\sum_{k=1}^N \langle i\hat{x}_k(t) \rangle_{\mathcal{Z}_k^{(0)}} = 0
\end{equation}
\begin{equation}
    \frac{\delta \mathcal{L}}{\delta i\hat{m}(t)} = m(t) - \frac1N \sum_{k=1}^{N} \langle x_k(t)\rangle_{\mathcal{Z}_k^{(0)}} = 0
    \label{eq:m_SP}
\end{equation}
where we have indicated with $\mathcal{Z}^{(0)}_k= \mathcal{Z}_k[0,0, C, \hat{C}, m, \hat{m}]$ {evaluated at the saddle point}.
Causality of the dynamics imposes that  $\langle i\hat{x}_k(t) i\hat{x}_k(t') \rangle_{\mathcal{Z}_k^{(0)}} =0$ and $\langle i\hat{x}_k(t) x_k(t) \rangle_{\mathcal{Z}_k^{(0)}} = 0$. This implies that $\hat{C}(t,t')=0$. On the same lines, causality implies that $\langle \hat{x}_k(t)\rangle_{\mathcal{Z}_k^{(0)}}=0$ and therefore $\hat m(t)=0$.
Therefore the saddle point equations reduce to Eqs.~\eqref{SP_C} and~\eqref{eq:m_SP}.

\subsubsection{Self-consistent stochastic process}
The crucial observation is that $\mathcal{Z}_k^{(0)}=1$ for $\bb,\hat\bb=0$. This means that $\mathcal{Z}_k^{(0)}$ describes the dynamical partition function of an autonomous stochastic process.
This is given by
\begin{equation}
  \partial_t x(t) = -\mu(t) x(t) + \eta(t) + J_0 m(t) + h
  \label{eq: effective single-site dynamic}
\end{equation}
where the noise $\eta(t)$ is a Gaussian process with zero mean and variance given by
\begin{equation}
    \langle \eta(t) \eta(t') \rangle_\eta = g^2f(C(t,t'))\:.
    \label{self_consisten_noise}
\end{equation}
The confining term in Eq.\eqref{eq: effective single-site dynamic} is given by $\mu(t) = \hat \mu(C(t,t))$.
Eqs.~\eqref{SP_C} and \eqref{eq:m_SP} imply that
\begin{equation}
\begin{split}
    C(t,t') &= \langle x(t)x(t')\rangle_{\eta}\\
    m(t)&=\langle x(t)\rangle_{\eta}
\end{split}
\label{eq: closure}
\end{equation}
Eqs.~\eqref{eq: effective single-site dynamic}-\eqref{eq: closure} are the dynamical mean field theory equations.
Therefore, solving the original dynamical system of ODEs for $N\to \infty$ corresponds to solving the self-consistent stochastic process in Eq.~\eqref{eq: effective single-site dynamic}. 
In principle, this is a complex task to do: a standard solver would start from a guess for the correlation function $C(t,t')$ and the magnetization $m(t)$, sample the stochastic process and use Eq.~\eqref{eq: closure} to update the guess for the correlation function \cite{eissfeller1992new, mignacco2020dynamical}. This numerical procedure works in practice but the sampling step makes the algorithm slow and limits the time window where the numerical integration can be carried out. {Crucially, this sampling step will not be needed for the family of simple models considered in this work (see below).}

\subsubsection{The self-consistent stochastic process via the dynamical cavity method}
In the previous section we have shown that in the high-dimensional limit, the dynamics can be reduced to the study of a self-consistent stochastic process. We arrived to this conclusion by transforming the dynamical system into a path integral and showing that in the $N\to \infty$ limit, this is dominated by a saddle point. In this section, we want to emphasize that the same conclusions can be reached in a equally simple way by using the dynamical cavity method \cite{mezard1987spin}. 
The idea of the derivation can be summarized as follows. 
Consider a system with $N$ degrees of freedom and add an additional one, call it $x_0(t)$.
Correspondingly, generate the corrections to the random processes $\br $. For example, in the case of the simplest dynamical system of Sec.\ref{sec_simplest}, generate the additional couplings $\{J_{0}^{(1)i},J_{i}^{(1)0}\}_{i=1,\ldots,N}$ and the corresponding ones in the tensor $J_{i}^{(2)jk}$.

The new extended dynamical system gets modified in the following way.
First there is an additional equation for $x_0$ given by
\begin{equation}
\begin{split}
    \frac{\de x_0(t)}{\de t} &= -\mu(t)x_0(t)+r_0(\bx(t))+\frac 1{N+1}\sum_{i=0}^{N}x_i(t)+h\\
    &\simeq -\mu(t)x_0(t)+r_0(\bx(t))+\frac 1{N}\sum_{i=1}^{N}x_i(t)+h
\end{split}
\end{equation}
where we neglected corrections of order $1/N$.
The random field $r_0$ is uncorrelated with $r_{i>0}$. Therefore the dynamics of the additional spin $x_0$ coincides with the one of the self-consistent stochastic process in Eq.~\eqref{eq: effective single-site dynamic}. To conclude the argument, it is very easy to show that the correction to the dynamics of the remaining degrees of freedom $x_{i>0}$ due to the addition of $x_0$ are subleading for $N\to \infty$.

\subsubsection{Correlation and response functions}\label{DMFT_C_R_m}
While in generic dynamical systems such as standard recurrent neural networks \cite{sompolinsky1988chaos} the DMFT equations reduce to the self-consistent stochastic process, for the class of models that we study, one has additional simplifications.
Taking the average of Eq.~\eqref{eq: effective single-site dynamic} with respect to $\eta$ we get
\begin{equation}
    \partial_t m(t) = -\mu(t) m(t) + J_0 m(t) + h\:.
    \label{eq_final_m}
\end{equation}
Instead, multiplying  Eq.~\eqref{eq: effective single-site dynamic} by $x(t')$ and averaging over $\eta$ we get that for $t\geq t'$
\begin{equation}
      \partial_t C(t,t') = -\mu(t) C(t,t') + g^2\int_0^{t'}ds\, f(C(t,s)) R(t',s) + \left[ J_0 m(t) + h \right]m(t')\:.
\label{eq_final_C}
\end{equation}
In Eq.~\eqref{eq_final_C} we have defined the response function given by
\begin{equation}
    R(t,t')=  \left\langle \frac{\delta x(t)}{\delta \eta(t')} \right\rangle_\eta \:.
\end{equation}
The definition of $R$ can be used to derive the corresponding equation
\begin{equation}
    \partial_t R(t,t') = -\mu(t) R(t,t') + \delta(t-t')\:.
    \label{eq_R}
\end{equation}
Note that causality implies that $R(t,t')=0$ for $t\leq t'$.
Eq.~\eqref{eq_R} is linear and therefore can be integrated explicitly. Its solution is
\begin{equation}
    R(t,t') =\theta(t-t')\exp\left[-\int_{t'}^t\de s \mu(s)\right]\:.
\end{equation}
Eq.~\eqref{eq_final_C} allows also to derive an equation for the evolution of $C(t,t)$. Indeed we get
\begin{equation}
    \frac {\de C(t,t)}{\de t} = 2 \left[-\mu(t) C(t,t) + g^2\int_0^{t}ds\, f(C(t,s)) R(t,s) + \left[ J_0 m(t) + h \right]m(t')\right]\:.
    \label{eq_final_Cd}
\end{equation}
Finally, the initial condition for the correlation function and the magnetization are
\begin{equation}
    \begin{split}
        C(0,0)&=\Ctt\\
        m(0)&=\mtt\:.
    \end{split}
\end{equation}
This concludes the derivation of the DMFT equations.

It is important to stress that for this particular class of models the self-consistent stochastic process can be solved efficiently by solving numerically Eqs.~\eqref{eq_final_m}, \eqref{eq_final_C} and \eqref{eq_final_Cd}.
The reason why this simplification appears is related to two facts:
\begin{itemize}
    \item the forcing fields $\br$ have Gaussian statistics;
    \item the covariance structure of the forcing fields can be expressed explicitly in terms of the dynamical correlation function $C$.
\end{itemize}

\subsection{Discrete time dynamical system}
It is useful to detail how the analysis carried out up to now can be extended to the discrete time dynamical system in Eq.~\eqref{discrete_time}.
It is easy to show that the self-consistent stochastic process in Eq.~\eqref{eq: effective single-site dynamic} can be replaced by
\begin{equation}
     x(t+\de t) =x(t)+\de t\left[ -\mu(t) x(t) + \eta(t) + J_0 m(t) + h\right]
  \label{eq: effective single-site dynamic_dt}
\end{equation}
Taking the same steps as before, we obtain the following equations for the evolution of correlation and response function
\begin{equation}
    \begin{split}
        C(t+\de t,t') &= C(t,t') + \de t \left[ - \mu(t) C(t,t') + g^2\de t \sum_{n=0}^{t'/\de t} f(C(t,n\de t))R(t',n\de t) \right.\\
        &\left. + (J_0 m(t)+h) m(t')\right]\ \ \ \ \ \ \ t>t'\\
        C(t+\de t,t+\de t) &= C(t,t)+ 2\de t \left[- \mu(t) C(t,t) + g^2\de t\sum_{n=0}^{t/\de t} f(C(t,n\de t))R(t,n\de t)\right] \\
        &+ 2\de t (J_0 m(t)+h) m(t)+\de t^2 L(t)\\
        R(t+\de t,t')&=R(t,t')-\de t \mu(t)R(t,t') + \delta_{t,t'}\\
        \mu(t) &= \hat \mu(C(t,t))\\
        m(t+\de t)&=m(t) + \de t\left[(J_0- \mu(t))m(t)+h\right]\\
        L(t)&= \mu^2(t) C(t,t) +(J_0  m(t)+h)^2-\mu(t)(J_0 m(t)+h)m(t)\\
        &- 2g^2  \mu(t)\sum_{n=0}^{t/\de t}f(C(t,n\de t))R(t,n\de t) {+ g^2f(C(t,t))}
    \end{split}
    \label{DMFT_discrete_C_R}
\end{equation}
The correction term given by $L(t)$ contributes by a quantity of order $\de t^2$ in the final dynamical equations. This term represents the finite time discretization correction to the dynamics and it becomes irrelevant in the $\de t\to 0$ limit.

Eqs.~\eqref{DMFT_discrete_C_R} track the dynamics for any finite value of $\de t$ and are exact in the large $N$ limit.
In the $\de t\to 0$ case they reduce to the non-linear PDEs derived in the previous section.
Therefore, for $\de t$ sufficiently small these equations are an approximation of the numerical solution of the continuous time equations.

\subsection{Numerical integration of the DMFT equations}
The DMFT equations of the dynamical correlation function and the magnetization can be efficiently solved numerically.
The simplest way is to iterate Eqs.~\eqref{DMFT_discrete_C_R}. For small values of $\de t$ they either give the exact evolution for the discrete time case or a very good approximation of the continuous time case.
The numerical solution of these equations is particularly easy because they have a causal structure: for any time $t$, the rhs of the equations for the correlation and response function depends only  on quantities defined at times smaller than $t$. Therefore the full profile of $C$, $R$ and $m$ can be constructed step-by-step in time.

\section{Steady state dynamics}\label{sec_steady_State}
In this section we report the phase diagram of the class of models defined in Sec.~\ref{sec_models}.
The derivation relies entirely on the analysis of the DMFT equations.
Given that the systems that we consider are out of equilibrium, it is important to make precise that the phase diagrams that we are interested in refer to the non-equilibrium stationary states of the dynamical systems themselves. In other words, we are interested in understanding the behavior of the dynamical systems at long times, as a function of the control parameters of the models, namely the strength $g$ of the non-linear forcing, the strength $J_0$ of the ferromagnetic interactions, and the amplitude of the external field $h$.

The properties of the steady state dynamics can be obtained following a route first developed in the context of random recurrent neural networks \cite{sompolinsky1988chaos}. 
However, we show that some instances of the models that we consider, in particular the simplest model of Sec.\ref{sec_simplest}, contain additional simplifications which allow to make further progress and obtain the exact expressions for the steady state form of the dynamical correlation function.

To simplify the derivation, in this section, we  mainly focus on continuous time dynamical systems.

\subsection{Phenomenology of steady states}
Before discussing the stationary solutions of the DMFT equations, it is convenient to describe how they look like when obtained from the numerical integration of the DMFT equations.
We find that the simple models that we consider, when not subjected to periodic time-dependent external fields, can give rise to two types of asymptotic solutions: fixed points and chaotic attractors.
In this section we discuss this phenomenology in a set of simple examples. This allows to understand the behavior of the dynamical correlation functions. 

\subsubsection{Fixed points}
\begin{figure}
\centering
\includegraphics[scale=0.645]{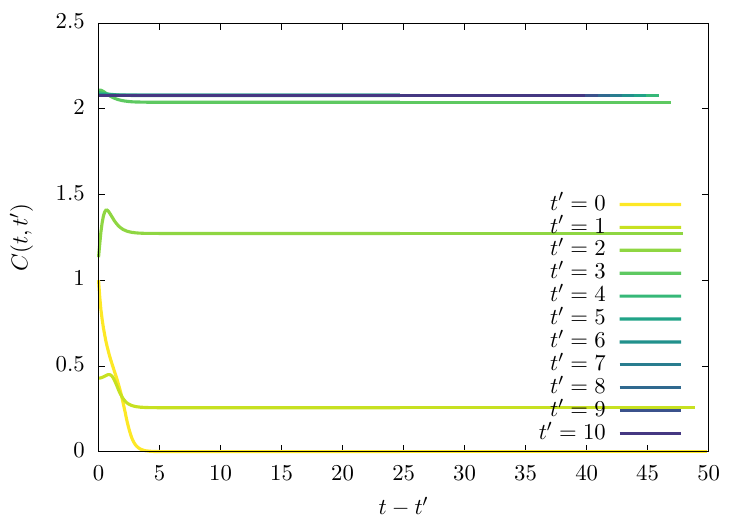}
\includegraphics[scale=0.645]{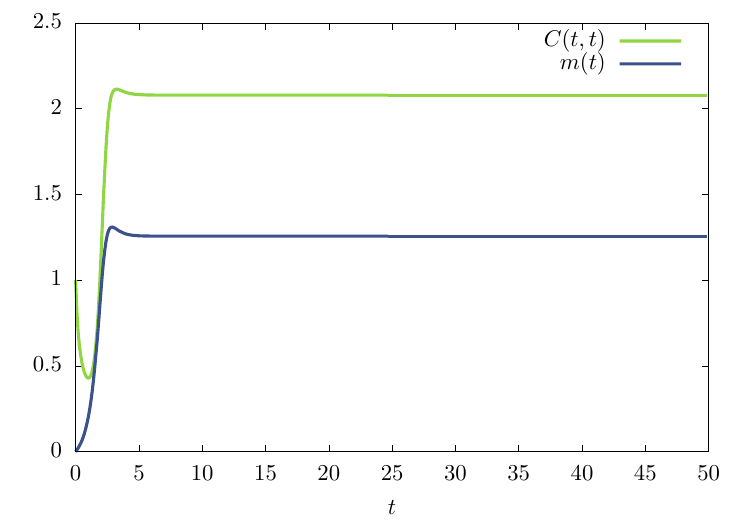}
\caption{Phenomenology of fixed point attractors. The results of the numerical integration of the DMFT equations for $g=0.5$, $g_1=0$, $g_2=1$, $J_0=2$, $h=0.1$. The initial condition for the dynamics is $\Ctt=1$ and $\mtt=0$. We used an integration timestep $\de t=0.1$ and integrate the corresponding discrete time DMFT equations. Left panel: the dynamical correlation function as a function of $t-t'$ for increasing values of $t'$. For large enough $t'$, $C$ becomes a constant. Right panel: the behavior of $C(t,t)$ and $m(t)$. Both quantities reach a constant asymptotic value.}
\label{fig_fixed_point_pheno}
\end{figure}

Fixed points are stationary solutions characterized by the following asymptotic behavior of the dynamical order parameters
\begin{equation}
    \begin{split}
        \lim_{t'\to \infty} C(t>t',t')&=\tilde C\\
        \lim_{t\to \infty}m(t)&=m_\infty\:.
    \end{split}\label{FP_generic}
\end{equation}
The first of Eqs.~\eqref{FP_generic} says that for $t,t'\to \infty$, the correlation function $C(t,t')$ becomes a constant independent of time. 
This implies that for $t\to \infty$, $\bx(t)\to\bx^*$ being $\bx^*$ the fixed point. The asymptotic value $\tilde C$ gives the square of the length of the vector $\bx^*$. The second of Eqs.~\eqref{FP_generic} instead provides the magnetization of the fixed point vector $\bx^*$.
In Fig.\ref{fig_fixed_point_pheno} we show an example of the behavior of the order parameters in the case in which the dynamics reaches a fixed point.

\subsubsection{Chaotic attractors}
\begin{figure}
\centering
\includegraphics[scale=0.645]{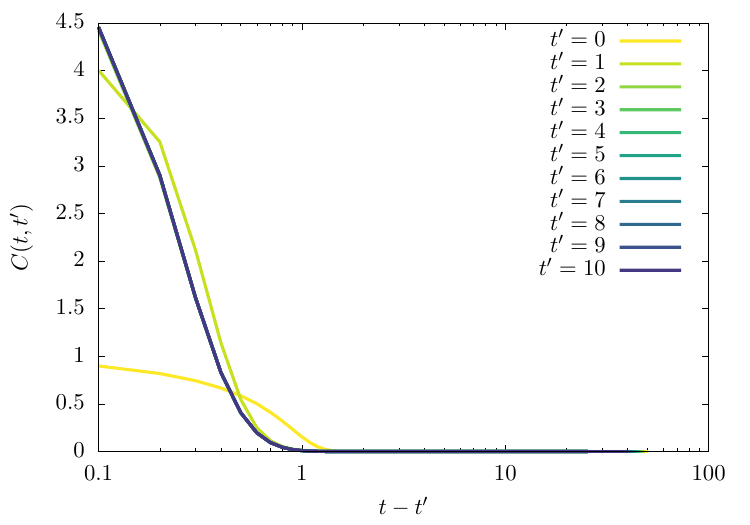}
\includegraphics[scale=0.645]{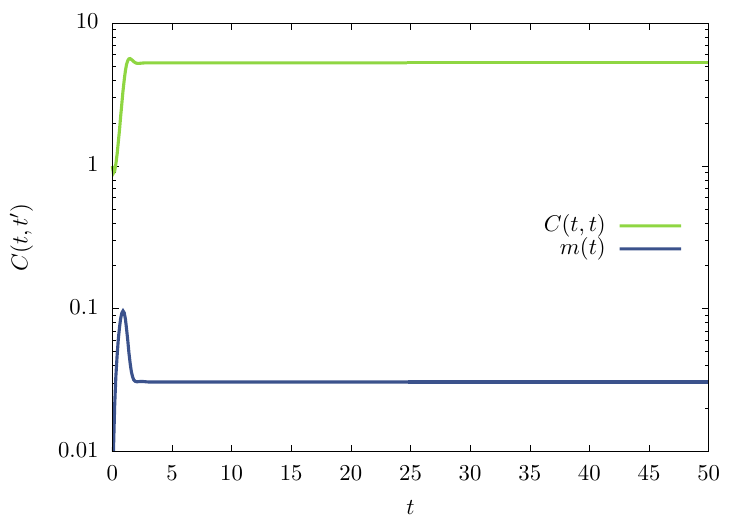}
\caption{Phenomenology of chaotic attractors. The results of the numerical integration of the DMFT equations for $g=2$, $g_1=0$, $g_2=1$, $J_0=2$, $h=0.1$. The initial condition for the dynamics is $\Ctt=1$ and $\mtt=0$. We used an integration timestep $\de t=0.1$ and integrate the corresponding discrete time DMFT equations. Left panel: the dynamical correlation function as a function of $t-t'$ for increasing values of $t'$. For large enough $t'$, $C(t,t')$ becomes a function of $t-t'$ only. Right panel: the behavior of $C(t,t)$ and $m(t)$. Both quantities reach a constant asymptotic value.}
\label{fig_chaos_point_pheno}
\end{figure}

When the non-linear drive in the dynamics becomes sufficiently strong, there is no fixed point that is reached at long times  and  the dynamics converges to a chaotic attractor. In this case, endogenous fluctuations of the degrees of freedom never stop and the dynamics reaches a steady state.
We find that the steady state is characterized by time-reversal symmetry and time translational invariance.
This means that
\begin{equation}
    \begin{split}
        \lim_{t'\to \infty;\,t-t'=\tau}C(t,t')&=c(\tau)\\
        c(\tau)&=c(-\tau)\\
        \lim_{t\to \infty}m(t)&=m_\infty
    \end{split}
    \label{Chaos_generic}
\end{equation}
The physical meaning of Eq.~\eqref{Chaos_generic} is that all one-time dynamical order parameters go to a stationary value that defines a shell in phase space where chaotic motion takes place. 
Instead, having a non-constant $c(\tau)$ implies that, at long times, the stationary state is dynamical: the degrees of freedom never stop fluctuating. However, their motion is time translational invariant on average. 
Note that fixed point solutions are a special case of chaotic attractors with constant $c(\tau)$.
In Fig.\ref{fig_chaos_point_pheno}, we plot an example of the behavior of the correlation function and magnetization as a function of time at a point in control parameter space where one finds a chaotic attractor. 
We note that the fact that $c(\tau)$ appears at long time does not automatically mean that the dynamical system is chaotic. Indeed, a standard definition of chaos involves the computation of the Lyapunov exponents. We postpone this analysis to Sec.\ref{Sec_MLE}.

The main goal of the next sections is to derive either the values of $\tilde C$ and $m_\infty$ in the case of the fixed point attractors, or the function $c(\tau)$ for chaotic attractors. We are also interested in understanding the phase transition between different phases and/or the possibility of having a coexistence of different attractors at the same point of the phase diagram.

\subsection{Steady-state solution of the DMFT equations}
The phase diagram describing the possible asymptotic states of the dynamical systems as a function of their control parameters can be obtained by looking for the steady state solutions of the DMFT equations. 
Given that, formally, fixed point solutions are special cases of the chaotic attractors, we look for solutions of the type described in Eq.~\eqref{Chaos_generic}.
It is convenient to introduce the following notation \begin{equation}
    \begin{split}
        C_0&=c(0)\\
        C_\infty&=\lim_{\tau\to \infty}c(\tau)\\
        \mu_\infty&=\hat \mu(C_0)\\
        r(\tau)&=\lim_{t'\to \infty; t-t'=\tau} R(t,t')
    \end{split}
\end{equation}
We first note that the equation for $R(t,t')$ guarantees that as far as for $t\to \infty$, $\mu(t)\to \mu_\infty$, the asymptotic function $r(\tau)$ is a pure exponential function, namely
\begin{equation}
    r(\tau)=e^{-\mu_\infty \tau}\ \ \ \ \tau>0\:.
\end{equation}
This implies that any integral where $R(t,t')$ appears in conjunction with a bounded kernel, can be evaluated only looking at the corresponding integrand at long times.
For example
\begin{equation}
    \lim_{t,t'\to\infty;\, t-t'=\tau}\int_0^{t'}\de s f(C(t,s))R(t',s) = \int_0^\infty \de s f(c(\tau+s))r(s)\:.
\end{equation}
This is because we expect that for small enough stepsize $\de t$, $C(t,t')$ is not diverging\footnote{Note that if $\de t$ is sufficiently large, even if the dynamical system is properly confined, one can have a situation where the dynamics self-excite and explodes at long times. This is due to the fact that if $\de t$ is sufficiently large, the dynamics jumps from one side to the other of the confining potential defined as $\int_0^C\de \hat c \hat \mu(\hat c)\hat c$. At each of this jumps the time derivative of each degree of freedom gets a larger and larger value and eventually this leads to an explosion of the dynamical system. In general, the explosion is not an instability of the steady state; rather, it happens at short time and therefore, to characterize it, one needs to study the transient behavior.}.
This implies that the asymptotic solution of the DMFT equations is described by
\begin{align}
\label{eq: steady-state of m}
&0 = (-\mu_{\infty} + J_0) m_{\infty} + h\\
&\partial_{\tau} c(\tau) = -\mu_{\infty} c(\tau) + g^2 \int_0^{\infty}ds\, f(c(\tau+s)) r(s) + \left[ J_0 m_\infty + h \right] m_\infty \label{steady_c_tau}\:.
\end{align}
Furthermore, in the steady-state, $c(\tau)=c(-\tau)$ which implies that $\partial_{\tau} c(\tau)
|_{\tau=0}=0$. Therefore, the equation for the dynamical correlation function can be simplified in two limiting cases
\begin{align}
\tau \to 0: \;\;\; 0 &= -\mu_{\infty} C_0 +  g^2 \int_0^{\infty}ds\, f(c(s)) e^{-\mu_{\infty}s} + \left[ J_0 m_\infty + h \right] m_\infty\\
\label{eq: C for tau to infinity}
\tau \to \infty: \;\;\; 0 &= -\mu_{\infty} C_{\infty} + g^2 \frac{f(C_{\infty})}{\mu_{\infty}} + \left[ J_0 m_\infty  + h \right] m_\infty.
\end{align}
These equations can be further simplified by taking an additional derivative of the Eq.~\eqref{steady_c_tau} with respect to $\tau$. This gives
\begin{equation}
\label{eq: second dmft of C}
(\mu_{\infty}^2 - \partial^2_{\tau}) c(\tau) = g^2 f( c(\tau)) + \left( J_0 m_{\infty} + h \right)^2\:.
\end{equation}
The main advantage of Eq.~\eqref{eq: second dmft of C} is that this is an autonomous ODE. Indeed, the dependence on $\tau$ is contained only in $c(\tau)$.
It follows that Eq.~\eqref{eq: second dmft of C} can be rewritten in terms of an effective potential $V(c;\mu_{\infty}, m_{\infty}, C_0)$, see \cite{sompolinsky1988chaos, fournier2025non}, so that
\begin{equation}
\label{eq: particle moving in a potential}
\partial^2_{\tau} c(\tau) = - \frac{\partial V}{\partial c},
\end{equation}
where
\begin{equation}
\begin{split}
  V(c;\mu_{\infty}, m_{\infty}, C_0) &= g^2 F(c) -\frac12 \mu_{\infty}^2 (c^2-C_0^2) + \left( J_0 m_{\infty} + h \right)^2 (c-C_0)\\
  F(c) &= \int_{C_0}^c\de \tilde c\, f(\tilde c)\:.
 \end{split}
 \label{def_eff_V}
\end{equation}
Note that the potential $V$ is fixed up to a constant. In Eq.~\eqref{def_eff_V} we fixed such constant so that $V(C_0; \mu_{\infty}, m_{\infty}, C_0)=0$.

Physical solutions $c(\tau)$ must be bounded $|c(\tau)| \leq C_0$. This is a reasonable assumption given that, in the stationary state, the forcing field tends to decorrelate configurations on the same shell $c(0)=C_0$. Furthermore, given that Eq.~\eqref{eq: particle moving in a potential} describes a conservative dynamical system, the corresponding solution must conserve the effective energy
\begin{equation}
E = \frac12 \left( \partial_{\tau} c(\tau) \right)^2 + V(c;\mu_{\infty}, m_{\infty}, C_0)\:.
\end{equation}
Given that $\partial_{\tau} c(\tau)|_{\tau=0} = \partial_{\tau} c(\tau)|_{\tau \to \infty} = 0$ we have that
\begin{equation}
    V(C_0;\mu_{\infty}, m_{\infty}, C_0) = V(C_{\infty};\mu_{\infty}, m_{\infty}, C_0)\:.
\end{equation}
Lastly, we can derive a relation between $C_0$ and $C_\infty$.
The chaotic or fixed point solution must land at stationary points in the landscape of $V$. Therefore, we expect that  
\begin{equation}
  \left. \frac{\partial V}{\partial c} \right\vert_{C_\infty} = 0 \:.
\end{equation}
This equation is nothing but Eq.~\eqref{eq: C for tau to infinity}.
Note that if the steady state solution is a fixed point attractor, then we must have that $C_0=C_\infty$ which implies that $V$ has to have a critical point also for $c=C_0$.

\paragraph{On limit cycles --} The analysis above suggests  that the DMFT equations could also accomodate limit cycles as stationary states.
Indeed, when chaotic attractors are present, the potential $V$ has at least a well. The chaotic trajectory is such that one starts from the appropriate initial condition $c(0)=C_0$ to land, for $\tau \to \infty$, in $C_\infty$. The potential $V(c)$ has at least a well for $c\in[C_\infty, C_0]$. Therefore the DMFT equations can also accomodate a periodic solution where the trajectory of $c(\tau)$ starts inside the well and oscillate with time, within the well. However, such solutions are unstable: an infinitesimal perturbation added to the dynamics makes the trajectory leave the limit cycle and land on the chaotic attractor. One way to see this is by performing the analysis in Sec.~\ref{Sec_MLE}. Otherwise, one can just integrate the DMFT equations at a given point in control parameter space, and discover, by direct inspection through the numerical integration of the DMFT equations, that limit cycles do not appear.  

\subsubsection{Summary of the asymptotic DMFT equations: \texttt{SteadyDMFT}}
We summarize the DMFT equations that we have to solve to obtain the phase diagram of the models. 

Both chaotic and fixed point attractors are defined by the following set of equations
\begin{equation}
\begin{split}
    0 &= (-\mu_{\infty} + J_0) m_{\infty} + h\\
    \mu_\infty&=\hat \mu(C_0)\\
    0&=V(C_0;\mu_\infty,m_\infty,C_0)-V(C_\infty;\mu_\infty,m_\infty,C_0)\\
    0&=\left.\frac{\partial V}{\partial c}\right|_{c=C_\infty}\:.
\end{split}
\label{Eq_C_NESS}
\end{equation}
For chaotic attractors we have $C_\infty<C_0$ while for fixed point attractors we have $C_\infty=C_0$.

It is important to stress that the potential $V$ is fixed self-consistently, namely it depends on $\mu_\infty$, $m_\infty$ and $C_0$ that have to be determined from the solution of the DMFT equations.
We call Eqs.~\eqref{Eq_C_NESS} the \texttt{SteadyDMFT} equations to underline that they describe the asymptotic steady state of the $N$-dimensional dynamical equations in the large $N$ limit.

It turns out that at any given point in control parameter space, namely for any value of $J_0$, $g$ and $h$, these equations can have multiple solutions. For example, one could have that a fixed point attractor coexist with a chaotic attractors and that they have disconnected basin of attractions. 
In the following, we will show that this can happen even in the simplest dynamical system of Eq.~\eqref{def_simplest}. We are interested in understanding where coexistence can take place and what is the limit of stability of the corresponding attractors.

\subsection{Steady state dynamics of the simplest model}
In this subsection we focus our analysis on the simplest dynamical system described by the set of ODEs in Eq.~\eqref{def_simplest}.
We recall that $\mu_\infty=\hat\mu(C_0)$. 
The effective potential reads
\begin{equation}
    V(c;\mu_{\infty}, m_{\infty}, C_0) = g^2\left(\frac {g_1^2}2(c^2-C_0^2) +\frac {2g_2^2}3  (c^3-C_0^3)\right) -\frac12 \mu_{\infty}^2 (c^2-C_0^2) + \left( J_0 m_{\infty} + h \right)^2 (c-C_0)\:.
\end{equation}
Furthermore, $m_\infty$ satisfies the equation
\begin{equation}
    0 = (-\mu_{\infty} + J_0) m_{\infty} + h\:.
\end{equation}
The \texttt{SteadyDMFT} equations also give
\begin{equation}
\begin{split}
    0&=g^2\left(\frac {g_1^2}2(C_0^2-C_\infty^2) +\frac {2g_2^2}3  (C_0^3-C_\infty^3)\right) -\frac12 \mu_{\infty}^2 (C_0^2-C_\infty^2) + \left( J_0 m_{\infty} + h \right)^2 (C_0-C_\infty)\\
    0&=\left( J_0 m_{\infty} + h \right)^2 -\left( \mu_\infty^2 C_\infty - g^2\left(g_1^2C_\infty +2g_2^2C_\infty^2\right)\right)
\end{split}
\end{equation}
The first equation comes from $V(C_0;\mu_\infty,m_\infty, C_0)=V(C_\infty;\mu_\infty,m_\infty, C_0)$ while the second one comes from the stationarity of $V$ at $c=C_\infty$.
These equations can be rewritten as
\begin{equation}
\label{eq: steady state for C0 fixed point or chaos}
     (C_0-C_\infty)^2\left[ g^2\left(\frac {g_1^2}2 + \frac{2 g_2^2}{3}(C_0+2C_\infty)\right)-\frac 12 \mu_\infty^2 \right] = 0\:.
\end{equation}
This equation admits two type of solutions. 

\paragraph{Fixed point solutions --}
Fixed points are characterized by $C_0=C_\infty$. Plugging this ansatz in the \texttt{SteadyDMFT} we get 
\begin{equation}
\label{eq: steady state eqs for fixed point solution}
    \begin{split}
        0 &= (-\mu_\infty + J_0) m_{\infty} + h\\
        0&=\left( J_0 m_{\infty} + h \right)^2 -\left( \mu_\infty^2 C_0 - g^2\left(g_1^2C_0 +2g_2^2C_0^2\right)\right)\:.
    \end{split}
\end{equation}
A numerical solution of these equations can be easily found.

\paragraph{Chaotic solutions --} In the case of chaotic attractors, $C_\infty<C_0$. 
Therefore the relevant set of equations read
\begin{equation}
\label{eq: steady state eqs for chaotic solution}
    \begin{split}
    0 &= (-\mu_\infty + J_0) m_{\infty} + h\\
    0&= g^2\left(\frac {g_1^2}2 + \frac{2 g_2^2}{3}(C_0+2C_\infty)\right)-\frac 12 \mu_\infty^2  \\
    0&=\left( J_0 m_{\infty} + h \right)^2 -\left( \mu_\infty^2 C_\infty - g^2\left(g_1^2C_\infty +2g_2^2C_\infty^2\right)\right)
    \end{split}
\end{equation}
Again, a numerical solution of these equations gives the corresponding chaotic phase.

\subsubsection{Expression for $c(\tau)$}
In the chaotic phase, additionally to $m_\infty$, $C_\infty$ and $C_0$, it could be interesting to derive the full profile of $c(\tau)$. This function starts at $C_0$ and asymptotically approaches $C_\infty$. 
Eq.~\eqref{eq: particle moving in a potential} fixes the form of $c(\tau)$ and we show that it can be solved explicitly. We assume that $C_0$ and $C_\infty$ have been already determined. The effective potential $V(c)$ has a single root in $c=C_0$ and a double root in $c=C_\infty$. Therefore it can be rewritten as
\begin{equation}
  {V}(c,\mu_\infty,m_\infty) = A \left(c-C_0\right)\left( c-C_\infty\right)^2
\end{equation}
where the constant $A=2(gg_2)^2/3$.
Energy conservation for the mechanical system in Eq.~\eqref{eq: particle moving in a potential} reads
\begin{equation}
\begin{split}
  &\partial_\tau c = \sqrt{-2{V}(c,\mu_\infty,m_\infty)} \\
  &\implies \tau = \int_{C_0}^{c} \frac{dC}{\sqrt{-2{V}(c,\mu_\infty,m_\infty)}} = \int_{C_0}^{c} \frac{dC}{\sqrt{2A \left(C_0-C\right)\left(C-C_\infty\right)^2}}
\end{split}
\end{equation}
so 
\begin{equation*}
  \begin{split}
    \tau &= \frac{1}{\sqrt{2AC_0}} \int_{1}^{c/C_0} \frac{du}{\left(u-\frac{C_\infty}{C_0}\right)\sqrt{1-u}} = - \frac{2}{\sqrt{2AC_0}} \int_{0}^{\sqrt{1-c/C_0}} \frac{dv}{1-\frac{C_\infty}{C_0}-v^2}\\
    &= -\frac{1}{\sqrt{2A(C_0-C_\infty)}} \int_{0}^{\sqrt{1-c/C_0}} \left[ \frac{dv}{\sqrt{1-\frac{C_\infty}{C_0}}-v} + \frac{dv}{\sqrt{1-\frac{C_\infty}{C_0}}+v} \right]\\
    &= -\frac{1}{\sqrt{2A(C_0-C_\infty)}} \ln \left( \frac{\sqrt{C_0-C_\infty}+\sqrt{C_0-c}}{\sqrt{C_0-C_\infty}-\sqrt{C_0-c}} \right)
  \end{split}
\end{equation*}
where we first made the change of variables $u=C/C_0$ and then $v=\sqrt{1-u}$. Inverting this relation, we get
\begin{equation}
  c(\tau) = C_0 - (C_0-C_\infty) \tanh^2 \left( \frac12 \sqrt{2A(C_0-C_\infty)}\, \tau \right)
  \label{solution_C_tau}
\end{equation}
which provides an explicit expression for the dynamical correlation function in the steady state.
It it interesting to note that the dependence on $g_1$ in Eq.~\eqref{solution_C_tau} is only through $C_0$ and $C_\infty$.

\section{Phase diagrams}
In the next sections, we extensively explore the phase diagram of the models as a function of the control parameters.
We focus on the simplest dynamical system of Eq.~\eqref{def_simplest}.

We generically denote by paramgnetic phases, steady states of the dynamics for which $m_\infty=0$. Instead, whenever $m_\infty\neq 0$, we denote such phases as ferromagnetic.
It is important to stress here that these attributes are not of equilibrium-thermodynamic origin: indeed, the dynamical system are not sampling any Gibbs measure.

\subsection{Absence of the external field}
We discuss the phase diagram in absence of the external field, $h=0$. We first determine the steady-states and then discuss the  transition lines between the different phases. Part of the analysis presented here follows and generalizes previous work, see \cite{fournier2025non}.

\subsubsection{Steady state phases}\label{sec_steady_state_absence_simplest}

\paragraph{Paramagnetic fixed point phase --}
In this case, the dynamical system in Eq.~\eqref{def_simplest} has always a paramagnetic fixed point $\bx=0$. This corresponds to $0=m_\infty=C_0=C_\infty$.
Such fixed point can be either stable or unstable. 
This depends on the control parameters in the model.
The simplest way to understand the stability of this fixed point is to linearize the dynamical system in Eq.~\eqref{def_simplest} around such point.
Calling $\delta \bx$ the displacement from the fixed point, for $\delta\bx$ small enough we have that the dynamical system in Eq.~\eqref{def_simplest} can be approximated by
\begin{equation}
    \frac{\de \delta \bx}{\de t} = \HH\delta\bx
    \label{linear_x_zero}
\end{equation}
where the matrix $\HH$ is defined as
\begin{equation}
    \HH_{ij}=-\hat \mu(0)\delta_{ij}+gg_1J^{(1)j}_i\:.
\end{equation}
The matrix $\HH$ is non-symmetric due to the non-reciprocal nature of the interactions between the degrees of freedom.
However its spectrum is known explicitly. In the large $N$ limit, the eigenvalues are uniformly distributed on the disk in the complex plane, centered at $-\hat \mu(0)$ and with radius $gg_1$. This implies that, with probability one for $N\to \infty$, as soon as $\hat\mu(0)<gg_1$, part of the spectrum of $\HH$ has eigenvalues with positive real part, implying that the linearized system in Eq.~\eqref{linear_x_zero} is unstable. Therefore, only for $\hat \mu(0)>gg_1$ the fixed point solution is a stable attractor of the dynamics in the high-dimensional limit.

It is important to stress that this analysis heavily relies on the existence of a non-vanishing $\HH$: this may be not always the case. In particular, as soon as both $\hat\mu(0)$ \emph{and} $g_1$ vanish, the analysis of the stability of the fixed point $\bx=0$ becomes non-linear. While this is a rather fine-tuned situation and we expect it to be rare among generic physical systems, we give a brief treatment of this case.

\paragraph{The case of $\hat \mu(z)=z$ and $g_1=0$ --} The linear stability analysis is not sufficient to determine whether in this case the fixed point $\bx =0$ is stable or not. 
One way to understand this point consists in initializing the dynamics of the system very close to $\bx=0$ and check whether the dynamics lands back on $\bx=0$ or not. 
Therefore, we consider the continuous time DMFT equations initialized with $\Ctt=\Ctt_0\ll 1$.
Let us denote the solution of the DMFT equations in this case as $C(t,t'; \Ctt_0, g)$ where we have indicated explicitly the dependence on the initial condition and the value of $g$.
It can be shown by direct inspection that $C(t,t'; \Ctt_0, g)=C(t \Ctt_0,t'\Ctt_0; 1, g/\sqrt{\Ctt_0})$. In other words, the solution of the DMFT equations initialized with arbitrarily positive $\Ctt_0$ can be obtained directly from the solution of the same equations for $\Ctt=1$ and a re-scaled coupling constant $g$.

Therefore rather than studying the solution of the equations for $\Ctt_0\ll 1$ we can fix $\Ctt=1$ and investigate what happens as $g$ is changed. We find that for any $g$ {(with $J_0=h=0$)}, the dynamics lands on a paramagnetic chaotic attractor.

\paragraph{Paramagnetic chaotic phase --}
The paramagnetic chaotic phase is characterized by a never ending motion with vanishing magnetization. In this case, $C_\infty=m_\infty=0$ and $C_0$ must be determined by solving the DMFT equations
\begin{equation}
    \frac12 (\hat\mu(C_0))^2 = g^2 \left( \frac12 g_1^2 + \frac23 g_2^2C_0 \right).
\end{equation}
The form of $c(\tau)$ is given by Eq.~\eqref{solution_C_tau}.

The paramagnetic chaotic phase appears when the ferromagnetic coupling $J_0$ is sufficiently small as compared to the non-reciprocal drive $g$ and when the latter is large compared to the confining term in the dynamics.
The simplest setting consists in having $J_0=0$. When $\hat\mu(0)$ is sufficiently large as compared to $gg_1$, the paramagnetic fixed point is stable. For $gg_1>\hat\mu(0)$ it becomes unstable and the dynamics lands on a chaotic attractor.

\begin{figure}
    \centering
    \includegraphics[scale=0.645]{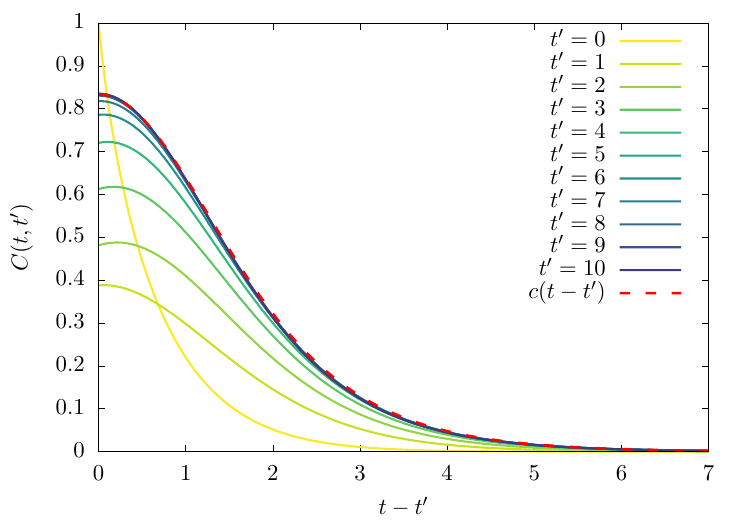}
    \includegraphics[scale=0.645]{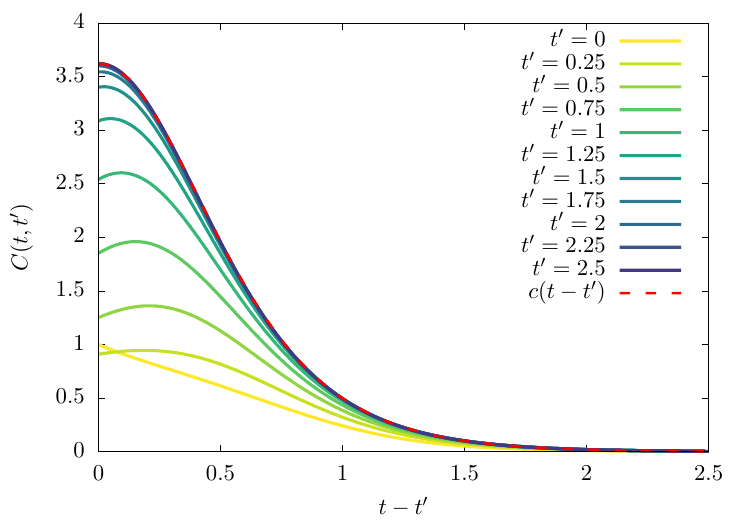}
\caption{Correlation function $C(t,t')$ for increasing values of $t'$ as a function of $t-t'$. The curves converge to a master curve which coincides with the one predicted by Eq.~\eqref{solution_C_tau}. Left panel: Model with $g=g_2=1$, $g_1=1.5$, $J_0=h=0$, starting from $\Ctt=1$ and $\mtt=0$ and integrated with $\de t=0.02$. The confining term is of the form $\hat \mu(z)=1+z$. Right panel: Same plot for $g=1$ and $g_1=g_2=1.5$, $J_0=h=0$ and $\de t=0.0025$ and same starting point as in the left panel. The confining potential has been chosen to be $\hat \mu(z)=z$.}
    \label{fig:example_para_chaos_C}
\end{figure}

In Fig.\ref{fig:example_para_chaos_C} we plot the numerical integration of the DMFT equations against the prediction coming from Eq.~\eqref{solution_C_tau}. The agreement is excellent.

\paragraph{Ferromagnetic fixed point phase --} For $J_0$ large enough the system develops a ferromagnetic solution. In particular, if the ferromagnetic term in the dynamical equations is much larger than the forcing field, the ferromagnetic phase contains two fixed point attractors with finite magnetization. The corresponding order parameters solve the equations
\begin{equation}
    \begin{split}
    \mu_\infty=\hat \mu(C_0)= J_0, \;\;\; C_{\infty}= C_0, \;\;\; m_{\infty} 
  = \pm \sqrt{C_0-\frac{g^2}{J_0^2}\left(g_1^2C_0+2g_2^2C_0^2\right)}.
    \end{split}
\end{equation}
It is important to note that such fixed points are locally stable in the sense that any small perturbed dynamics around them will eventually collapse on them. However, they are reached only if the initial condition has a small $\mathbb Z_2$ breaking field. In other words, in order to land on one of the two attractors we need to initialize the dynamical equations with $\mtt$ finite but arbitrarily small. 
In practice, the small $\mathbb Z_2$ symmetry breaking field is provided by the finite size fluctuations of $m(0)$: if we consider $\mtt=0$, for any finite $N$, $m(0)$ will be of order $1/\sqrt N$.
Whenever we will talk about ferromagnetic phases we will always have in mind that they can be accessed by adding a small initial condition that breaks the statistical spin flip symmetry of the original dynamical system.

\paragraph{Ferromagnetic chaotic phase --}
For intermediate values of $J_0$ as compared to $g$ the ferromagnetic solution becomes chaotic. This means that the magnetization of the system is different than zero on average but the dynamics never stops fluctuating. The corresponding solution for the order parameters is
\begin{equation}
  \mu_\infty = \hat\mu(C_0) = J_0, \;\;\; C_\infty = \frac12 \left(\frac{3J_0^2}{4g_2^2g^2} -\frac{3g_1^2}{4g_2^2}-C_0\right), \;\;\; m_\infty = \pm \sqrt{C_\infty - \frac{g^2}{J_0^2}(g_1^2 C_\infty + 2g_2^2 C_\infty^2)}\:.
\end{equation}

\begin{figure}
    \centering
    \includegraphics[scale=0.8]{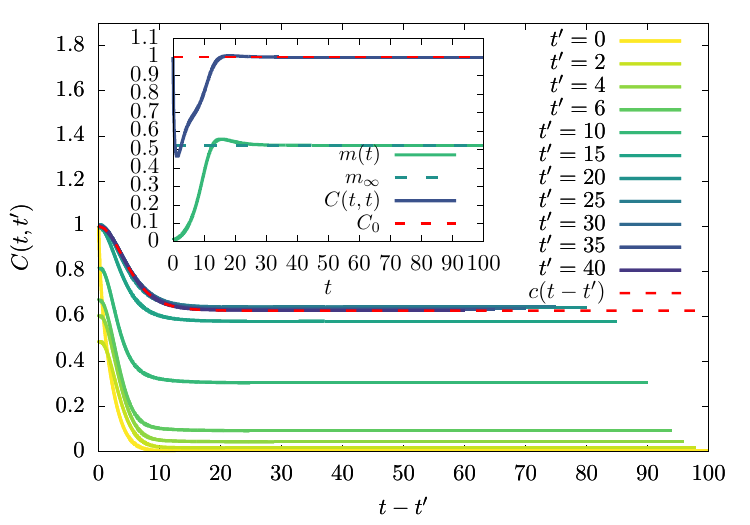}
    \caption{The behavior of the order parameters in the ferromagnetic chaotic phase. The dynamical system is characterized by $\hat \mu(z)=z$ and $1=J_0=g_1=g_2$, $g=0.5$, $\de t=0.01$, $\Ctt=1$, $\mtt=0.01$ and $h=0$. Main figure: the dynamical correlation function $C(t,t')$ for increasing values of $t'$. The curves collapse on the master curve $c(\tau)$ predicted by Eq.~\eqref{solution_C_tau}. Inset: the behavior of $C(t,t)$ and $m(t)$. At long times these curves converge to the asymptotic values $C_0$ and $m_\infty$.}
    \label{fig: example Chaos Ferro}
\end{figure}

\subsubsection{Phase transition lines}

\paragraph{Transition from the fixed point paramagnet to the chaotic paramagnet --} When the strength of the non-linear drive in the dynamical equations is sufficiently large, one can go from a fixed point paramagnetic solution to a chaotic paramagnetic phase. The critical point when this happens can be found in two ways. As discussed in Sec.~\ref{sec_steady_state_absence_simplest}, the critical value of the non-linear strength is given by $g^2g_1^2=\hat \mu(0)$. 
This condition is equivalent to say that at the critical point one has
\begin{equation}
    \left.\frac{\partial^2 V}{\partial c^2}\right|_{c=C_0=0} = 0
\end{equation}
as in a typical bifurcation scenario.
Note that this transition point depends only on $g_1$ and $\hat \mu(0)$. Indeed, both the non-linear term proportional to $g_2$ and higher order terms in $\hat \mu$ are irrelevant close to $\bx=0$.

\paragraph{Transition from the paramagnetic fixed point phase to the fixed point ferromagnetic phase --} This happens when \texttt{SteadyDMFT} equations with $C_0=C_\infty$ satisfy both the paramagnetic condition $C_0=0$ and the ferromagnetic one $\mu_\infty=J_0$. The corresponding transition line is given by the condition
\begin{equation}
    \hat\mu(0) = J_0.
\end{equation}

\paragraph{Line between the ferromagnetic fixed point phase and the ferromagnetic chaotic phase --}
The same argument applies for the phase transition line between the ferromagnetic fixed point phase and the ferromagnetic chaotic phase. This transition can be induced by increasing the value of the non-linear drive.
The critical point is determined by the condition 
\begin{equation}
\left. \frac{\partial^2 V}{\partial C^2}\right\vert_{C=C_0=C_\infty} = 0
\end{equation}
which in the case of the simplest model gives $J_0^2 = g^2 (g_1^2 + 4g_2^2 C_0)$, where $C_0$ is fixed by $\hat\mu(C_0) = J_0$.

\paragraph{Line between the paramagnetic chaotic phase and the ferromagnetic chaotic phase --}
This transition line can simply be found using the relation $V(C_\infty;\mu_\infty,m_\infty,C_0)=V(C_0;\mu_\infty,m_\infty,C_0)=0$ and imposing that $m_\infty=C_\infty=0$ and $\mu_\infty=J_0$. For the simplest model of Eq.~\eqref{def_simplest}, this condition reads $J_0^2 = g^2\left( g_1^2+\frac43g_2^2C_0\right)$, where $C_0$ is again fixed by $\hat\mu(C_0) = J_0$.

\begin{figure}
  \centering
  \includegraphics[width=1\textwidth]{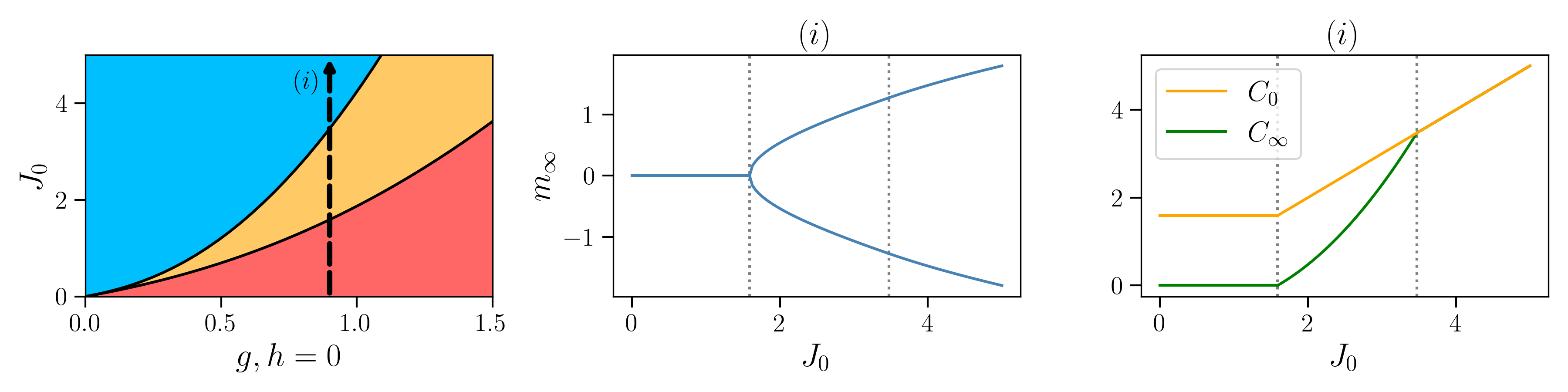}
  \includegraphics[width=1\textwidth]{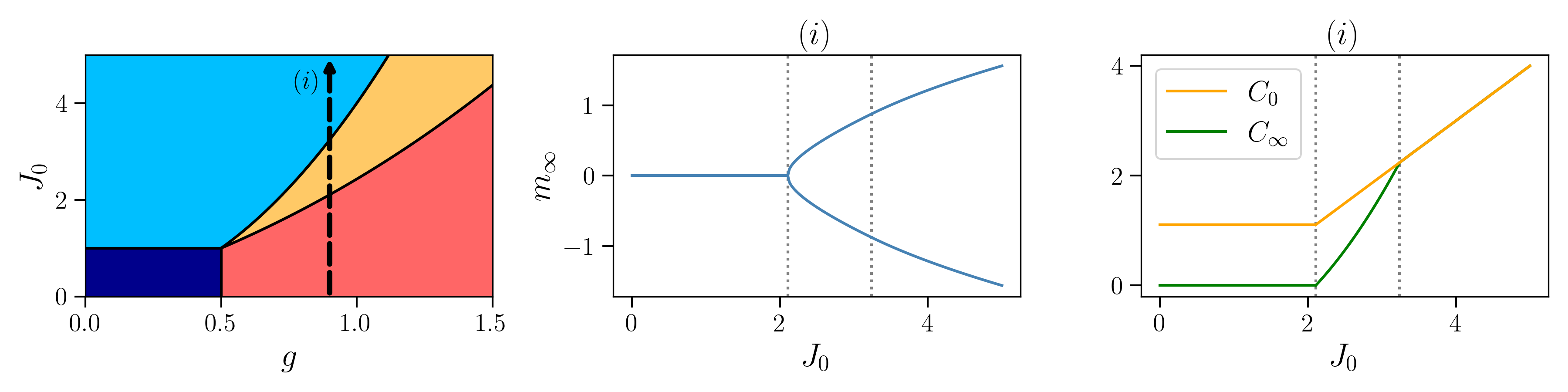}
  \caption{Phase diagrams and order parameters for $\hat\mu(z)=z,\,g_1=1,\,g_2=1$ (Upper panel) and $\hat\mu(z)=1+z,\, g_1=2,\, g_2=1$ (Lower panel). (Left) Phase diagrams with the paramagnetic chaotic phase in \textit{red}, ferromagnetic chaotic phase in \textit{orange}, ferromagnetic fixed point phase in \textit{light blue} and paramagnetic fixed point phase in \textit{dark blue}. (Middle and Right) Steady-state order parameters as a function of $J_0$ for fixed $g=0.9$ (the path in the phase diagrams along which the order parameters are plotted is represented by a \textit{black dashed} arrow). The \textit{gray dotted} lines denote phase transitions. Note that similar phase diagrams have been found in \cite{fournier2025non}.}
  \label{fig: phase diagrams of simplest model with linear mu and no external field}
\end{figure}

\subsubsection{Summary of phase diagrams}
In Fig.~\ref{fig: phase diagrams of simplest model with linear mu and no external field} we show the phase diagrams of two realisations of the simplest model: one with $\hat\mu(z)=z$ and $g_1=g_2=1$, and the other one with $\hat\mu(z)=1+z$ and $g_1=2,g_2=1$. In the second case, the system has a stable paramagnetic fixed point phase shown in \textit{dark blue}.
In the ferromagnetic phases, the basins of attraction of the two solutions with $m_\infty$ positive or negative are separated by the initial condition $\mtt=0$. We show here only cases of linear $\hat\mu$ and discuss the case of non-linear $\hat\mu$ in Sect.~\ref{sect: co-existence of chaotic attractors} to demonstrate how one can have multiple stable chaotic attractors which are not related by a $\mathbb{Z}_2$ symmetry.

\begin{figure}
  \centering
  \includegraphics[width=1\textwidth]{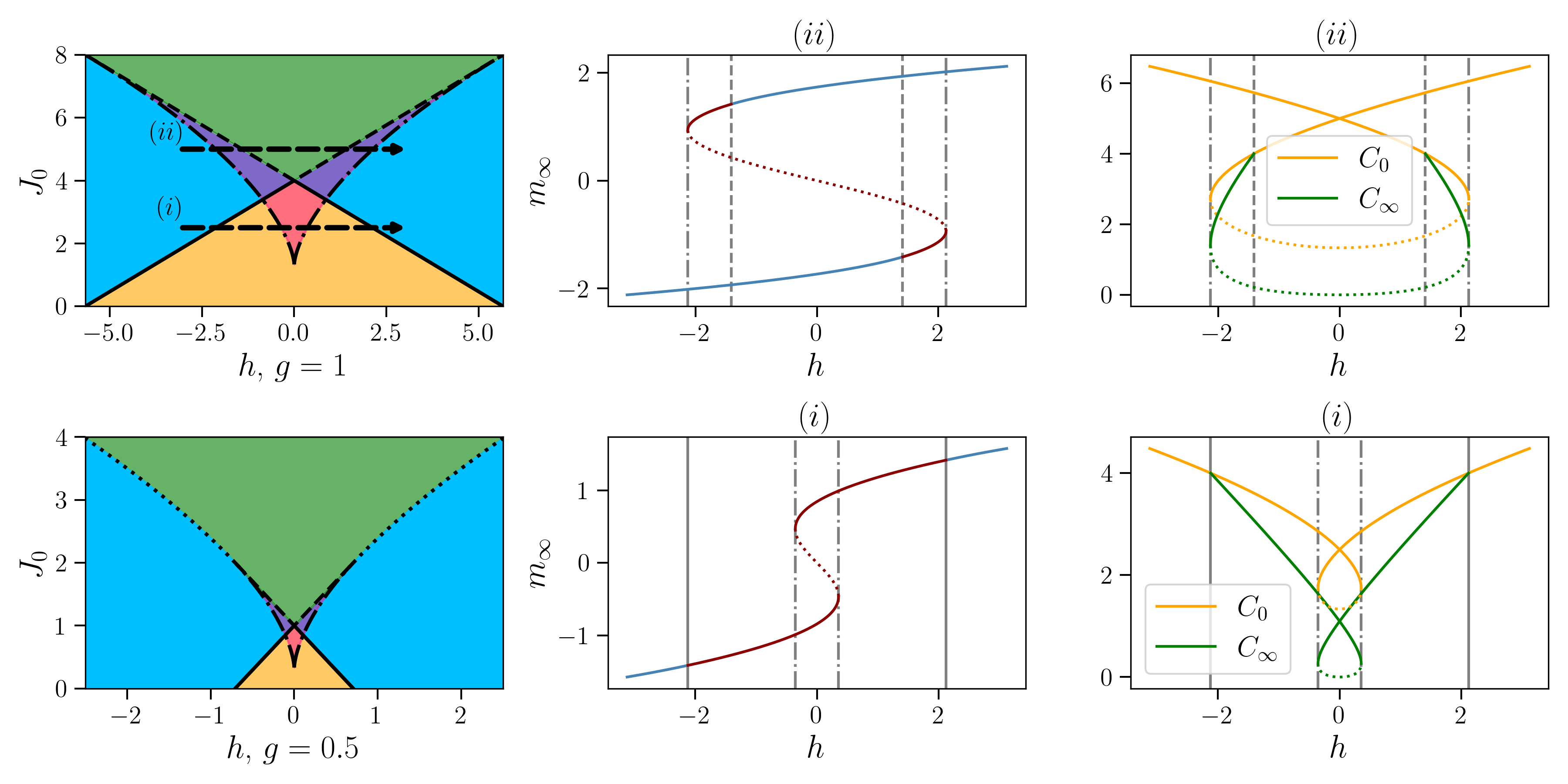}
  \caption{Phase diagram of the model with $\hat\mu(C_0)=C_0$, $g_1=0$, $g_2=1$ in the presence of an external field $h$. Phase diagram in the plane $(h,J_0)$ at fixed $g=1$ (Upper Left) and $g=0.5$ (Lower Left). Different colors are used to distinguish between the number of co-existing phases: one chaotic solution (\textit{orange}), two chaotic solutions (\textit{pink}), one fixed point solution (\textit{blue}), one chaotic and one fixed point solution (\textit{purple}), two fixed point solutions (\textit{green}). (Right) Stationary order parameters in the slices $(i)$ and $(ii)$ indicated by two arrows in the phase diagram. The vertical gray lines denote phase transitions. For $m_\infty$, we distinguish fixed point solutions (\textit{blue}) from chaotic ones (\textit{dark red}). The dotted solutions are unstable attractors that are not seen as steady states.}
  \label{fig: steady state with external field as a function of h}
\end{figure}
\subsection{External field and hysteresis loop}
In this section, we study the dynamical systems when a constant external field $h$ is added to the equations of motion.

Having $h\neq0$ has two consequences: first, all paramagnetic phases disappear in the sense that, as soon as $h\neq 0$, one has $m_\infty\neq 0$; furthermore, one can observe hysteresis loops emerging from the ferromagnetic phases of the model with $h\neq0$. This can be understood as follows. Suppose that one starts at $h=0$ and with a sufficiently large $J_0$. Assume that the system has landed on an attractor (either fixed point or chaotic) with a positive magnetisation. Suppose then that the external field $h$ is changed adiabatically. This means that at each infinitesimal change in $h$, the dynamics has time to reach the steady state. If $h$ is decreased, it is reasonable to expect that there exists a critical value $h_-$ where the ferromagnetic attractor with positive magnetization is not stable anymore and the dynamics jumps to another attractor characterized by a negative magnetization. Conversely, if now one starts from this last attractor and increases the external field $h$, there exists a critical value $h_+$ where the system will go back to an attractor with a positive magnetisation. If $J_0$ is sufficiently large, the two critical points $h_-$ and $h_+$ are different, with $h_-<h_+$ and $h_-=-h_+$.

This hysteresis loop mirrors the one that is found in standard equilibrium statistical mechanics. Obviously, it implies a coexistence of attractors as soon as $h_-<h_+$. Note that there is nothing that prevents the two attractors to be chaotic in nature. The precise form and properties of the hysteresis loop depend in general on the details of the dynamical systems. In the following, we will focus on the case where $\hat\mu(z)=z$, $g_1=0$ and $g_2=1$, and we will describe the corresponding phenomenology as we change $h$.

\begin{figure}
  \centering
  \includegraphics[width=1\textwidth]{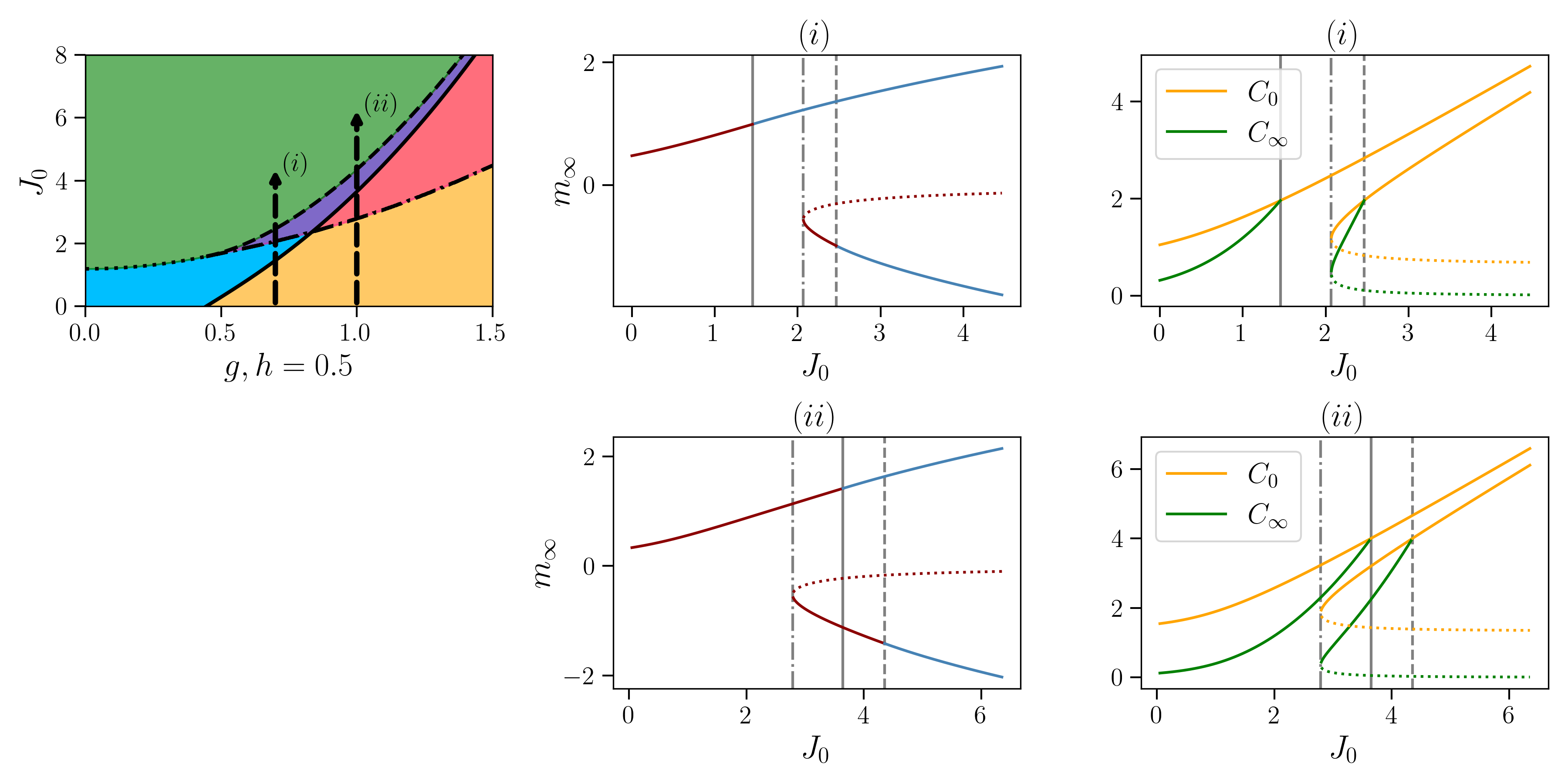}
  \caption{Phase diagram of the model with $\hat\mu(C_0)=C_0$, $g_1=0$, $g_2=1$ in the presence of an external field $h$. (Left) Phase diagram in the plane $(g,J_0)$ at fixed $h=1$. The color code and linestyle follow the same rules as in Fig.~\ref{fig: steady state with external field as a function of h}.}
  \label{fig: steady state with external field as a function of J0}
\end{figure}

\paragraph*{Fixed points --} The equations that determine the values of the order parameters in the case of fixed point attractors are given by eqs.~\eqref{eq: steady state eqs for fixed point solution} and, in the present case with $h\neq 0$, they read
\begin{equation}
\begin{split}
  C_\infty&=C_0\\
  C_0 &= J_0 + \frac{h}{m_\infty}\\
  0&=(J_0m_\infty+h)^2 \left[m_\infty^3 - (J_0 - 2g^2)m_\infty - h\right]\:.
 \end{split}
 \label{eq: steady state fp with external field}
\end{equation}
The last equation for $m_\infty$ is non-linear and therefore can have multiple solutions. One solution is always $m_\infty = -h/J_0$ which implies $C_0=0$. This solution is not consistent as soon as $g\neq0$ because $C_0=0$ implies that the dynamics land on $\bx =0$ which is not possible for $h\neq 0$. 
The second set of solutions are the roots $m_\infty$ that solve
\begin{equation}
    m_\infty^3 - (J_0 - 2g^2)m_\infty - h= 0\: .
    \label{cubic_eq_hyst}
\end{equation}
This is a depressed cubic equation whose solutions can be found analytically. Depending on $\Delta = \left((J_0-2g^2)/3\right)^3 - \left(h/2\right)^2 $, one can have either one or three solutions.
For $\Delta<0$ there is only one solution
\begin{equation}
    m_\infty = \sqrt[3]{\frac{h}{2} + \sqrt{-\Delta}} + \sqrt[3]{\frac{h}{2} - \sqrt{-\Delta}}\,.
\end{equation}
Otherwise, if $\Delta>0$, there are three real solutions that can be written as
\begin{equation}
\begin{split}
    m_\infty^{(i)} &= 2 \sqrt{\frac{J_0-2g^2}{3}} \cos\left(\frac{\theta}{3} + i\frac{2\pi}{3}\right), \;\;\;  \;\; i=0,1,2 \\
    \cos \theta &= \frac{h}{2}\left(\frac{3}{J_0-2g^2}\right)^{3/2}\,.
    \end{split}
\end{equation}
Therefore the case in which $\Delta>0$ correspond to coexistence of attractors.

\paragraph*{Chaotic attractors --} When the dynamics is chaotic, the stationary order parameters are given by eq.~\eqref{eq: steady state eqs for chaotic solution} which in the present model read
\begin{equation}
\label{eq: steady state chaotic with external field}
  \begin{split}
    C_\infty&=\frac12\left( \frac{C_0^2}{\frac43 g^2} - C_0 \right)\\
    C_0 &= J_0 + \frac{h}{m_\infty},\\
    0&=(J_0m_\infty+h)^2 \left[ 4m_\infty^4 - \left(\frac{3J_0^2}{8g^2}+J_0-2g^2\right)m_\infty^2 
    -\left(\frac{3hJ_0}{4g^2}+h\right)m_\infty - \frac{3h^2}{8g^2} \right] \:.
  \end{split}
\end{equation}
Again, the polynomial equation fixing $m_\infty$ has a solution $m_\infty = -h/J_0$, $C_0=0$ that is unstable for $g\neq0$. The other solutions are the roots of a quartic equation which can be found numerically.

\paragraph*{Transitions lines between fixed points and chaotic attractors --} 
When the external field is changed, it is possible that the fixed point attractors turn into chaotic ones, or vice versa.
The instability condition that fixes where these transitions take place is given by $\partial^2 V/\partial C^2 \vert_{C=C_0} = 0$. This implies that on the transition line $m_\infty = h/4g^2-J_0$. Plugging this expression into the depressed cubic equation in \eqref{eq: steady state fp with external field}, we get that the transition lines are found when 
\begin{equation}
\label{eq:Stability transition of a fixed point solution with external field}
  h = \pm \sqrt{2g^2}(J_0-4g^2)\:.
\end{equation}

\paragraph*{Spinodal transitions in the hysteresis loop --} 
Hysteresis loops are characterized by spinodal transitions. These happen when an attractor is not stable anymore and further changing the external field makes the magnetization to jump. 
The critical points where this happens are the ones where $\Delta>0$ but two solutions of Eqs.~\eqref{eq: steady state fp with external field} collide. We denote by $h_s$ the critical values of the external field where spinodal transitions happen.
They correspond to the points where $\left.\partial m_\infty/\partial h \right|_{h\to h_s} \to \infty$.
Taking a derivative with respect to $h$ of Eq.~\eqref{cubic_eq_hyst}, we get that
\begin{equation}
    \frac{\partial m_\infty}{\partial h} \left[ 3m_{\infty}^2 - (J_0 - 2g^2)  \right] = 1\,.
\end{equation}
Therefore, at the spinodal transitions, we must have that $3m_{\infty,s}^2 - (J_0 - 2g^2) = 0$ which means that the value of $m_\infty$ at the spinodal in the fixed point phase is given by
\begin{equation}
  m_{\infty,s} = \pm \sqrt{\frac{J_0 - 2g^2}{3}}.
\end{equation}
Plugging this result back into the cubic equation in \eqref{eq: steady state fp with external field}, we obtain the corresponding value of the external field at the spinodal in the fixed point phase
\begin{equation}
\label{eq: spinodal line in fp phase}
  h_s = \mp \frac{2}{3\sqrt{3}} \left(J_0 - 2g^2\right)^{3/2}.
\end{equation}
A similar procedure in the chaotic phase given by eq.~\eqref{eq: steady state chaotic with external field} yields
\begin{align}
  &m_{\infty,s} = \pm \frac1{4} \sqrt{  \frac{3J_0^2}{16g^2} + \frac{J_0}2 - 5g^2 + \frac{g}4 \left( \frac{3J_0}{4g^2}+1 \right) \sqrt{\frac{J_0^2}{g^2} + \frac{8J_0}3 + \frac{176g^2}{3}} }\\
  \label{eq: spinodal line in chaotic phase}
  &h_s = \frac{m_{\infty,s}}{\frac{3J_0}{4g^2}+1} \left( 16m_{\infty,s}^2  -\frac{3J_0^2}{4g^2}-2J_0+4g^2 \right).
\end{align}

Eqs.~(\ref{eq:Stability transition of a fixed point solution with external field}, \ref{eq: spinodal line in fp phase}, \ref{eq: spinodal line in chaotic phase}) can be used to construct the  phase diagram of the models. We report it in Figs.~(\ref{fig: steady state with external field as a function of h}, \ref{fig: steady state with external field as a function of J0}) in some specific cases. The stability lines \eqref{eq:Stability transition of a fixed point solution with external field} are denoted by \textit{full} and \textit{dashed black} lines in the phase diagrams, while the spinodal lines eqs.~(\ref{eq: spinodal line in chaotic phase}, \ref{eq: spinodal line in fp phase}) are denoted respectively by \textit{dash dotted} and \textit{dotted black} lines.

\begin{figure}
    \centering
    \includegraphics[width=\textwidth]{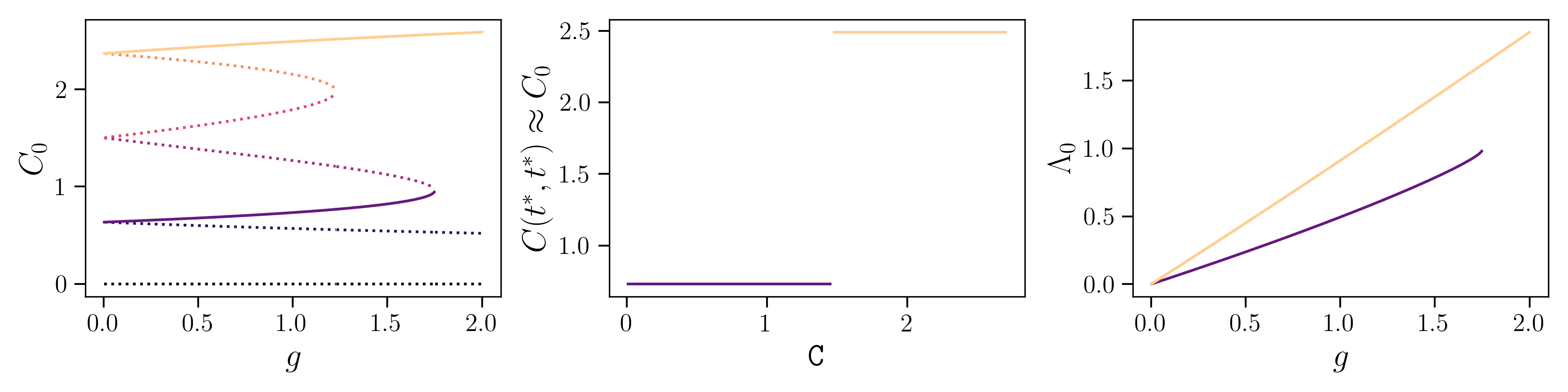}
    \includegraphics[width=\textwidth]{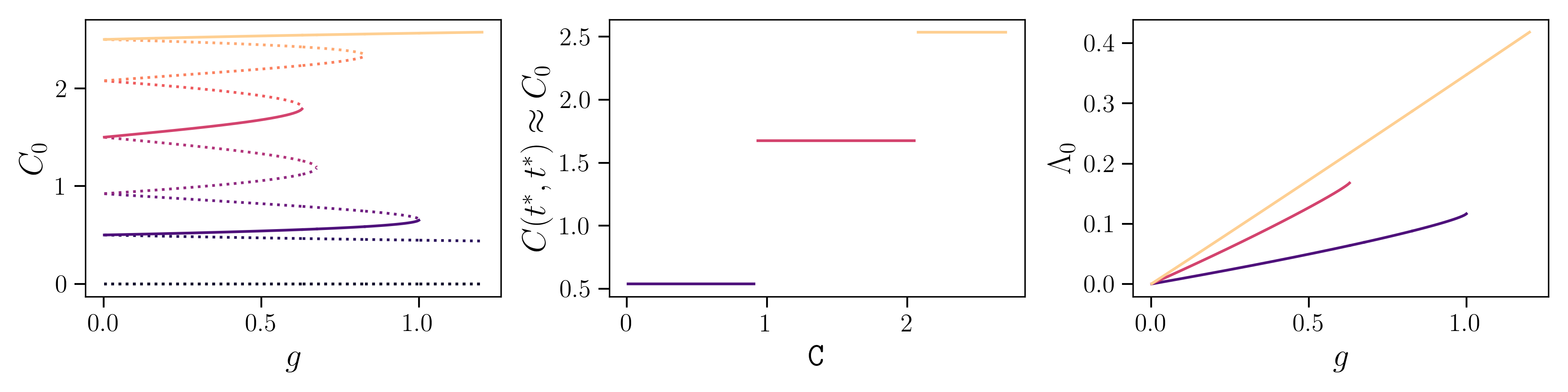}
    \caption{The behavior of the order of parameters when $\hat\mu$ is non-linear and $J_0=h=0$. (Upper panel) $\hat\mu$ is given by Eq.~\eqref{eq: confining potential for 2 chaotic attractors} and the choice of hyperparameters is $g_1=0$, $g_2=1$, $\alpha=1$, $\beta=3/2$, $d=3/2$. (Lower panel) $\hat\mu$ is given by Eq.~\eqref{eq: confining potential for 3 chaotic attractors} and the choice of hyperparameters is $g_1=1$, $g_2=1/2$, $a=1/2$, $b=3/2$, $c=5/2$. (Left) Numerical solutions of Eq.~\eqref{eq: eq for C0 in coexistence of multiple chaotic attractors} as a function of the control parameter $g$. The dotted lines represent unstable solutions. 
    (Middle) Basin of attraction of the dynamics at $g=1$ in the Upper panel and $g=0.5$ and the Lower panel. The solutions are found by integrating the DMFT equations
    starting from different initial conditions $\mtt=0$ and $\Ctt\in[0,2.7]$. The numerical integration was done using a timestep $dt=0.01$ until $t^*=10$, which is sufficient to reach the steady-state. (Right) Maximal Lyapunov exponent of the stationary dynamics computed via the techniques developed in Sec.\ref{Sec_MLE}.}
    \label{fig: 2 and 3 co-esxisting chaotic attractors}
\end{figure}

\subsection{Coexistence of chaotic attractors}
\label{sect: co-existence of chaotic attractors}
Up to now, we have studied particular instances of the dynamical systems where $\hat \mu$ is in general a convex function. 
In these cases, the coexistence of attractors is mainly due to the emergence of ferromagnetic phases linked to a $\mathbb{Z}_2$ statistical symmetry and its breaking. 
In this section, we want to investigate what happens when $\hat \mu$ is non-convex and there is no ferromagnetic term driving the dynamics. We want to show that in this case, the confining potential can induce multiple chaotic attractors that can coexist, and which are selected by the initial conditions of the dynamics. 
For simplicity, we will consider the simplest model of Eq.~\eqref{def_simplest} when $J_0=h=0$ but the results could easily be extended to more complicated cases. The absence of a ferromagnetic term in the dynamics implies that $m_\infty=0$ and $C_\infty=0$. Therefore, the chaotic attractors can be only distinguished from their corresponding value of $C_0$.
The equation fixing this order parameter  is given by Eq.~\eqref{eq: steady state for C0 fixed point or chaos}
\begin{equation}
\label{eq: eq for C0 in coexistence of multiple chaotic attractors}
    -\frac12 \hat\mu(C_0) C_0^2 + g^2\left( \frac12g_1^2 C_0^2 + \frac23 g_2^2 C_0^3 \right) = 0
\end{equation}
where we explicitly replaced $\mu_\infty$ by its expression $\hat\mu(C_0)$. Eq.~\eqref{eq: eq for C0 in coexistence of multiple chaotic attractors} is in general non-linear and therefore can have multiple solutions depending on the function $\hat\mu$ and on the value of the control parameters $g,g_1,g_2$. 
As in the other cases with coexisting attractors, not all solutions are physical and can be seen as asymptotic dynamical states.
In the following, we discuss a few cases of coexistence of chaotic paramagnetic attractors.

\paragraph*{Two chaotic attractors --} The mechanism to induce multiple chaotic attractors can be ascribed to the form of the confining potential. Define
\begin{equation}
    V_\mu(C) = \int_{C_0}^C\de c \hat \mu(c) 
\end{equation}
then the confining term in Eq.~\eqref{generic_eq} can be written as
\begin{equation}
    -\mu(t)x_i(t) = -\frac N2 \frac{\partial V_\mu(|\bx|^2/N)}{\partial x_i}\:.
\end{equation}
This implies that, as soon as $V_\mu$ has several minima, these define phase space shells where the dynamics can land.

Therefore, we can consider the case where $V_\mu$ has a double well shape
\begin{equation}
\label{eq: confining potential for 2 chaotic attractors}
    V_\mu(C_0) = \alpha(C_0-d)^4 - \beta(C_0-d)^2
\end{equation}
where $\alpha,\beta,d$ are hyperparameters. This confining potential has two symmetric wells located at $C_0 = d\pm \frac12\sqrt{2\beta/\alpha}$. The steady state of the dynamics can then be trapped in one of the two wells, which corresponds to different shells in phase space. The phase coexistence of the chaotic attractors and the corresponding order parameters can be found in Fig.\ref{fig: 2 and 3 co-esxisting chaotic attractors}

\paragraph*{Three chaotic attractors --} Choosing $V_\mu(C_0)$ with three wells can induce the coexistence of three chaotic attractors. For instance
\begin{equation}
\label{eq: confining potential for 3 chaotic attractors}
    V_\mu(C_0) = (C_0-a)^2(C_0-b)^2(C_0-c)^2
\end{equation}
where $a,b,c$ are again hyperparameters. This confining potential has three wells located at $C_0=a,b,c$.
The behavior of the order parameters in the different chaotic attractors is shown in Fig.\ref{fig: 2 and 3 co-esxisting chaotic attractors}.

\section{Steady state under periodic drive}\label{sec periodic drive}
In this section, we study the phase diagram of dynamical systems when a time-dependent periodic external perturbation is added to the equations of motion.
This problem has been first studied in \cite{rajan2010stimulus} in the case of dynamical systems describing simple models of random recurrent neural networks. This work shows that, depending on the amplitude and frequency of the periodic perturbation, one can have:
\begin{itemize}
    \item a chaotic phase which partially retains  the periodic nature of the perturbation in the power specturm of the dynamical correlation function;
    \item a fully periodic phase where the perturbation is so strong that the dynamical system lands on a periodic attractor whose base frequency (first harmonic) is the one of the perturbation. 
\end{itemize}
In this section, we study this setting in the context of the dynamical systems that we are considering.
We anticipate that we obtain the same conclusions as in \cite{rajan2010stimulus}.

Consider the dynamical system
\begin{equation}
        \partial_t x_i(t) = -\mu(t) x_i(t) + g r_i(\underline x(t)) +H_i(t)
    \label{generic_eq_periodic}
\end{equation}
where 
\begin{equation}
\begin{split}
    H_i(t)&=I\cos\left(\omega t + \theta_i\right)\\
    \theta_i&\sim \Unif\left([0,2\pi]\right)\:.
\end{split}
\end{equation}
For the simplest model of Eq.~\eqref{def_simplest}, we just need to set the random field $\br$ with a covariance given by $f(z)=g_1^2z+2g_2^2x^2$.
The phases $\theta_i$ are added randomly to prevent a trivial in-phase global oscillation of the system.
In Eq.~\eqref{generic_eq_periodic}, we did not add a ferromagnetic term and a constant external field.
The computation that we are going to detail can be extended without particular problems also to these situations.

\subsection{DMFT with periodic drive}
In order to analyze the dynamical system in the large $N$ limit we follow the same strategy as in \cite{rajan2010stimulus}. 

Consider the decomposition
\begin{equation}
    \bx(t) = \bx^{(0)}(t)+\bx^{(1)}(t)
    \label{decomposition}
\end{equation}
where $\bx^{(0)}$ and $\bx^{(1)}$ satisfy
\begin{equation}
\begin{split}
    \partial_t x_i^{(0)}(t)&=-\mu(t) x_i^{(0)}(t)+H_i(t)\\
    \partial_t x_i^{(1)}(t)&=-\mu(t) x_i^{(1)}(t)+gr_i(\bx(t))\:.
\end{split}
\end{equation}
The initial condition for the dynamical system in Sec.~\ref{sec_init} leaves an arbitrary choice for the initialization of $\bx^{(0)}$ and $\bx^{(1)}$. 
We choose
\begin{equation}
    \begin{split}
        \bx^{(0)}(0)&=0\\
        \bx^{(1)}(0)&={\cal N}(0,1)\sqrt{\Ctt}\:.
    \end{split}
\end{equation}
The dynamics of $\bx^{(0)}$ and $\bx^{(1)}$ are coupled in two ways: first, the confining potential $\mu(t)$ depends on $C(t,t)$ which takes contributions both from $\bx^{(0)}$ and $\bx^{(1)}$. Second, the random driving term $\br$ has a convariance structure that depends on $C(t,t')$ and therefore has to be computed using both the dynamics of $\bx^{(0)}$ and $\bx^{(1)}$.

The advantage of using the decomposition in Eq.~\eqref{decomposition} is that it allows to treat separately the effect of the periodic and chaotic drive.
Furthermore, the dynamics of the degrees of freedom $\bx^{(0)}$ is almost decoupled\footnote{Except for the confining potential.}.
Indeed, the formal solution of the dynamics of $\bx^{(0)}$ is given by
\begin{equation}
    x_i^{(0)}(t) = I\int_0^t\de s \cos(\omega s+\theta_i)e^{-\int_s^t\de \hat s \mu(\hat s)}\:.
\end{equation}
Furthermore, we can define the dynamical correlation function for $\bx^{(0)}$ given by
\begin{equation}
    C^{(0)}(t,t')=\frac 1N \bx^{(0)}(t)\cdot \bx^{(0)}(t')\:.
\end{equation}
In the large $N$ limit, this function concentrates on its mean and therefore
\begin{equation}
    C^{(0)}(t,t') = I^2\int_0^t\de s\int_0^{t'}\de s' \E\left[\cos(\omega s +\theta)\cos(\omega s' +\theta)\right]e^{-\int_s^t\de \hat s \mu(\hat s)-\int_{s'}^{t'}\de \hat s' \mu(\hat s')}\:.
\end{equation}
Using that
\begin{equation}
    \E\left[\cos(\omega s +\theta)\cos(\omega s' +\theta)\right] = \frac 12 \cos(\omega(s-s')),
\end{equation}
we get that
\begin{equation}
    C^{(0)}(t,t')= \frac {I^2}2\left(\overline c(t)\overline c(t')+\overline s(t) \overline s(t')\right) 
\end{equation}
with the following definitions
\begin{equation}
    \begin{split}
        \overline c(t) &= \int_0^t\de s \cos(\omega s) e^{-\int_s^t\de \hat s \mu(\hat s)}\\
        \overline s(t) &= \int_0^t\de s \sin(\omega s) e^{-\int_s^t\de \hat s \mu(\hat s)}\:.
    \end{split}
    \label{cs_hat}
\end{equation}
Note that $\overline c$ and $\overline s$ satisfy the following simple ODEs
\begin{equation}
    \begin{split}
        \frac{\de \overline c(t)}{\de t} &= -\mu(t)\overline c(t) +\cos(\omega t)\\
        \frac{\de \overline s(t)}{\de t} &= -\mu(t)\overline s(t) +\sin(\omega t)
    \end{split}
\end{equation}
with initial conditions given by $\overline c(0)=\overline s(0)=0$.

The DMFT analysis for $\bx^{(1)}$ can be easily derived following the computations of the previous sections.
Denoting by $C^{(1)}(t,t')=\bx^{(1)}(t)\cdot \bx^{(1)}(t')/N$ the dynamical correlation function for the $\bx^{(1)}$ degrees of freedom and $R^{(1)}(t,t')$ the corresponding response function, we have 
\begin{equation}
\begin{split}
    \partial_tC^{(1)}(t,t') &= -\mu(t) C^{(1)}(t,t') +g^2\int_0^{t'}\de s f(C(t,s))R^{(1)}(t',s)\\
    \partial_tR^{(1)}(t,t')&-\mu(t)R^{(1)}(t,t')+\delta(t-t')\:.
\end{split}
\label{DMFT_periodic}
\end{equation}
The global correlation function $C(t,t')$ can be obtained from $C^{(0)}$ and $C^{(1)}$  as
\begin{equation}
    C(t,t')=C^{(0)}(t,t')+C^{(1)}(t,t')\:.
\end{equation}
because in the $N\to \infty$ limit, $\bx^{(0)}(t)\cdot \bx^{(1)}(t')/N=0$. This is due to the fact that the dynamics of $\bx^{(0)}$ is autonomous (it does not depend on $\bx^{(1)}$ and the dynamics on $\bx^{(1)}$ depends on $\bx^{(0)}$ only through $\br$. When taking the large $N$ limit one can reduce the dynamics of $\bx^{(1)}$ to an  effective single site degree of freedom $x^1(t)$ that obeys a self-consistent stochastic equation which depends on $\bx^{(0)}$ only through $C(t,t')$. Therefore there is  no correlation between these two sets of degrees of freedom.
The same property was observed in \cite{rajan2010stimulus}.

\subsection{Phenomenology of the steady state}\label{sec_pheno_periodic}
The DMFT equations \eqref{cs_hat} and \eqref{DMFT_periodic} can be integrated numerically in the same way as it has been done in the previous sections.
Here, we investigate what happens to the steady state in the simplest dynamical system where we choose $\hat \mu(z)=z$ and $f(z)=g_1^2z+2 g_2^2z^2$.
We start by noting that this particular model has a chaotic phase in the absence of the external periodic perturbation. Given that we want to study the effect of the external drive, we fix $g=g_1=g_2=1$ and explore what is the resulting steady state as a function of the two control parameters $I$ and $\omega$.

\begin{figure}
\centering
\includegraphics[scale=0.8]{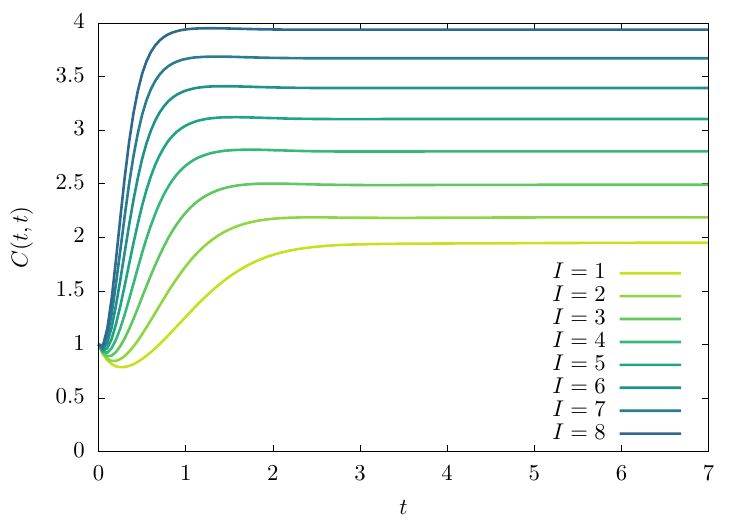}
\caption{The behavior of $C(t,t)$ as a function of $t$ for different values of $I$ at fixed $\omega=1$. Data for $g=g_1=g_2=1$.}
\label{fig_periodic_C0}
\end{figure}

We first study the behavior of one time quantities as a function of time. In particular, we focus on $C(t,t)$ which defines also the long time behavior of $\mu(t)$ through the function $\hat \mu$.
In Fig.\ref{fig_periodic_C0}, we plot the behavior of $C(t,t)$ as a function of $t$ for different values of $I$ at fixed angular frequency $\omega$. In all cases, we get that $C(t,t)$ approaches a constant.

\begin{figure}
    \centering
    \includegraphics[scale=0.645]{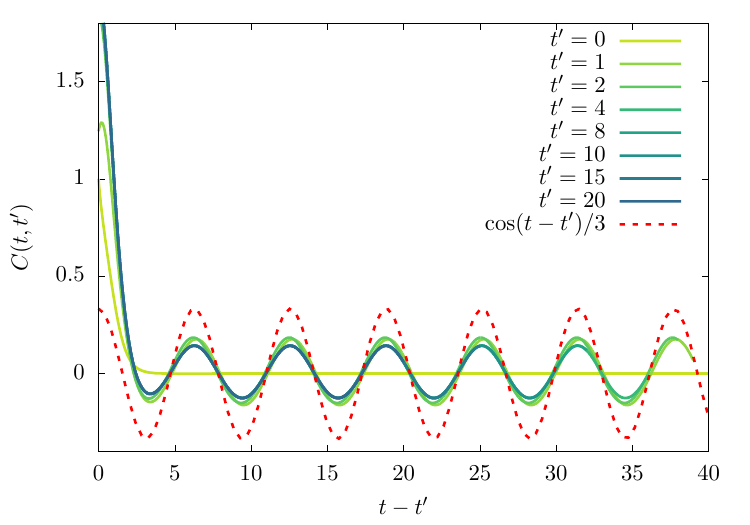}
    \includegraphics[scale=0.645]{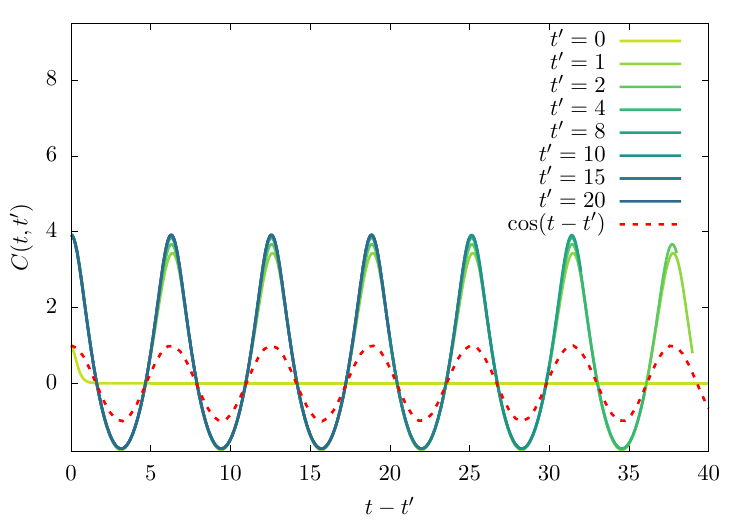}
    \caption{The dynamical correlation function $C(t,t')$ as a function of $t-t'$ for increasing values of $t'$. Left panel: data for $I=1$ and $\omega=1$; right panel: data for $I=8$ and $\omega=1$. In both cases, the steady state dynamics converges to a time translational invariant function. In red, we plot the periodic term of the forcing field. Both plots show that the power spectrum of the long time limit of the dynamical correlation function is supported on the base frequency of the forcing term and higher harmonics. Data for $g=g_1=g_2=1$.}
    \label{fig_periodic_Ctau}
\end{figure}

To understand further the steady state, in Fig.\ref{fig_periodic_Ctau}, we plot $C(t,t')$ as a function of $t-t'$ for increasing values of $t'$. If $t'$ is large enough such that, correspondingly, $C(t',t')$ is stationary, the function $C(t,t')$ is manifestly a periodic function of $t-t'$. This happens regardless of the value of $I$ and $\omega$, and shows that, in all these cases, the dynamics lands on steady states that are time translational invariant.  This is an important point: even if the dynamics is subjected to a periodic drive with angular frequency $\omega$, the resulting steady state is time translational invariant. This is due to the fact that the periodic drive is not in phase with the different degrees of freedom because the $\theta_i$ are structureless. Therefore, the asymptotic steady state is described by $\mu_\infty$, $c(\tau)$.

Interestingly, we find that the dynamics can land on two types of steady states.
Fixing $\omega$, one can distinguish to cases:
\begin{itemize}
\item at low values of $I$, the form of $c(\tau)$ suggests that the dynamics lands on a chaotic attractor: however, $c(\tau)$ shows traces of the periodic forcing given that it oscillates  at the same frequency of the the periodic drive for $\tau\gg 1$.
\item at high values of $I$, the forcing term is so strong that the dynamics lands on limit cycles where the steady state is fully periodic and the dynamics oscillates at the same angular frequency as the drive. Note that the form of $c(\tau)$ does not coincide with the periodic forcing. The power spectrum of $c(\tau)$ has support both on the fundamental frequency $\omega$ and higher harmonics.
\end{itemize}
The order parameter that distinguish the two phases can be easily obtained from these considerations.
Define
\begin{equation}
\Delta C = \lim_{t'\to \infty}\left[C(t',t')-C(t'+2\pi/\omega,t')\right]=c(0)-c(2\pi/\omega)\:.
\end{equation}
In the chaotic phase, $\Delta C>0$ while limit cycles are characterized by $\Delta C=0$.
Therefore, monitoring the point where $\Delta C$ becomes strictly positive defines the transition line between the two phases.

\subsection{DMFT for the steady state}
One way to plot the phase diagram of the model is to develop a direct DMFT treatment for the steady state of the dynamics.
This can be done following the same strategy of the previous sections.
Assuming that the steady state is time translation invariant, we get
\begin{equation}
\begin{split}
    \partial_\tau^2c(\tau) &= \mu_\infty^2 c(\tau) -g^2 f(c(\tau)) -\frac{I^2}{2}\cos(\omega \tau)\\
    \lim_{\tau\to 0}\frac{\de c}{\de \tau}&=0\\
    \mu_\infty&=\hat \mu(c(0))
\end{split}
    \label{steady_periodic_DMFT}
\end{equation}
Therefore, computing the steady state solution is equivalent to finding self-consistent solutions of Eq.~\eqref{steady_periodic_DMFT}. The corresponding ODE is a non-linear Duffing equation for which there are no general analytic solutions. Self-consistent solutions are either fully periodic (limit cycles) or partially periodic (chaotic phase).
Finding such solutions from the numerical integration of Eq.~\eqref{steady_periodic_DMFT} is rather complicated because the non-linearity makes the solution of the ODE rather unstable upon changing slightly the initial condition $c(0)$.

 \begin{figure}
    \centering
    \includegraphics[scale=0.645]{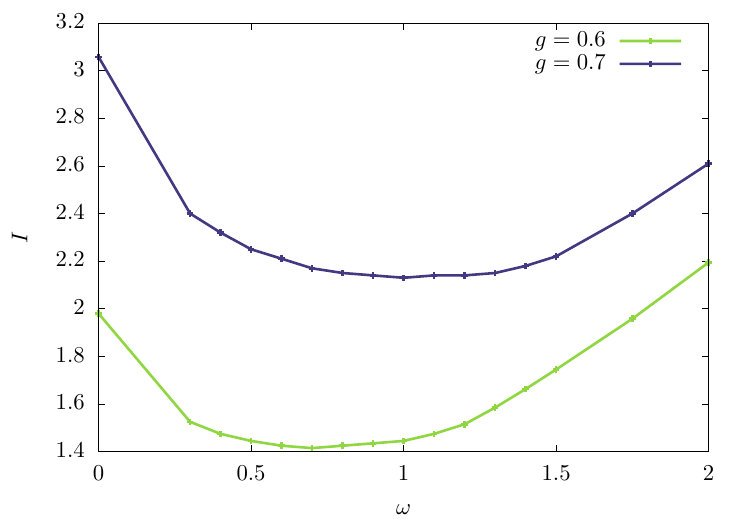}
    \includegraphics[scale=0.645]{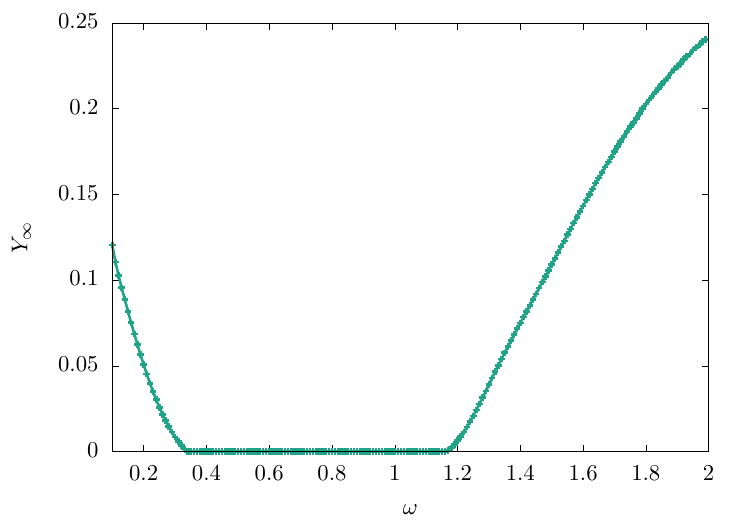}
    \caption{Left Panel: the phase diagram under periodic drive. Data for $g_1=g_2=1$ and for two values of $g$. Above the transition line, the periodic drive induces limit cycles in the steady state. Below the lines, the activity remains chaotic. The phase diagram shows a $g$-dependent frequency range where the phase-incoherent drive synchronizes the activity of the dynamical system. Right Panel: the behavior of $Y_\infty$ as a function of $\omega$ for $g=0.6$ and $I=1.5$.}
    \label{Phase-diagram-periodic}
\end{figure}

\subsection{Phase diagram of the simplest model}
In order to obtain the phase diagram, we avoid solving Eq.~\eqref{steady_periodic_DMFT} and we implement an alternative procedure.
At fixed $\omega$ we start from  $I=0$ and increase $I$ progressively. For each value of $I$ we solve numerically Eqs.~\eqref{DMFT_periodic} and compute
\begin{equation}
    Y(t)=\frac 1{C\left(t,t\right)}\left[C\left(t,t\right)-C\left(t+\frac{2\pi}{\omega},t\right)\right]
\end{equation}
In this way we  have access to $Y(t)$ only on a fixed time interval.
However, we can interpolate numerically this function to extrapolate its behavior at $t\to \infty$. Define
\begin{equation}
    Y_\infty=\lim_{t\to \infty}Y(t)\:,
\end{equation}
we expect that
\begin{equation}
    Y_\infty =\frac{\Delta C}{c(0)} \:.
\end{equation}
Therefore, the point where $Y_\infty$ vanishes corresponds to the phase transition from chaotic to limit cycle phases.
In practice, we define the phase transition point as the point where $Y_\infty$ becomes smaller than a threshold value ($10^{-4}$ in our numerics).

This procedure is rather stable as soon as $\omega$ is not too small, in which case we need to wait a long time before reaching the steady state. However for $\omega\to 0$, the self-consistent equation \eqref{steady_periodic_DMFT} can be easily solved due to the fact that the ODE can be written again as a descent on an effective potential and therefore one can re-use the analysis done in the previous sections.

The result of this analysis is plotted in Fig.\ref{Phase-diagram-periodic}.
The phase diagram shows that the transition line from chaotic to limit cycles is a non-monotonous function of the angular frequency of the forcing term in the dynamics. This result  reflects qualitatively the same behavior found in standard recurrent neural networks \cite{rajan2010stimulus}.

This implies that dynamical systems can synchronize with a phase-incoherent external forcing but this happens only in a selected frequency range whose location depends on the dynamical system itself.

\subsection{Periodic drive with broad frequency spectrum}
\subsubsection{Superposition of out-of-phase periodic signals}
The computation above can be generalized to the case in which the forcing signal has a broader frequency and amplitude spectrum.
We  briefly detail the main changes to the analysis that has been already done.
Consider a forcing term of the form
\begin{equation}
    H_i(t)=\sum_{k=1}^KI_k \cos(\omega_k t +\theta_i^{(k)})
\end{equation}
where $K$ is the number of frequency modes, and $\omega_k$ and $I_k$ the corresponding frequencies and amplitudes. We assume that these $K$ periodic signals are all incoherent. This means that $\btheta^{(k)}$ are random vectors independent and identically distributed as $\btheta^{(k)}\sim \Unif\left([2\pi]^N\right)$.
Furthermore, we assume that the frequencies $\omega_k$ are commensurate to each other so that the resulting signal is periodic in time.

The DMFT equations in Eq.~\eqref{DMFT_periodic} can be generalized easily in this case. Indeed the only change that has to be made concerns $C^{(0)}(t,t')$ that is now given by
\begin{equation}
    C^{(0)}(t,t') = \frac 12 \sum_{k=1}^K I_k^2 \left(\overline c_k(t)\overline c_k(t')+\overline s_k(t)\overline s_k(t')\right)
\end{equation}
with
\begin{equation}
    \begin{split}
        \overline c_k(t)&=\int_0^t\de s \cos(\omega_kt)e^{-\int_s^t\de s'\mu(s')}\\
        \overline s_k(t)&=\int_0^t\de s \sin(\omega_kt)e^{-\int_s^t\de s'\mu(s')}\ \ \ k=1,\ldots,K
    \end{split}
\end{equation}
We find that the numerical solution of the DMFT equations suggest that the stationary state of this dynamics is time-translational invariant in the sense that $C(t,t)$ converges to a constant at long times and two-time quantities are functions of the time difference only.
The corresponding DMFT for the steady state dynamics is then given by
\begin{equation}
\begin{split}
    \partial_\tau^2c(\tau) &= \mu_\infty^2 c(\tau) -g^2 f(c(\tau)) -\frac 12 \sum_{k=1}^K{I^2_k}\cos(\omega_k \tau)\\
    \lim_{\tau\to 0}\frac{\de c}{\de \tau}&=0\\
    \mu_\infty&=\hat \mu(c(0))
\end{split}
    \label{steady_periodic_DMFT_broad}
\end{equation}
The DMFT equations can be integrated numerically as we did in the previous sections.

\begin{figure}
    \centering
    \includegraphics[scale=0.645]{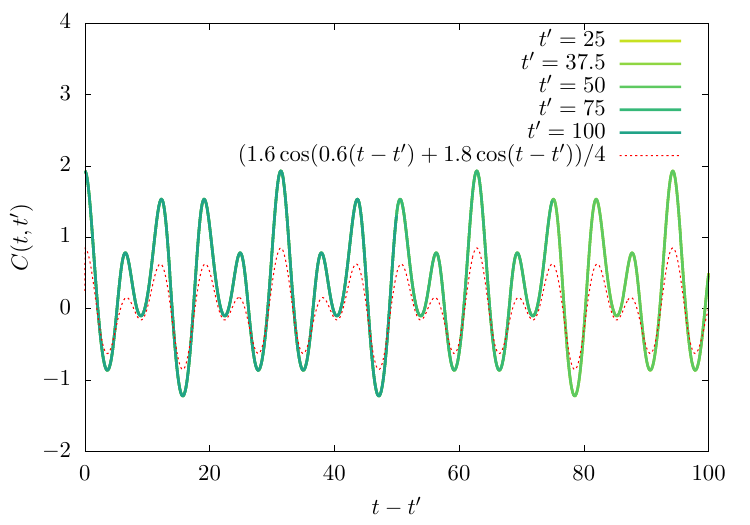}
    \includegraphics[scale=0.645]{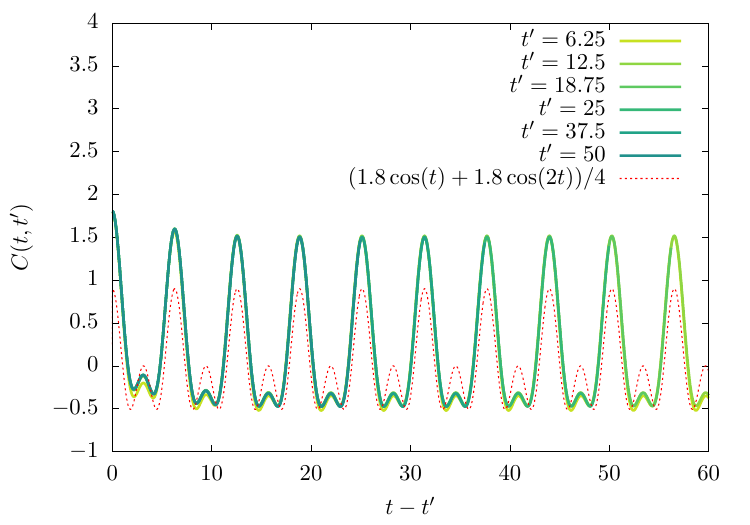}
    \caption{The response of the simplest dynamical system in Eq.~\eqref{def_simplest} to a periodic drive with support on two frequency modes with possibly different amplitudes. The details of the dynamical system coincide with the same used in Fig.~\ref{Phase-diagram-periodic} with $g=0.6$. Here we plot the dynamical correlation function $C(t,t')$ as a function of $t-t'$ for increasing values of $t'$. In both cases, the dynamics land on a time-translational invariant attractor. (Left) $I_1=1.6$, $I_2=1.8$, $\omega_1=0.6$ and $\omega_2=1$. This corresponds to a case where both frequencies and amplitudes are such, that individually, each term of the forcing field would lead to a limit cycle. The superposition leads to a limit cycle too. (Right) $I_1=I_2=1.8$, $\omega_1=1$ and $\omega_2=2$. This situation corresponds to the case where the lowest frequency mode of the forcing field would lead to a limit cycle if used individually, and the second one would lead to a chaotic attractor. We find that the superposition of the two leads to a chaotic attractor. }
    \label{fig:two_frequency_periodic_h}
\end{figure}

A natural question we can ask is how the system responds to a periodic forcing field that has two frequency modes.
We have explored this question in a rather empirical way and the results of this analysis are summarized in Fig.\ref{fig:two_frequency_periodic_h}.

Let us denote by $\Omega$ the portion of the phase diagram in Fig.~\ref{Phase-diagram-periodic}-left where the steady state of the system syncronizes with the external drive when this is composed by a single signal with a given frequency. 
Let us suppose that $K=2$ and $(I_i,\omega_i)_{i=1,2}\in \Omega$.
We find that the steady state of the system is synchronous with the injected signal. Conversely we empirically see that when one of the two modes has an amplitude-frequency pair that does not belong to $\Omega$, the resulting steady state is chaotic.

We remark that these are empirical observations but we do not try any attempt to characterized the corresponding phase diagram and to check whether the phase transition lines of Fig.~\ref{Phase-diagram-periodic} change when the signal has a broader frequency spectrum.

\subsubsection{Superposition of in-phase periodic signals}
Up to now, we have investigated the possibilities of a dynamical system with different uncorrelated periodic input sources and non-in-phase periodic signals. 
In this section, we consider the case in which the periodic signal is a superposition of different frequency modes, all in phase with each other and still out-of-phase on single dynamical degrees of freedom.
In other words, we choose
\begin{equation}
    H_i(t)=\sum_{k=1}^KI_k \cos(\omega_k t +\theta_i)\:.
\end{equation}
As before, we assume that the angular frequencies $\omega_k$ are commensurate to each other so that the resulting drive is periodic.
The theory we developed so far can be extend in the following way.
Define
\begin{equation}
    \begin{split}
        \tilde c(t)&=\sum_{k=1}^K I_k\int_0^t\de s \cos(\omega_k t) e^{-\int_s^t\de s' \mu(s')}\\
        \tilde s(t)&=\sum_{k=1}^K I_k\int_0^t\de s \sin(\omega_k t) e^{-\int_s^t\de s' \mu(s')}
    \end{split}
\end{equation}
Then
\begin{equation}
    C^{(0)} (t,t') = \frac 12 \left[\tilde c(t)\tilde c(t')+\tilde s(t)\tilde s(t')\right]
    \label{C0_coherent}
\end{equation}
Note that 
\begin{equation}
    \begin{split}
        \frac{\de \tilde c}{\de t} & = -\mu(t)\tilde c(t) +\sum_{k=1}^KI_k\cos(\omega_k t)\\
        \frac{\de \tilde s}{\de t} & = -\mu(t)\tilde s(t) +\sum_{k=1}^KI_k\sin(\omega_k t)
    \end{split}
\end{equation}
with initial condition given by $\tilde c(0)=\tilde s(0)=0$.
The DMFT equations for this case coincide with the ones in Eq.~\eqref{DMFT_periodic} where $C^{(0)}(t,t')$ is given by Eq.~\eqref{C0_coherent}.

\begin{figure}
    \centering
    \includegraphics[scale=0.645]{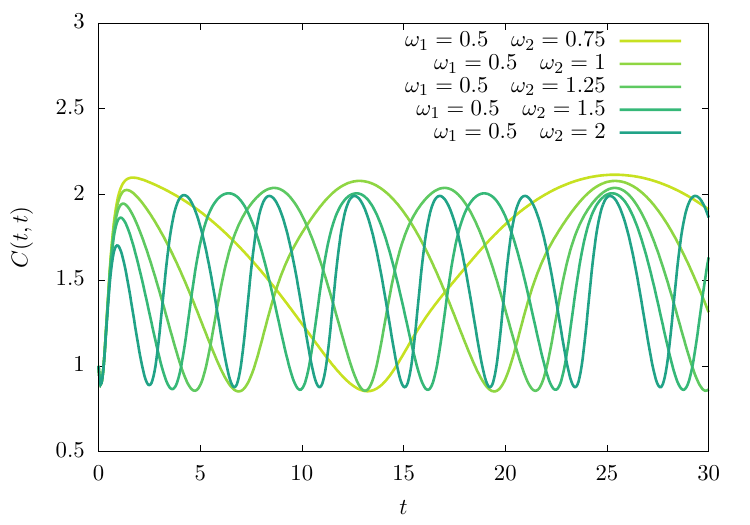}
    \caption{The dynamical correlation function $C(t,t)$ as a function of time for several values of the two frequencies defining the time-dependent external field that is driving the dynamics. The external drive is a coherent superposition of two pure periodic signals. Data comes from a model with $J_0=0$, $g_1=g_2=1$, $g=0.6$ and $\hat \mu(z)=z$.}
    \label{fig:Coherent_drive_Cdiag}
\end{figure}

\begin{figure}
    \centering
    \includegraphics[scale=0.645]{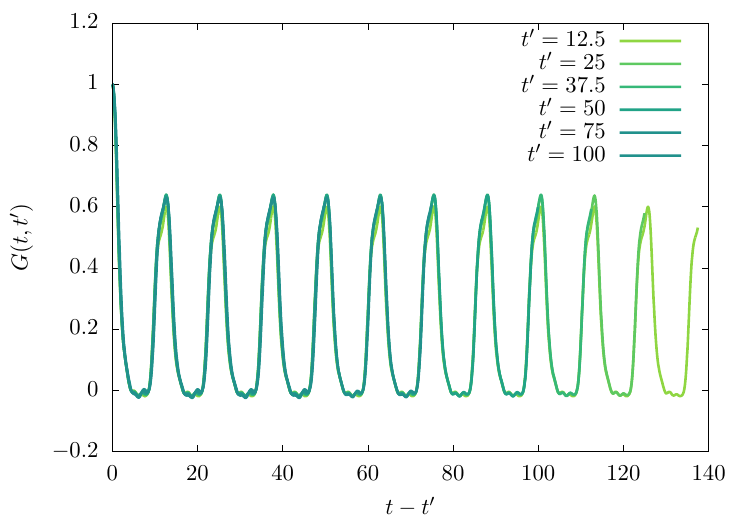}
    \includegraphics[scale=0.645]{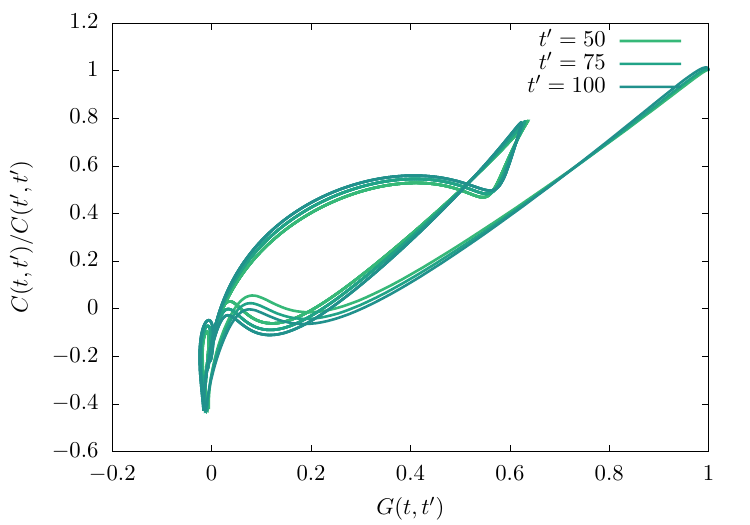}
    \caption{Superposition of two coherent periodic signals with $I_1=I_2=1$, $\omega_1=0.5$, $\omega_2=2$, $J_0=0$, $g=0.6$, $g_1=g_2=1$ and the confining term is given by $\hat \mu(z)=z$. Left panel represents the behavior of $G(t,t')$ as a function of $t-t'$ and for increasing values of $t'$. The right panel represents the behavior of $C(t,t')/C(t',t')$ as a function of $t$ plotted parametrically as a function of $G(t,t')$. The curves do not close the cycle at (1,1) and therefore the dynamics is not a limit cycle.}
    \label{periodic_choerent_chaotic}
\end{figure}

A numerical integration of such equations shows that, as soon as the support of the spectrum of the external drive is not a single point, the dynamics never reaches a time-translational invariant state.
The asymptotic dynamics instead lands on a periodic attractor. 
This can be seen in Fig.~\ref{fig:Coherent_drive_Cdiag} where we plot $C(t,t)$ for $K=2$ and $I_1=I_2$ and changing the values of the drivining frequencies. 
In all cases, the asymptotic dynamics is periodic at the level of $C(t,t)$. However, this only does not allow to understand whether the corresponding attractor is a limit cycle or it is chaotic. In order to clarify this, we define the normalized dynamical correlation function defined as
\begin{equation}
    G(t,t')=\frac{C(t,t')}{\sqrt{C(t,t)C(t',t')}}=\frac{\bx (t)\cdot \bx(t')}{\sqrt{|\bx(t)|^2|\bx(t')|^2}}
\end{equation}
For $t>t'\gg 1$ the dynamics lands on its asymptotic steady state but $G$ does not become time-translational invariant. Therefore we expect that plotting $G$ as a function of $t-t'$ for increasing values of $t'$ does not lead to a data collapse for the curves of $G(t,t')$. However, by definition $G(t',t')=1$ for all $t'$ and if the dynamics does not land on a limit cycle, $G(t,t')<1$ for $t>t'$.
We find that if $I_1$ and $I_2$ are sufficiently small the resulting dynamics is chaotic. This is shown in Fig.\ref{periodic_choerent_chaotic}

\begin{figure}
    \centering
    \includegraphics[scale=0.645]{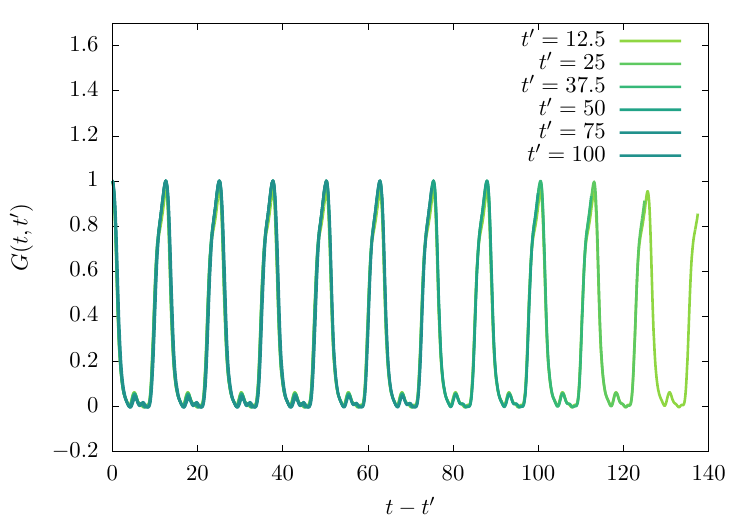}
    \includegraphics[scale=0.645]{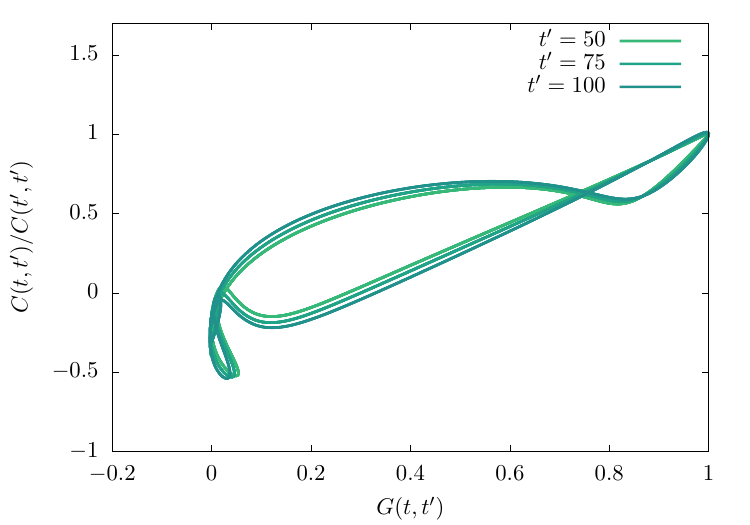}
    \caption{Superposition of two coherent periodic signals with $I_1=I_2=1.5$, $\omega_1=0.5$, $\omega_2=2$, $J_0=0$, $g=0.6$, $g_1=g_2=1$ and the confining term is given by $\hat \mu(z)=z$. Left panel represents the behavior of $G(t,t')$ as a function of $t-t'$ and for increasing values of $t'$. The right panel represents the behavior of $C(t,t')/C(t',t')$ as a function of $t$ plotted parametrically as a function of $G(t,t')$. The curves do close the cycle at (1,1) and therefore the dynamics is a limit cycle.}
    \label{fig:coherent_cycle}
\end{figure}

Conversely, having $G(t,t')$ reaching one when $t>t'$ does not allow to conclude that the dynamics lands on a limit cycle. 
To fix this point, we plot $C(t,t')/C(t',t')$ as a function of $G(t,t')$ for $T>t'$. If the dynamics is on a limit cycle, the corresponding curve should pass through the point (1,1) multiple times\footnote{Actually, given that the dynamics is deterministic, if the point (1,1) is crossed once it is crossed a infinite (numerable) number of times.}.

In Fig.\ref{fig:coherent_cycle} we show that this situation happens at least if $I_1$ and $I_2$ are sufficiently large.

All in all these results show that as soon as the dynamics is forced via a coherent superposition of periodic signals with commensurate frequencies, increasing the overall amplitude of the external drive can induce limit cycles. 
Conversely when the external drive is not strong enough the asymptotic dynamics breaks time-translational invariance with a chaotic periodic motion with $|\bx|^2/N \in [C_0^\textrm{ min},C_0^\textrm{ max}]$.

\section{Maximum Lyapunov exponent}\label{Sec_MLE}
Chaotic attractors can be characterized by their maximal Lyapunov exponent (MLE), which describes how trajectories are sensible to small perturbations in their initial condition~\cite{Ott_2002}.

The computation of the MLE in random recurrent neural networks  via DMFT was first done in a foundational work by Sompolinsky, Crisanti and Sommers \cite{sompolinsky1988chaos,crisanti2018path}. 
However, their calculations rely on field theory methods that presents complications when one wants to extend the computation to cases where one has more complex order parameters (ferromagnetic phases for example). Moreover, their formalism and model does not allow for exact solution: they arrive to a set of equations for the MLE that involve the diagonalization of a differential operator. Finally, their work heavily relies on the assumption that the steady state is stationary.

In this section, we study the problem of determining the MLE for the class of models we are focusing on. We develop a formalism that is alternative to the one of \cite{sompolinsky1988chaos} and that partially avoids the field theory techniques that becomes more complicated when dealing with several order parameters. Our technique is similar to \cite{helias2020statistical, schuecker2018optimal} and it leads to equations that are close to those presented in \cite{sompolinsky1988chaos}. In some of the models that we analyze we can fully compute the analytical expression of the MLE and the asymptotic divergence of the distance between two closely initialized trajectories in terms of the control parameters of the dynamical systems. In the cases in which this is not possible we can still access the MLE by fitting accurately the exponential divergence of two closely initialized trajectories that we can determine via DMFT.

In the next sections, we define the MLE and we outline its computation via DMFT. We finally compare the results of this analysis with numerical simulations.

\subsection{Definition of the maximal Lyapunov exponent}
When the system is in a chaotic phase, one can study what happens to dynamical trajectories when they are slightly perturbed at a given time and measure their sensitivity to this perturbation.
Consider the dynamical systems in Eq.~\eqref{generic_eq}. Suppose that two trajectories of the system, $\bx_1(t)$ and $\bx_2(t)$, are evolved up to time $t_0$ starting from the same initial condition at $t=0$. Then, they are slightly perturbed as
\begin{equation}
\label{eq: random perturbation}
    \bx_\alpha (t_0^+) = \bx(t_0)+ \eps \bb_\alpha\ \ \ \ \ \alpha=1,2\:.
\end{equation}
The two vectors $\bb_1$ and $\bb_2$ are in principle generic, but without loosing generality, we can assume that $\bb_{1,2}$ are two uncorrelated random vectors with Gaussian independent entries with zero mean and unit variance.
Therefore the two trajectories are the same up to $t_0$ and start to become different for $t>t_0$.

In a chaotic system the distance between the two trajectories, $\delta\bx(t)=\bx_1(t)-\bx_2(t)$ grows exponentially at long times
\begin{equation}
    |\delta \bx(t)| \sim \eps |\bb_1-\bb_2|\, e^{\Lambda_0 (t-t_0)}\ \ \ \ t\gg t_0
    \label{def_lyap_simple}
\end{equation}
The argument of the exponential factor defines the MLE, $\Lambda_0$.
In a random dynamical system as the class of models that we are considering, the distance $\delta \bx(t)$ has both positive and negative entries. 
Therefore, it is useful to define the norm of $\delta\bx$
\begin{equation}
\label{eq: D from simulations}
    D(t,t_0)= \frac 1N \sum_{i=1}^N \delta x_i(t)^2
\end{equation}
where we made explicit that $D$ is a function of the time at which the norm is measured $t$ and the time at which the perturbation is done, $t_0$. By construction, $D(t<t_0,t_0)=0$.
The MLE can then be computed as
\begin{equation}
\Lambda_0 = \lim_{t-t_0\to \infty}\frac 1{2(t-t_0)} \lim_{\eps\to 0} \ln\frac{D(t,t_0)}{2\eps^2}\:.
\end{equation}
Therefore, $D(t,t_0)$ gives access to the MLE.

\subsection{DMFT analysis}
The distance $D(t,t_0)$ is a self-averaging quantity \cite{crisanti2018path,schuecker2018optimal}, indeed one can evaluate it directly by decomposing it into the dynamical correlation functions between the two copies of the system and developing a DMFT treatment.

The two copies are initialized at the same initial point, perform the same trajectory up to time $t_0$, get perturbed by the addition of two infinitesimally small random vectors, and then are let free to evolve.

It is important to remark that at all times in this protocol, the two trajectories do not influence each other. They perform a dynamical path described by the same original dynamical system (except for the perturbation at time $t_0)$ and there is no interaction between them.

\subsubsection{Two replicas formalism}
The computation of $D(t,t_0)$ can be done by extending the path integral formalism developed in Sec.~\eqref{path_integral_derivation} to the two trajectories (or replicas).
Given that the they are not interacting, their dynamics can be reduced to two self-consistent stochastic processes
\begin{equation}
    \frac{\de x_\alpha(t)}{\de t} = -\mu_\alpha(t)x_\alpha(t) + \eta_\alpha(t) + J_0 m_\alpha(t)+h \ \ \ \ \alpha=1,2\:.
\end{equation}
The noise $\eta_\alpha(t)$ is Gaussian with the following structure
\begin{equation}
    \begin{split}
        \langle\eta_\alpha(t) \rangle &= 0\\
        \langle \eta_\alpha(t)\eta_\beta(t')\rangle &= g^2 f(C_{\alpha\beta}(t,t'))
    \end{split}
\end{equation}
where the correlation function $C_{\alpha\beta}(t,t')$ represent the large $N$ limit of
\begin{equation}
    C_{\alpha\beta}(t,t')=\lim_{N\to \infty}\frac 1N \sum_{i=1}^N x_{\alpha,i}(t)x_{\beta,i}(t')\:.
\end{equation}
The magnetization $m_\alpha(t)$ is defined as
\begin{equation}
    m_\alpha(t) = \lim_{N\to \infty}\frac 1N \sum_{i=1}^N x_{\alpha,i}(t)\:.
\end{equation}
It is important to stress that even if the two trajectories are non-interacting, they are subjected to the same random forcing field $\br$ and therefore their motion is correlated.

It is useful for numerical purposes also to write the self-consistent stochastic process that corresponds to the discrete time dynamical system. This is given by
\begin{equation}
     x_\alpha(t+\de t) =x_\alpha(t)+\de t\left[ -\mu_\alpha(t)x_\alpha(t) + \eta_\alpha(t) + J_0 m_\alpha(t)+h\right]
     \label{eq_discrete_2}
\end{equation}

The DMFT analysis can be easily extended to the introduction of the perturbation at time $t_0$.
We assume that the two replicas are initialized at the same point, characterized by $\Ctt$ and $\mtt$. 
Up to time $t\leq t_0$, their evolution is described by Eq.~\eqref{eq_discrete_2}. At time $t_0$, the two trajectories get perturbed so that 
\begin{equation}
    x_\alpha(t_0+\de t) = x_\alpha(t_0) +\de t \left[ -\mu_\alpha(t)x_\alpha(t) + \eta_\alpha(t) + J_0 m_\alpha(t)+h\right] + \eps b_\alpha,
\end{equation}
where $b_\alpha$ are two uncorrelated Gaussian random variables with zero mean and unit variance.
This process can be described in the single equation
\begin{equation}
    x_\alpha(t+\de t) =x_\alpha(t)+\de t \left[ -\mu_\alpha(t)x_\alpha(t) + \eta_\alpha(t) + J_0 m_\alpha(t)+h\right] + H_\alpha(t),
\end{equation}
where
\begin{equation}
    \begin{split}
        H_\alpha(t)&=\begin{cases}
        \eps b_\alpha & t=t_0\\
        0 & t\neq t_0\:.
        \end{cases}
    \end{split}
\end{equation}
Note that one can take easily the continuous time limit by setting
\begin{equation}
    \hat H_\alpha(t) = \eps b_\alpha \delta(t-t_0)
\end{equation}
so that
\begin{equation}
        \frac{\de x_\alpha(t)}{\de t} = -\mu_\alpha(t)x_\alpha(t) + \eta_\alpha(t) + J_0 m_\alpha(t)+h+\hat H_\alpha(t)\:.
        \label{perturbed_SCSP}
\end{equation}
Eq.~\eqref{perturbed_SCSP} can be solved by projecting it onto the dynamics of the correlation and response functions.
We find
\begin{equation}
    \begin{split}
        \partial_t C_{\alpha\beta}(t,t') &= -\mu_\alpha(t) C_{\alpha\beta}(t,t') + g^2\sum_{\gamma=1,2}\int_0^{t'}\de s f(C_{\alpha\gamma}(t,s))R_{\beta\gamma}(t',s) \\
        &+ (J_0m_{\alpha}(t)+h)m_\beta(t')\ \ \ \ \ \ \ t>t'\\
        \frac{\de C_{\alpha\beta}(t,t)}{\de t} &= 2\left[-\mu_\alpha(t) C_{\alpha\beta}(t,t) + g^2\sum_{\gamma=1,2}\int_0^{t}\de s f(C_{\alpha\gamma}(t,s))R_{\beta\gamma}(t,s)\right] \\
        &+ 2(J_0m_{\alpha}(t)+h)m_\beta(t)+ \delta_{\alpha\beta}\eps^2\delta(t-t_0)\\
        \partial_tR_{\alpha\beta}(t,t')&=-\mu_\alpha(t)R_{\alpha\beta}(t,t') + \delta_{\alpha\beta}\delta(t-t')\\
        \mu_\alpha(t) &= \hat \mu(C_{\alpha\alpha}(t,t))\\
        \frac{\de m_\alpha(t)}{\de t}&=(J_0-\mu_\alpha(t))m_{\alpha}(t)+h
        \end{split}\label{2rep_DMFT}
\end{equation}
with initial conditions:
\begin{equation}
    \begin{split}
        C_{\alpha\beta}(0,0)&=\Ctt\\
        m_\alpha(0)&=\mtt \ \ \ \ \forall \alpha,\beta=1,2\:.
    \end{split}
    \label{init_2rep}
\end{equation}
Finally we note that $C_{\alpha\beta}(t,t')=C_{\beta\alpha}(t',t)$.
We call Eq.~\eqref{2rep_DMFT} the two-replicas DMFT equations.

\subsubsection{Statistical symmetry for the solution of the two-replicas DMFT equations}
The two-replicas DMFT equations admit a simple solution. Indeed, given the initial conditions in Eq.~\eqref{init_2rep}, and given that the perturbation of the trajectories affects the term in the evolution of $C_{\alpha\beta}(t,t)$ only by a diagonal term, it follows that
\begin{equation}
C_{\alpha\beta}(t,t')=\delta_{\alpha\beta}C_d(t,t')+(1-\delta_{\alpha\beta})C_o(t,t') \:.    
\end{equation}
Furthermore, the solution for $R_{\alpha\beta}(t,t')$ has the following diagonal structure
\begin{equation}
    R_{\alpha\beta}(t,t') = \delta_{\alpha\beta}R_d(t,t') =  \delta_{\alpha\beta}\theta(t-t') \exp\left[-\int_{t'}^t\de s \mu_\alpha(s)\right]\:.
\end{equation}
Finally, given that the two replicas are statistically identical for $t>t_0$ and are exactly the same for $t\leq t_0$, we have that $m_\alpha(t)=\tilde m(t)$ independently on $\alpha=1,2$. The same argument applies to $\mu_\alpha(t) = \tilde \mu(t) =\hat \mu(C_d(t,t))$ for all $\alpha=1,2$.

\subsubsection{Symmetric two replicas DMFT equations}
\label{subsubsect: Cd, Co DMFT formalism}
The structure of the solutions of the two-replicas DMFT equations can be used to reduce their complexity to
\begin{equation}
    \begin{split}
        \partial_t C_d(t,t') &= -\tilde \mu(t) C_d(t,t') + g^2\int_0^{t'}\de s f(C_d(t,s))R_d(t',s) \\
        &+ (J_0\tilde m(t)+h)\tilde m(t')\ \ \ \ \ \ \ t>t'\\
        \partial_t C_o(t,t') &= -\tilde \mu(t) C_o(t,t') + g^2\int_0^{t'}\de s f(C_o(t,s))R_d(t',s) \\
        &+ (J_0\tilde m(t)+h)\tilde m(t')\ \ \ \ \ \ \ t>t'\\
        \frac{\de C_{d}(t,t)}{\de t} &= 2\left[-\tilde \mu(t) C_d(t,t) + g^2\int_0^{t}\de s f(C_d(t,s))R_d(t,s)\right] \\
        &+ 2(J_0\tilde m(t)+h)\tilde m(t)+ \eps^2\delta(t-t_0)\\
        \frac{\de C_{o}(t,t)}{\de t} &= 2\left[-\tilde \mu(t) C_o(t,t) + g^2\int_0^{t}\de s f(C_o(t,s))R_d(t,s)\right] \\
        &+ 2(J_0\tilde m(t)+h)\tilde m(t)\\
        \partial_tR_d(t,t')&=-\tilde \mu(t)R_d(t,t') + \delta(t-t')\\
        \tilde \mu(t) &= \hat \mu(C_d(t,t))\\
        \frac{\de \tilde m(t)}{\de t}&=(J_0-\tilde \mu(t))\tilde m(t)+h \:.
    \end{split}
    \label{Symm_DMFT_2rep}
\end{equation}
The initial conditions for such equations are
\begin{equation}
    \begin{split}
    &C_d(0,0)=C_o(0,0)=\Ctt\\
    &\tilde m(0)=\mtt\:.
    \end{split}
    \label{init_symm_2DMFT}
\end{equation}
Eqs.~\eqref{Symm_DMFT_2rep} have a set of important properties:
\begin{itemize}
    \item Among all the unknowns, $C_d$, $R_d$, $\tilde m$ and $\tilde \mu$ obey a set of closed, autonomous PDEs. We call the corresponding equations the diagonal subset of the equations. This property follows directly from the fact that the two replicas are non-interacting.
    \item By construction, for all times $t,t'\leq t_0$, we have  that $C_d(t,t')=C_o(t,t')$.
    \item For $\eps\to 0$, the diagonal subset of the symmetric DMFT equations converges to the single replica DMFT equations of Sec.\ref{DMFT_C_R_m}. It follows that the asymptotic steady state solution is characterized in the same way: for example, for $t$ and $t'$ large, we have $\lim_{\eps\to 0} C_d(t,t') = c(t-t')$.
\end{itemize}
The distance between the two perturbed replicas can be computed as
\begin{equation}
    D(t,t_0) = 2(C_d(t,t)-C_o(t,t))
\end{equation}
Note that for $t\leq t_0$, one has $D(t,t_0)=0$, while $D(t_0^+,t_0)=2\eps^2$.

\paragraph{Discrete time version of the symmetric two-replicas DMFT equations --} It is useful, when it comes to numerically integrate the DMFT equations, to write the corresponding ones for the discrete time version of the dynamical system.
Here, we report the equations under the symmetric ansatz directly.
\begin{align}
    \label{2Rep_discrete beginning}
    C_d(t+\de t,t') &= C_d(t,t') + \de t \left[ -\tilde \mu(t) C_d(t,t') + g^2\de t \sum_{n=0}^{t'/\de t} f(C_d(t,n\de t))R_d(t',n\de t) \right.\\
    &\left. + (J_0\tilde m(t)+h)\tilde m(t')\right]\ \ \ \ \ \ \ t>t' \nonumber\\
    C_o(t+\de t,t') &= C_o(t,t') + \de t \left[ -\tilde \mu(t) C_o(t,t') + g^2\de t \sum_{n=0}^{t'/\de t} f(C_o(t,n\de t))R_d(t',n\de t) \right.\\
    &\left. + (J_0\tilde m(t)+h)\tilde m(t')\right]\ \ \ \ \ \ \ t>t'\nonumber\\
    C_{d}(t+\de t,t+\de t) &= C_d(t,t)+ 2\de t \left[-\tilde \mu(t) C_d(t,t) + g^2\de t\sum_{n=0}^{t/\de t} f(C_d(t,n\de t))R_d(t,n\de t)\right] \\
    &+ 2\de t (J_0\tilde m(t)+h)\tilde m(t)+ \eps^2\delta_{t,t_0}+\de t^2 L_d(t)\nonumber\\
    C_{o}(t+\de t,t+\de t) &= C_o(t,t)+ 2\de t \left[-\tilde \mu(t) C_o(t,t) + g^2\de t\sum_{n=0}^{t/\de t} f(C_o(t,n\de t))R_d(t,n\de t)\right] \\
    &+ 2\de t (J_0\tilde m(t)+h)\tilde m(t)+\de t^2 L_o(t)\nonumber\\
    R_d(t+\de t,t') &= R_d(t,t')-\de t \tilde \mu(t)R_d(t,t') + \delta_{t,t'}\\
    \tilde \mu(t) &= \hat \mu(C_d(t,t))\\
    \tilde m(t+\de t)&=\tilde m(t) + \de t\left[(J_0-\tilde \mu(t))\tilde m(t)+h\right]\\
    L_d(t)&=\tilde \mu^2(t) C_d(t,t) +(J_0 \tilde m(t)+h)^2-\tilde\mu(t)(J_0 m(t)+h)\tilde m(t)\\
    &- 2g^2 \tilde \mu(t)\sum_{n=0}^{t/\de t}f(C_d(t,n\de t))R_d(t,n\de t)  {+ g^2f(C_d(t,t))} \nonumber \\
    \label{2Rep_discrete end}
    L_o(t)&=  \tilde \mu^2(t) C_o(t,t) +(J_0 \tilde m(t)+h)^2-\tilde\mu(t)(J_0 m(t)+h)\tilde m(t)\\
    &- 2g^2 \tilde \mu(t)\sum_{n=0}^{t/\de t}f(C_o(t,n\de t))R_d(t,n\de t) {+ g^2f(C_o(t,t))} \nonumber \:.  
\end{align}
Note that we have used the convention that $R_d(t,s)=0$ if $s\geq t$.
These equations have to be integrated with the initial condition given by Eq.~\eqref{init_symm_2DMFT}.

\subsubsection{Phenomenology of exponential divergence of chaotic trajectories}
\begin{figure}
    \centering
    \includegraphics[width=\textwidth]{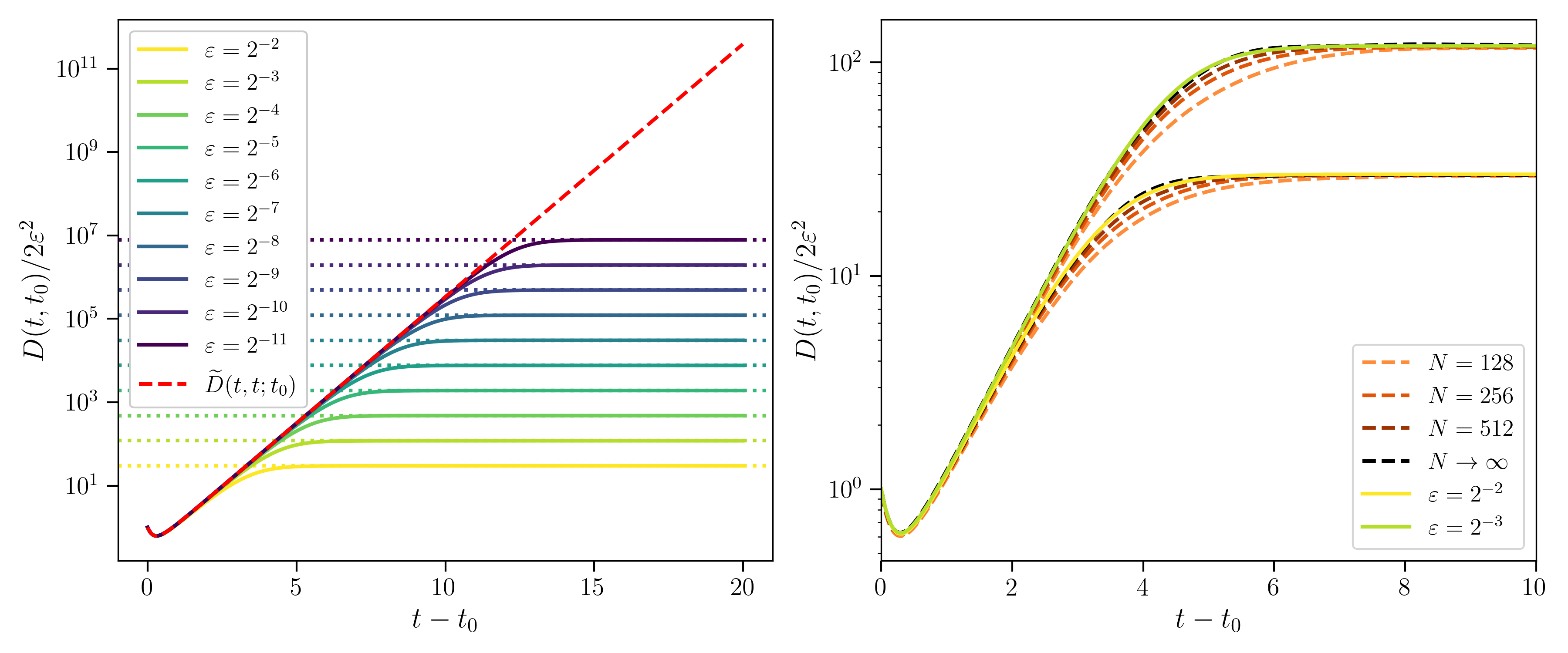}
    \includegraphics[width=\textwidth]{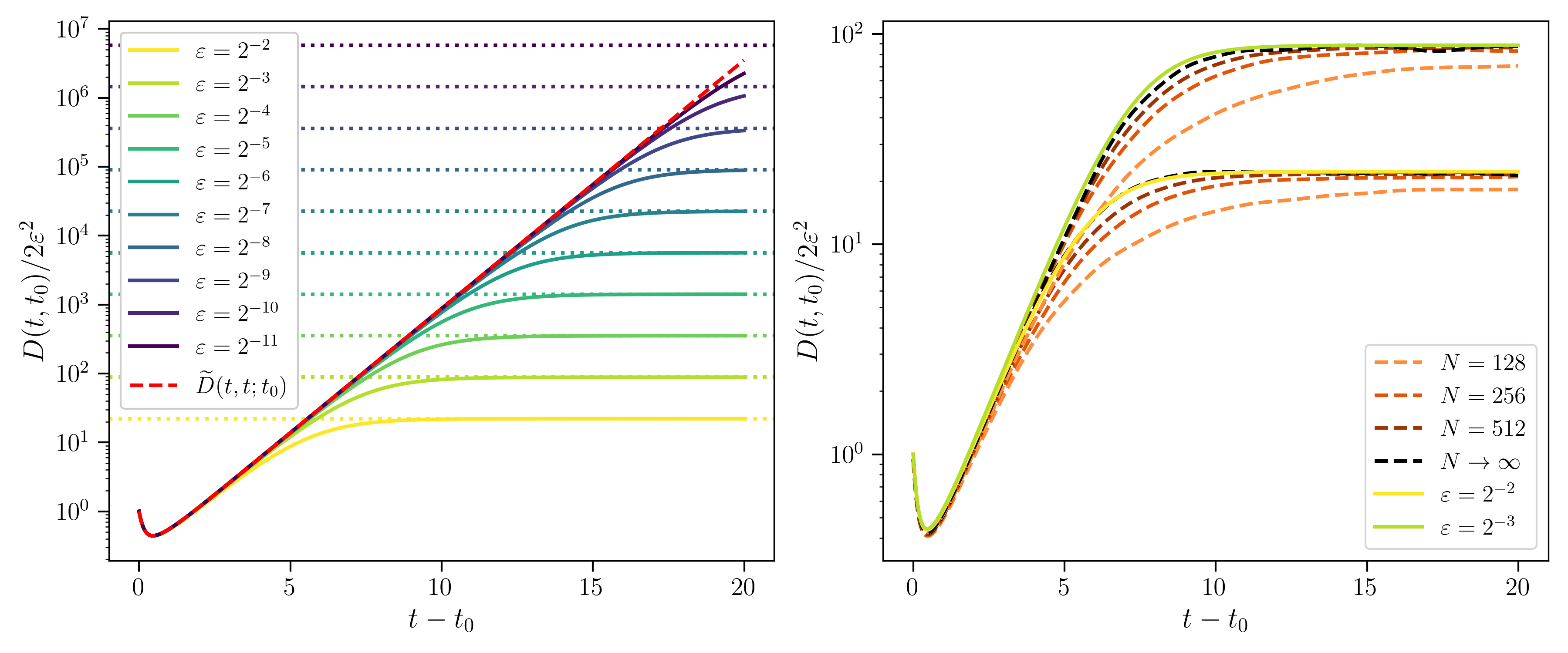}
    \caption{The distance between two replicas as a function of time after the perturbation. We integrated numerically the discrete time version of the two-replicas DMFT equations for decreasing values of $\eps$ (\textit{full} lines). At large times, each curve converge to a plateau which is given by $(C_0-C_\infty)/\eps^2$ (\textit{dotted} lines). For $\eps\to 0$, these curves superimpose on a master curve  which corresponds to the one obtained by integrating numerically Eqs.~\eqref{DMFT_MLE_Dtilde} (\textit{red dashed} line). The result of DMFT is compared to simulations of increasing system-size $N$ (\textit{orange dashed} lines), see sect.~\ref{sect: simulations for lyapunov}.
    Data shown for the simplest model of Eq.~\eqref{def_simplest} with (Upper panel) $\hat \mu(z)=z$, $g=g_1=g_2=1$, $J_0=h=0$ and (Lower panel) $\hat \mu(z)=1+z$, $g_1=2$, $g_2=g=1$, $J_0=h=0.5$, and in both cases $\de t=0.01$.}
    \label{fig: comparison Dtilde D}
\end{figure}

The Symmetric two-replicas DMFT equations can be integrated numerically.
In \ref{fig: comparison Dtilde D}-left, we plot the behavior of $D(t,t_0)/(2\eps^2)$ as a function of $t-t_0$ for decreasing values of $\eps$. The curves have an initial dip followed by a linear ramp in linear-log scale. This implies that at these intermediate times, $D$ grows exponentially with time. The exponential growth is saturated at later times and $D(t,t_0)$ converges to an $\eps$-dependent plateau. The value of the plateau can be understood in the following way. At late times, the two trajectories become completely uncorrelated. This implies that
\begin{equation}
    \lim_{t\to \infty}\frac{D(t,t_0)}{2\eps^2}=\frac{C_0-C_\infty}{\eps^2}
    \label{predicion_Asym_D}
\end{equation}
The prediction in Eq.~\eqref{predicion_Asym_D} corresponds to the horizontal dotted lines in Fig.\ref{fig: comparison Dtilde D}.
In Fig.\ref{fig: comparison Dtilde D}-right, we compare the result of the numerical integration of the DMFT equations with numerical simulations at finite but increasing system size. We get that the numerical simulations curve approach the theoretical one when $N$ increases and we estimate the infinite size limit of these curves via a polynomial fit. The agreement is excellent.

The results of this discussion is that $D(t,t_0)/(2\eps^2)$ shows an exponential divergence when $\eps\to 0$.

\subsection{Calculation of the maximum Lyapunov exponent}
The calculation of the MLE can be done by tracking where, in the DMFT formalism, the distance $D(t,t_0)$ develops an exponential divergence. In this section, we will focus on the continuous time case.
Define 
\begin{equation}
    \tilde D(t,t';t_0) = \lim_{\eps\to 0}\frac 1{\eps^2}\left[C_d(t,t')-C_o(t,t')\right]\:.
\end{equation}
The MLE is then given by
\begin{equation}
    \Lambda_0 = \lim_{t-t_0\to \infty}\frac 1{2(t-t_0)} \ln \tilde D(t,t;t_0)
\end{equation}
Note that $\tilde D(t,t';t_0) = 0$ as soon as either $t<t_0$ or $t'<t_0$.
The equation that fixes $\tilde D$ can be obtained directly from the DMFT equations. Indeed, taking the difference between the flow equation for $C_d$ and the one for $C_o$, multiplying the resulting equation by $1/\eps^2$ and taking the $\eps\to 0 $ limit, one gets
\begin{equation}
    \begin{split}
        \partial_t\tilde D(t,t';t_0) &= -\mu_\infty \tilde D(t,t';t_0) + g^2\int_{t_0}^{t'}\de s f'(c(t-s))\tilde D(t,s;t_0)r(t'-s) \ \ \ \ t>t'\\
        \frac {\de \tilde D(t,t;t_0)}{\de t} &= -2\mu_\infty \tilde D(t,t;t_0) + 2g^2\int_{t_0}^{t}\de s f'(c(t-s))\tilde D(t,s;t_0)r(t-s) + \delta(t-t_0)
    \end{split}
    \label{DMFT_MLE_Dtilde}
\end{equation}
Note that since $\tilde D$ vanishes as soon as one of its two time arguments is smaller than $t_0$, we can fix $t_0=0$ from now on without loss of generality. 
Eqs.~\eqref{DMFT_MLE_Dtilde} can be solved numerically.
In Fig.\ref{fig: comparison Dtilde D}-left, we plot the numerical integration of Eqs.~\eqref{2Rep_discrete beginning}-\eqref{2Rep_discrete end} for decreasing values of $\eps$ and compare the corresponding data with the numerical solution of Eq.~\eqref{DMFT_MLE_Dtilde}. The agreement is excellent.

An important remark is that, in Eqs.~\eqref{DMFT_MLE_Dtilde}, the dependence on the control parameters other than $g$ are all in the solution $c(\tau)$. In other words, the previous equations do not depend explicitly on $h$ and $J_0$ and correspondingly on $m_\infty$: the dependence on these quantities is entirely contained in $c(\tau)$ which solves a set of equations that do depend on the control and order parameters.
 
Taking a derivative with respect to $t'$ of Eq.~\eqref{DMFT_MLE_Dtilde} and taking into account the corresponding initial condition, one gets
\begin{equation}
    (\partial_t +\mu_\infty)(\partial_{t'}+\mu_\infty)\tilde D(t,t';t_0) =  g^2 f'(c(t-t'))\tilde D(t,t';t_0) + \delta(t-t_0)\delta(t'-t_0)\:.
    \label{close_to_sch}
\end{equation}
It is useful to perform the change of variables
\begin{gather*}
  \begin{cases}
    &T = t+t'\\
    &\tau = t-t'
  \end{cases}
  \;\;\;\;\;
  \begin{cases}
    &\partial_{t} = \partial_T + \partial_\tau\\
    &\partial_{t'} = \partial_T - \partial_\tau
  \end{cases}.
\end{gather*}
Therefore Eq.~\eqref{close_to_sch} can be rewritten as
\begin{equation}
    \left(\partial^2_T+2\mu_\infty \partial_T+\mu_\infty^2-\partial_\tau^2\right)\tilde D(T,\tau;t_0) =  g^2 f'(c(\tau))\tilde D(T,\tau;t_0)+2\delta(T-2t_0)\delta(\tau)
    \label{bethe_salpeter}
\end{equation}
and we used a simplified notation $\tilde D(T,\tau;t_0) \equiv \tilde D(T+\tau,T-\tau;t_0)$.
The solution of Eq.~\eqref{bethe_salpeter} can be written as
\begin{equation}
    \tilde D(T,\tau;t_0) = 2\sum_{n\geq 0}g_n(T,t_0)\varphi_n(\tau)\varphi_n^*(0)
    \label{general_sol}
\end{equation}
where $\varphi_n(\tau)$ are the orthonormal solutions of the Schr\"odinger problem
\begin{equation}
    \left[\mu_\infty^2-\partial_\tau^2-g^2f'(c(\tau))\right]\varphi_n(\tau)=\varepsilon_n\varphi_n(\tau)
    \label{schrodinger}
\end{equation}
Note that the sum in Eq.~\eqref{general_sol} can become an integral if the Schr\"odinger Hamiltonian as a continuous spectrum. 
Plugging the ansatz in Eq.~\eqref{general_sol} into Eq.~\eqref{bethe_salpeter}, one gets an equation for the growth factors $g_n$
\begin{equation}
    \left[\partial_T^2+2\mu_\infty\partial_T+\varepsilon_n\right]g_n(T,t_0)=\delta(T-2t_0)\:.
    \label{eq_G}
\end{equation}

Eqs.\eqref{bethe_salpeter}-\eqref{eq_G} appear also in more standard recurrent neural networks \cite{crisanti2018path}. 
Knowning the solution to Eq.~\eqref{schrodinger}, one can compute the MLE since
\begin{equation}
    \tilde D(t,t;t_0)=2\sum_{n\geq 0} g_n(2t,t_0)\varphi_n(0)\varphi^*_n(0) \:.
\end{equation}
Therefore, the MLE is related to the behavior of $g_n(t,t_0)$ for $t-t_0\to \infty$.

We will see that for the simplest dynamical system in Eq.~\eqref{def_simplest} the Schr\"odinger problem can be solved exactly.

\subsubsection{The maximum Lyapunov exponent in the simplest model}
We now specialize the computation to the simplest class of models in Eq.~\eqref{def_simplest}. 
We need to solve two equations: the Schr\"odinger problem in Eq.~\eqref{schrodinger} and Eq.~\eqref{eq_G}.
We analyze the two problem separately.

\paragraph{The growth factor problem --} In this section, we discuss the solution of Eq.~\eqref{eq_G}.
The solution for $g_n$ is
\begin{equation}
    g_n(T,t_0)=\frac{\theta(T-2t_0)}{\sqrt{\mu_\infty^2-\eps_n}}e^{-\mu_\infty(T-2t_0)}\sinh\left[{\sqrt{\mu_\infty^2-\eps_n}(T-2t_0)}\right]
\end{equation}
Remember that $T=t+t'$ and we are interested in the case where $t=t'$ so $T=2t$.
For $T-2t_0\to \infty$ and $\eps_n<\mu_\infty^2$, $g_n$ has an exponential divergence. In particular in this limit, we have
\begin{equation}
    g_n(T,t_0)\sim \frac 1{2\sqrt{\mu_\infty^2-\eps_n}}e^{\left(-\mu_\infty+\sqrt{\mu_\infty^2-\eps_n}\right)(T-2t_0)}\:.
    \label{asym_gn}
\end{equation}
This will be important for the evaluation of the maximal Lyapunov exponent.

\paragraph{The Schr\"odinger problem --}
The Schr\"odinger problem corresponding to the simplest dynamical system can be solved explicitly. 
Define
\begin{equation}
    V''(c) = f'(c)-\mu_\infty^2=g^2\left[g_1^2+4g_2^2c\right]-\mu_\infty^2\:.
\end{equation}
The eigenvalue problem becomes
\begin{equation}
    \left[-\partial_\tau^2-V''(c(\tau))\right] \psi(\tau)=\epsilon \psi(\tau) 
    \label{Sch_simp}
\end{equation}
Plugging the explicit form of $c(\tau)$ (see Eq.~\eqref{solution_C_tau}) inside Eq.~\eqref{Sch_simp}, we get
\begin{equation}
      \left[ \partial_\tau^2 + 4g_2^2g^2 C_0 - 4g_2^2g^2 (C_0-C_\infty) \tanh^2\left(\frac12w\tau\right) + g_1^2g^2- \mu_\infty^2 + \epsilon \right] \psi(\tau) =  0,
      \label{explicit_Sch}
\end{equation}
where 
\begin{equation}
w=\sqrt{\frac 43 (g g_2)^2(C_0-C_\infty)}\:.
\label{def_w}
\end{equation}
The potential $-V''(c)$ of the Schr\"odinger equation is a dip centered around $\tau=0$ and has two asymptotes at the same heigth for $\tau\to\pm \infty$. 
Therefore the spectrum of the corresponding Schr\"odinger operator has both a continuous and a discrete part.
Given that we need to look for the ground state, we will focus from now on on the discrete part.
This means that we will assume that
\begin{equation}
    \epsilon_n<\mu_\infty^2-g^2(g_1^2+4g_2^2C_\infty)\:.
    \label{contraint_en}
\end{equation}
Consider the change of variables
\begin{gather}
  x = \tanh\left(\frac12w\tau\right), \;\;\;\;\; \frac{\de}{\de\tau} = \frac12w(1-x^2) \frac{\de}{\de x}, \;\;\;\;\; \frac{\de^2}{\de \tau^2} = \frac14 w^2(1-x^2) \frac{\de }{\de x}\left((1-x^2)\frac{\de }{\de x} \right).
\end{gather}
Eq.~\eqref{explicit_Sch} can be rewritten as
\begin{equation}
  \frac14 w^2(1-x^2) \frac{\de }{\de x}\left((1-x^2)\frac{\de \psi(x)}{\de x} \right) + \left[ 4g^2 g_2^2 C_0 (1-x^2) + 4g^2 g_2^2 C_\infty x^2 + g^2 g_1^2- \mu_\infty^2 + \epsilon \right] \psi(x) = 0
\end{equation}
Divide the previous equation by $\frac14 w^2(1-x^2) = \frac 13 g^2g_2^2(C_0-C_\infty)(1-x^2)$. We get
\begin{equation*}
  \frac{\de }{\de x}\left((1-x^2)\frac{\de \psi(x)}{\de x} \right) + \left[ 12 \frac{C_0}{C_0-C_\infty} + 12 \frac{C_\infty}{C_0-C_\infty} \frac{x^2}{1-x^2} + \frac{3(\epsilon +g^2g_1^2- \mu_\infty^2)}{g^2g_2^2(C_0-C_\infty)(1-x^2)} \right] \psi(x) = 0
\end{equation*}
which can be re-written
\begin{equation}
  \frac{\de }{\de x}\left((1-x^2)\frac{\de \psi(x)}{\de x} \right) + \left[12 + \frac{12}{C_0-C_\infty}\frac{C_\infty - \frac{1}{4g^2g_2^2}(\mu_\infty^2-g^2g_1^2-\epsilon)}{(1-x^2)} \right] \psi(x) = 0.
\end{equation}
This is the equation for a generalized Legendre polynomial which in its canonical form reads
\begin{equation*}
  \frac{\de }{\de x}\left((1-x^2)\frac{\de \psi(x)}{\de x} \right) + \left[ S(S+1) - \frac{M^2}{1-x^2} \right] \psi(x) = 0,
\end{equation*}
where we can identify
\begin{align*}
  &S(S+1)=12 \implies S=3\\
  &M^2 = \frac{ \frac{3}{g^2g_2
  ^2}(\mu_\infty^2-g^2g_1^2-\epsilon)-12C_\infty}{C_0-C_\infty}.
\end{align*} 
Note that the condition $M^2>0$ can be satisfied only for $\epsilon_n$ that satisfies Eq.~\eqref{contraint_en}.
Then the spectrum of the Schr\"odinger problem can be found\footnote{see for example \cite{landau2013quantum}.} by setting $M-S=-n$ where $n \in \mathbb{N}$. The constraint on $n$ being a natural number imposes that $n<S=3$.
Therefore the energies $\eps_n$ are found to be
\begin{equation}
  \eps_n = \mu_\infty^2-g^2g_1^2 - g^2g_2^2  \left[4 C_\infty + \frac{(3-n)^2}{3}(C_0-C_\infty)\right]\ \ \ n<3\:.
\end{equation}
The eigenstates corresponding to these eigenvalues are Generalized Legendre polynomials. For example, setting $n=0$ we have
\begin{equation}
    \psi_0(x)=-15(1-x^2)^{3/2} \ \ \ \ x\in[-1,1]
\end{equation}
Therefore, going back to the $\tau$ variable, we have
\begin{equation}
    \varphi_0(x) = \sqrt{\frac{3w}{2(6!)}}\psi_0\left(\tanh\left(\frac12 w\tau\right)\right)\:.
\end{equation}

\paragraph{The Maximal Lyapunov Exponent --}
Eq.~\eqref{asym_gn} implies that the maximal Lyapunov exponent can be obtained by considering the ground state of the Schr\"odinger problem. This is obtained for $n=0$. Furthermore, we take $T=2t\gg 2 t_0$ so that
\begin{equation}
    \tilde D(t,t',t_0)\sim \tilde D_\infty(t,t',t_0)\equiv \frac{1}{\sqrt{\mu_\infty^2-\eps_0}} \varphi_0(0)\varphi^*_0(t-t')e^{2\left(-\mu_\infty+\sqrt{\mu_\infty^2-\eps_0}\right)(\frac 12(t+t')-t_0)}\ \ \ \ \ \ \ \ t-t_0\to \infty
    \label{full_prediction_Lambda}
\end{equation}
Therefore, the MLE can be evaluated by setting $t'=t$ and it is given by
\begin{equation}
  \Lambda_0 = -{\mu_\infty} +  \sqrt{\mu_\infty^2 - \eps_0} = -{\mu_\infty} + g\sqrt{g_1^2+g_2^2(C_\infty + 3 C_0)}\:.
  \label{Formula_Lyapunov}
\end{equation}
This formula was derived with standard methods and presented first in \cite{fournier2025non} for cases where $h=0$. Here we point out that such expression holds in full generality and can be used also to study metastable phases in the hysteresis loop.

The prefactor of the exponential growth of $\tilde D(t,t_0)$ can be also computed explicitly.
Indeed we have that 
\begin{equation}
    \varphi_0(0)\varphi_0^*(t-t') = \frac{15 w}{32}\left(1-\tanh^2\left(\frac 12 w (t-t')\right)\right)^{3/2}
\end{equation}
where $w$ is given by Eq.~\eqref{def_w}.

\begin{figure}
    \centering
    \includegraphics[scale=.645]{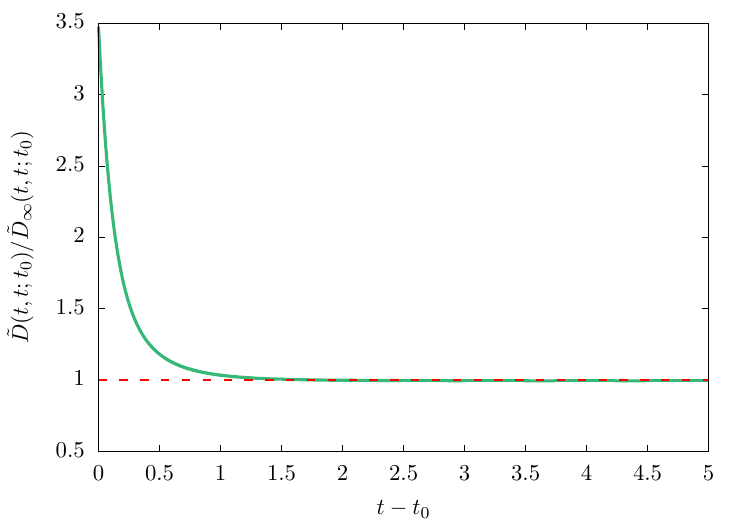}
    \includegraphics[scale=0.645]{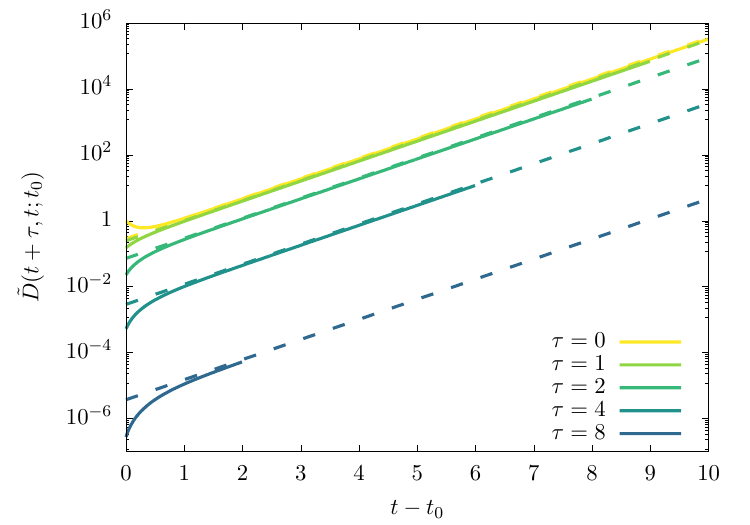}
    \caption{The growth rate of the distance between two perturbed replicas. Left panel: The prediction in Eq.~\eqref{full_prediction_Lambda} for $t=t'$. Left panel: the prediction from Eq.~\eqref{full_prediction_Lambda} for $t\neq t'$. Data comes from $\hat \mu(z)=z$, $g=g_1=g_2=1$ and $J_0=h=0$, same as in Fig.~\ref{fig: comparison Dtilde D}.}
    \label{check_full_prediction}
\end{figure}

We can test directly the prediction in Eq.~\eqref{full_prediction_Lambda}. In Fig.\ref{check_full_prediction}-left we plot $\tilde D(t,t;t_0)/\tilde D_\infty(t,t;t_0)$. For $t-t_0\to \infty$ this function approaches one, which confirms the prediction in Eq.~\eqref{full_prediction_Lambda}. In the right panel of the same figure, we evaluate $\tilde D(t,t'; t_0)$ for $t\neq t'$ and we test its asymptotic behavior against the prediction in Eq.~\eqref{full_prediction_Lambda}. The agreement is excellent.

\begin{figure}
    \centering
    \includegraphics[width=\textwidth]{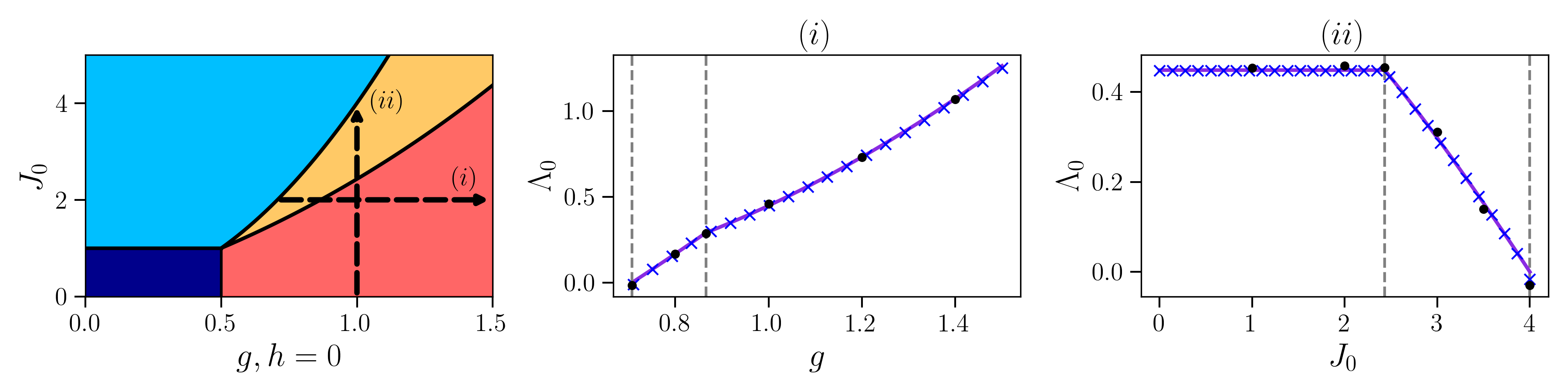}
    \includegraphics[width=\textwidth]{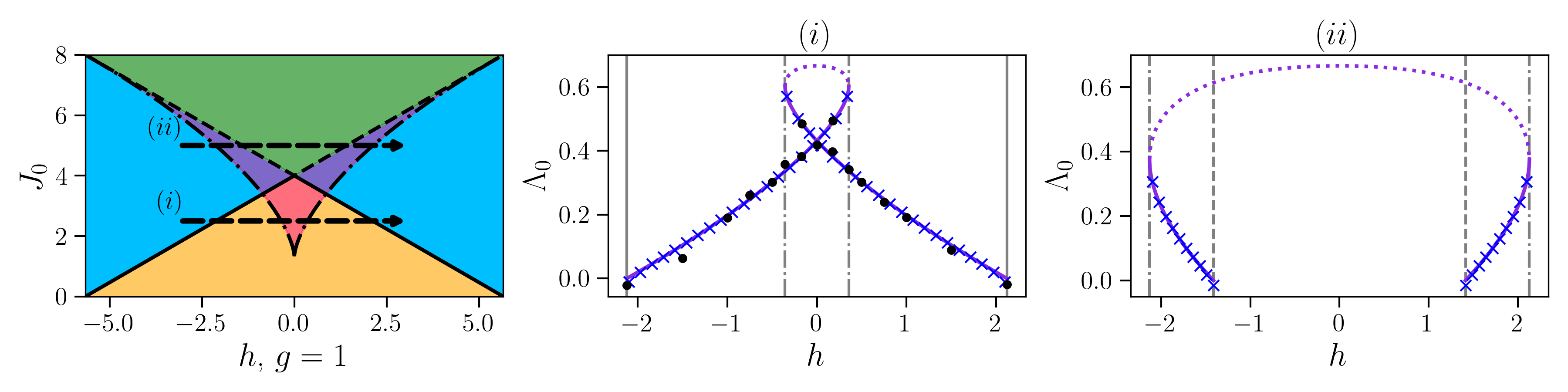}
    \includegraphics[width=\textwidth]{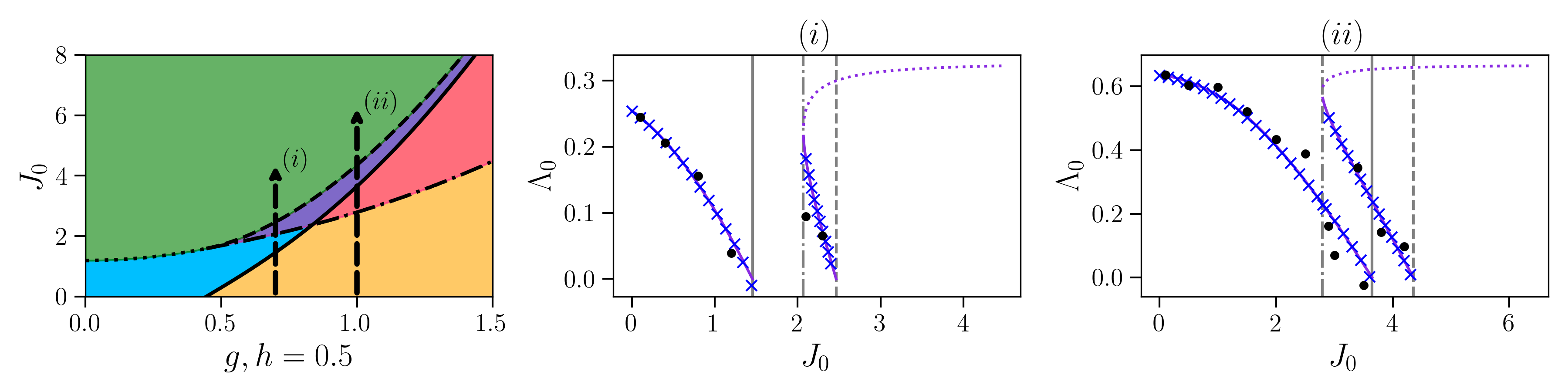}
    \caption{Maximum Lyapunov exponent across the phase diagrams. Upper panel: model with $\hat\mu(z)=1+z$, $g_1=2$, $g_2=1$ and $h=0$. This phase diagram is very similar to the one presented in \cite{fournier2025non}. Middle and lower panels: model with $\hat\mu(z)=z$, $g_1=0$ and $g_2=1$. \textit{Purple} lines are analytical results. \textit{Blue} x's are the result of fitting the exponential divergence of $\tilde{D}(t,t;t_0)$ with $dt=0.01$, $t_0=15$ and the fit was performed for times $t\in[t_0+15, t_0+20]$. \textit{Black} dots are predictions from finite size simulations (see sect.~\ref{sect: simulations for lyapunov}). The agreement between the analytical lines, the predictions from $\tilde{D}(t,t;t_0)$ and numerical simulations is excellent. Note that numerical simulations are hard to obtain in some regions of the phase diagrams since they require either a small integration timestep $dt$ to avoid numerical instabilities, or a large system size $N$ to land on the chaotic attractor.}
    \label{fig: MLE across the phase diagram and comparison with D_tilde and simulations}
\end{figure}


\subsection{The MLE across the phase diagram}
The MLE can be computed in the stationary phases in the chaotic regions of the phase diagram. Such analysis is presented in Fig.\ref{fig: MLE across the phase diagram and comparison with D_tilde and simulations} where we show  both the MLE computed with its explicit expression and a comparison with its value extracted from the fitting of the exponential divergence of $\tilde D$. The agreement between the predicition is excellent. We note that the MLE is in general non-divergent. In particular the MLE approaches a finite limit close to the spinodal transitions. This is due to the fact that it is computed by first taking $N\to \infty$ and then the long time limit. This order in the limits is responsible for stabilizying the steady state phases up to the spinodal points.

\subsubsection{Comparison with numerical simulations}
\label{sect: simulations for lyapunov}
\begin{figure}
    \centering
    \includegraphics[width=0.4\textwidth]{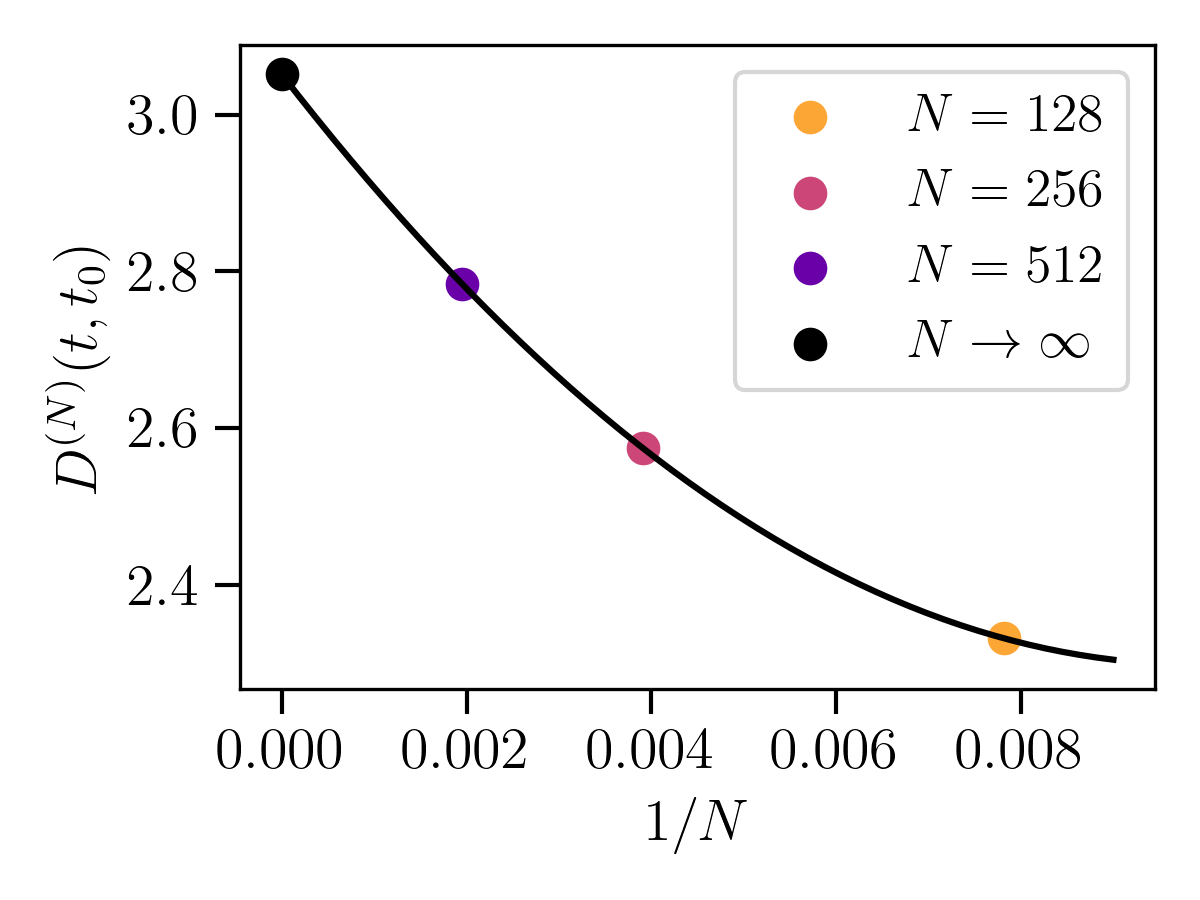}
    \includegraphics[width=0.5\textwidth]{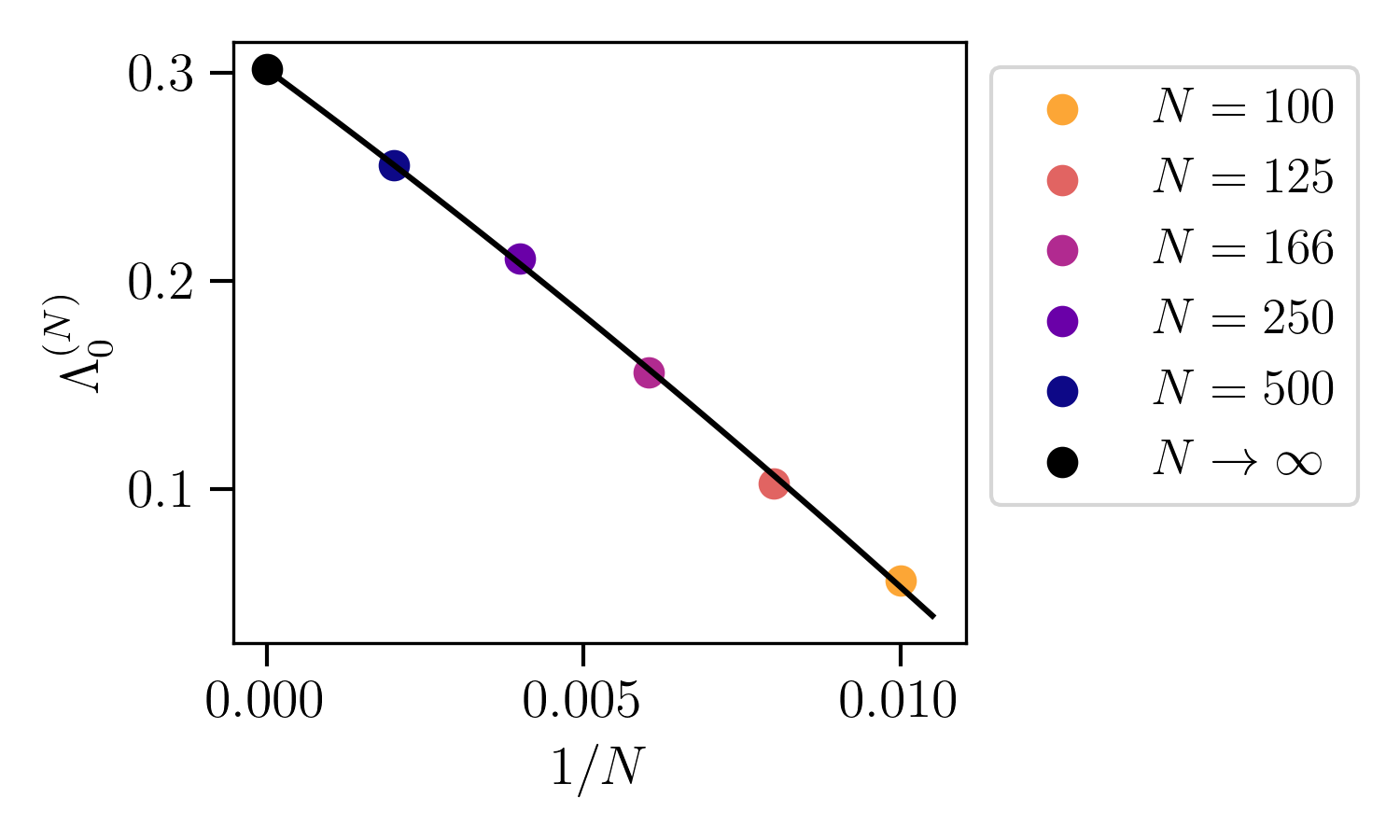}
    \caption{(Left) quadratic fit to extract $D(t,t_0)$ from finite-size simulations. The model is with $\hat\mu(z)=z$, $g_1=g_2=1$ and the other parameters are $g=1$, $J_0=h=0$, $dt=0.01$, $S=1000$ samples, $\varepsilon=2^{-2}$, $t_0=15$ and $t=t_0+4$. (Right) quadratic fit to extract $\Lambda_0$ from finite-size simulations. The model is with $\hat\mu(z)=z$, $g_1=0$, $g_2=1$ and the other parameters are $g=1$, $J_0=2.5$, $h=-0.5$, $S=100$ samples, $N_\mathrm{iter}=300/dt$, $N_\mathrm{transient}=100/dt$ and $dt=0.01$.}
    \label{fig: quadratic fit for infinite size predictions}
\end{figure}

The results developed so far can be also validated with numerical simulations. First, the finite-$\varepsilon$ DMFT formalism of \ref{subsubsect: Cd, Co DMFT formalism} can be checked by directly simulating the finite-size, two-replicas dynamical systems
\begin{equation}
    \bx_\alpha(t+dt) = \bx_\alpha(t) + \de t {\bf \mathcal{E}}(\bx_\alpha(t))\ \ \ \ \ \alpha=1,2\:
\end{equation}
where at $t_0$, each replica receives independently of one another a random perturbation of magnitude $\varepsilon$ as in \eqref{eq: random perturbation}. Then, for any finite $\varepsilon$ and $N$, we can define
\begin{equation}
    D^{(N)}(t,t_0) = \frac1N \sum_{i=1}^N \langle (\delta x_i(t))^2 \rangle_S, \;\;\;\;\; \delta \bx(t) = \bx_1(t) - \bx_2(t)
\end{equation}
where $\langle \cdot \rangle_S$ is an average over $S$ samples of the dynamical system taken with different initializations of the dynamics and different realisations of the randomness. We want to compare the results of finite-size simulations to the prediction of the theory which is valid in the infinite size limit $N\to\infty$. In order to extract this limit from the finite-$N$ simulations, we use the scaling (see Fig.~\ref{fig: quadratic fit for infinite size predictions})
\begin{gather}
   D^{(N)}(t,t_0) \simeq D(t,t_0) + \frac{d_1}{N} + \frac{d_2}{N^2} + \mathcal{O}\left(\frac 1{N^3}\right)\:.
    \label{scalin_D_N}
\end{gather}
In Fig.~\ref{fig: comparison Dtilde D}-Right, we computed $D^{(N)}(t,t_0)$ from numerical simulations with $dt=0.01$, $t_0=15$, $S=1000$, and for different values of $N\in[2^7, 2^8, 2^9]$ and $\varepsilon\in[2^{-2}, 2^{-3}]$.

It is also possible to obtain a direct numerical estimate of the MLE in the stationary state via the Bennetin-Wolf algorithm \cite{BennetinWolf1, BennetinWolf2}. The pseudo-code of the algorithm is given in Alg.~\ref{algo: simulations for lyapunov}. This algorithm computes a quantity $\gamma(t)$ that is related to the MLE by
\begin{gather}
    \Lambda_0^{(N)} = \left\langle \lim_{t \to \infty}  \frac{\gamma(t)}{t} \right\rangle_S,
    \label{finit_N_L}
\end{gather} 
where $\langle \cdot \rangle_S$ is an average over $S$ samples of the dynamical system taken with different initializations of the dynamics and different realisations of the randomness.

\begin{algorithm}
\caption{Bennetin-Wolf algorithm to compute the maximal Lyapunov exponent $\Lambda_0^{(N)}$ of the stationary dynamics}
\For{$\mathrm{samples} \; s=1,\ldots,S$}{
\KwInitialize{$t=0,\, \mathbf{u}=\frac1{\sqrt{N}}(1,\ldots,1)$}\;
\KwDraw{$\mathrm{random\; initialization}\; \bx_0, \; \mathrm{random\; disorder}\; J$}\;
\While{$\mathrm{step\; number} <N_\mathrm{iter}$}{
    $\textbf{x} \gets \textbf{x} + dt\left[-\mu({\textbf{x}}) \, \textbf{x}+\textbf{f}({\textbf{x}}) \right]$ \\
    $\tilde{\textbf{u}} \gets T(\textbf{x}) \textbf{u}$ with $T_{ij}(\textbf{x})=\delta_{ij} + dt \left[-\frac{\partial \mu({\bf x})}{\partial x_j}x_i -\mu({\bf x}) \delta_{ij} + \frac{\partial f_i({\bf x})}{\partial x_j} \right]$ $\forall i,j\in\{1,\ldots,N\}$\\
    $\textbf{u} \gets \tilde{\textbf{u}} /  \|\tilde{\textbf{u}}\|$\\
    \If{$\mathrm{step\; number} >N_\mathrm{transient}$}{
        $t \gets t+dt$\\
        $\gamma \gets \gamma + \log(\|\tilde{\textbf{u}}\|)$
    }
}}
\Return{$\gamma(t)/t$}
\label{algo: simulations for lyapunov}
\end{algorithm}

We then wish to take the infinite size limit
\begin{gather}
    \Lambda_0 = \lim_{N\to\infty}  \Lambda_0^{(N)}\:.
    \label{extrap_L}
\end{gather}
In order to extract this limit from the finite-$N$ simulations, we use the scaling (see Fig.~\ref{fig: quadratic fit for infinite size predictions})
\begin{gather}
   \Lambda_0^{(N)}\simeq \Lambda_0 + \frac{\ell_1}{N} + \frac{\ell_2}{N^2} + \mathcal{O}\left(\frac 1{N^3}\right)\:.
    \label{scalin_Lya_N}
\end{gather}
In Fig.~\ref{fig: MLE across the phase diagram and comparison with D_tilde and simulations}, we computed numerically $\Lambda_0^{(N)}$ for different values of $N=[100,\, 125,\, 166,\, 250,\, 500]$, using $S=100$ samples, $N_\mathrm{iter}=300/dt$, $N_\mathrm{transient}=100/dt$ and $dt=0.01$. $N_\mathrm{iter}$ is taken sufficiently large so that the dynamics reaches a stationary state.
In Fig.\ref{fig: MLE across the phase diagram and comparison with D_tilde and simulations} we compare the prediction of the MLE with numerical simulations and we show an overall good agreement.

\section{Conclusions}
In this work, we considered a simple class of high-dimensional dynamical systems exhibiting a rich variety of dynamical phases and responses to external perturbations. 
We followed closely the literature on random recurrent neural networks of the last forty years and showed that most (all) of the conclusions reached for standard models of random recurrent neural networks follow with the dynamical systems that we consider in this work.
The main advantage of the formalism developed in this manuscript is that the simple dynamical systems that we are considering can be treated more easily than standard recurrent neural networks. This advantage is not crucial for random recurrent neural networks where computations can also be done (albeit with more difficulties). However, it becomes important when it comes to understand what happens when neural networks are trained. This is because in this case it is important to be able to track the adaptive dynamics of the interactions between the degrees of freedom. The class of the dynamical systems considered in this work has been studied in this regard in \cite{fournier2023statistical} using a DMFT approach to track the dynamics of a particular training algorithm called FORCE \cite{sussillo2009generating}. We believe that one of the most interesting perspectives on the results developed in this work consist in applying them to study the stability of the dynamical attractors as a function of training time. This type of analysis is very challenging and it has been done only very recently via extensive numerical simulations \cite{engelken2023lyapunov}. We believe that our work represents a promising direction to make analytical progress.

\section*{Acknowledgments}
We would like to thank Alessia Annibale, Alessandro Pacco and Valentina Ros for discussions related to the topics discussed here.
PU acknowledges funding by the French government under the France
2030 program (PhOM - Graduate School of Physics) with
reference ANR-11-IDEX-0003.
This manuscript was partially written at Zif, in Bielefeld where PU was visiting the program \emph{The lush world of random matrices}. He thanks Zif for the kind hospitality.

\bibliography{refs.bib}

@book{Ott_2002,
place={Cambridge}, edition={2}, title={Chaos in Dynamical Systems}, publisher={Cambridge University Press}, author={Ott, Edward}, year={2002}}

@article{chen2018tap,
  title={On the TAP free energy in the mixed p-spin models},
  author={Chen, Wei-Kuo and Panchenko, Dmitry},
  journal={Communications in Mathematical Physics},
  volume={362},
  number={1},
  pages={219--252},
  year={2018},
  publisher={Springer}
}

@article{fournier2023statistical,
  title={Statistical physics of learning in high-dimensional chaotic systems},
  author={Fournier, Samantha J and Urbani, Pierfrancesco},
  journal={Journal of Statistical Mechanics: Theory and Experiment},
  volume={2023},
  number={11},
  pages={113301},
  year={2023},
  publisher={IOP Publishing}
}

@article{montanari2025dynamical,
  title={Dynamical decoupling of generalization and overfitting in large two-layer networks},
  author={Montanari, Andrea and Urbani, Pierfrancesco},
  journal={arXiv preprint arXiv:2502.21269},
  year={2025}
}

@article{mignacco2020dynamical,
  title={Dynamical mean-field theory for stochastic gradient descent in gaussian mixture classification},
  author={Mignacco, Francesca and Krzakala, Florent and Urbani, Pierfrancesco and Zdeborov{\'a}, Lenka},
  journal={Advances in Neural Information Processing Systems},
  volume={33},
  pages={9540--9550},
  year={2020}
}

@article{eissfeller1992new,
  title={New method for studying the dynamics of disordered spin systems without finite-size effects},
  author={Eissfeller, H and Opper, M},
  journal={Physical review letters},
  volume={68},
  number={13},
  pages={2094},
  year={1992},
  publisher={APS}
}

@article{fournier2025non,
  title={Non-reciprocal interactions and high-dimensional chaos: comparing dynamics and statistics of equilibria in a solvable model},
  author={Fournier, Samantha J and Pacco, Alessandro and Ros, Valentina and Urbani, Pierfrancesco},
  journal={arXiv preprint arXiv:2503.20908},
  year={2025}
}

@article{cugliandolo1993analytical,
  title={Analytical solution of the off-equilibrium dynamics of a long-range spin-glass model},
  author={Cugliandolo, Leticia F and Kurchan, Jorge},
  journal={Physical Review Letters},
  volume={71},
  number={1},
  pages={173},
  year={1993},
  publisher={APS}
}

@article{rieger1989solvable,
  title={Solvable model of a complex ecosystem with randomly interacting species},
  author={Rieger, H},
  journal={Journal of Physics A: Mathematical and General},
  volume={22},
  number={17},
  pages={3447},
  year={1989},
  publisher={IOP Publishing}
}

@article{rieger1993dynamical,
  title={DYNAMICAL MEAN-FIELD-THEORY FOR A COMPLEX ECOSYSTEM WITH RANDOMLY INTERACTING SPECIES},
  author={Rieger, Heiko},
  journal={Complex Dynamics},
  pages={207},
  year={1993},
  publisher={Nova Publishers}
}

@article{crisanti2004spherical,
  title={Spherical 2+ p Spin-Glass Model: An Exactly Solvable Model<? format?> for Glass to Spin-Glass Transition},
  author={Crisanti, Andrea and Leuzzi, Luca},
  journal={Physical review letters},
  volume={93},
  number={21},
  pages={217203},
  year={2004},
  publisher={APS}
}

@article{nieuwenhuizen1995exactly,
  title={Exactly solvable model of a quantum spin glass},
  author={Nieuwenhuizen, Th M},
  journal={Physical review letters},
  volume={74},
  number={21},
  pages={4289},
  year={1995},
  publisher={APS}
}

@article{schuecker2018optimal,
  title={Optimal sequence memory in driven random networks},
  author={Schuecker, Jannis and Goedeke, Sven and Helias, Moritz},
  journal={Physical Review X},
  volume={8},
  number={4},
  pages={041029},
  year={2018},
  publisher={APS}
}

@article{rajan2010stimulus,
  title={Stimulus-dependent suppression of chaos in recurrent neural networks},
  author={Rajan, Kanaka and Abbott, LF and Sompolinsky, Haim},
  journal={Physical Review E—Statistical, Nonlinear, and Soft Matter Physics},
  volume={82},
  number={1},
  pages={011903},
  year={2010},
  publisher={APS}
}

@article{cugliandolo1995full,
  title={Full dynamical solution for a spherical spin-glass model},
  author={Cugliandolo, Leticia F and Dean, David S},
  journal={Journal of Physics A: Mathematical and General},
  volume={28},
  number={15},
  pages={4213},
  year={1995},
  publisher={IOP Publishing}
}

@book{landau2013quantum,
  title={Quantum mechanics: non-relativistic theory},
  author={Landau, Lev Davidovich and Lifshitz, Evgenii Mikhailovich},
  volume={3},
  year={2013},
  publisher={Elsevier}
}

@book{katok1995introduction,
  title={Introduction to the modern theory of dynamical systems},
  author={Katok, Anatole and Katok, AB and Hasselblatt, Boris},
  number={54},
  year={1995},
  publisher={Cambridge university press}
}

@article{huang2025freezing,
  title={Freezing chaos without synaptic plasticity},
  author={Huang, Weizhong and Huang, Haiping},
  journal={arXiv preprint arXiv:2503.08069},
  year={2025}
}

@article{dembo2020dynamics,
  title={Dynamics for spherical spin glasses: disorder dependent initial conditions},
  author={Dembo, Amir and Subag, Eliran},
  journal={Journal of Statistical Physics},
  volume={181},
  number={2},
  pages={465--514},
  year={2020},
  publisher={Springer}
}

@article{cugliandolo2002dynamics,
  title={Dynamics of glassy systems},
  author={Cugliandolo, Leticia F},
  journal={arXiv preprint cond-mat/0210312},
  volume={11},
  year={2002},
  publisher={Springer}
}

@article{de1978field,
  title={Field-theory renormalization and critical dynamics above T c: Helium, antiferromagnets, and liquid-gas systems},
  author={De Dominicis, C and Peliti, L},
  journal={Physical Review B},
  volume={18},
  number={1},
  pages={353},
  year={1978},
  publisher={APS}
}

@article{martin1973statistical,
  title={Statistical dynamics of classical systems},
  author={Martin, Paul Cecil and Siggia, Eric D and Rose, Harvey A},
  journal={Physical Review A},
  volume={8},
  number={1},
  pages={423},
  year={1973},
  publisher={APS}
}

@article{martorell2024dynamically,
  title={Dynamically selected steady states and criticality in non-reciprocal networks},
  author={Martorell, Carles and Calvo, Rub{\'e}n and Annibale, Alessia and Munoz, Miguel A},
  journal={Chaos, Solitons \& Fractals},
  volume={182},
  pages={114809},
  year={2024},
  publisher={Elsevier}
}

@article{fyodorov2015large,
  title={Large time zero temperature dynamics of the spherical p= 2-spin glass model of finite size},
  author={Fyodorov, Yan V and Perret, Anthony and Schehr, Gr{\'e}gory},
  journal={Journal of Statistical Mechanics: Theory and Experiment},
  volume={2015},
  number={11},
  pages={P11017},
  year={2015},
  publisher={IOP Publishing}
}

@book{zinn2021quantum,
  title={Quantum field theory and critical phenomena},
  author={Zinn-Justin, Jean},
  volume={171},
  year={2021},
  publisher={Oxford university press}
}

@article{crisanti1987dynamics,
  title={Dynamics of spin systems with randomly asymmetric bonds: Langevin dynamics and a spherical model},
  author={Crisanti, Andrea and Sompolinsky, Haim},
  journal={Physical Review A},
  volume={36},
  number={10},
  pages={4922},
  year={1987},
  publisher={APS}
}

@article{crisanti2018path,
  title={Path integral approach to random neural networks},
  author={Crisanti, A and Sompolinsky, H},
  journal={Physical Review E},
  volume={98},
  number={6},
  pages={062120},
  year={2018},
  publisher={APS}
}

@book{mezard1987spin,
  title={Spin glass theory and beyond: An Introduction to the Replica Method and Its Applications},
  author={M{\'e}zard, Marc and Parisi, Giorgio and Virasoro, Miguel Angel},
  volume={9},
  year={1987},
  publisher={World Scientific Publishing Company}
}

@article{qiu2024optimization,
  title={An optimization-based equilibrium measure describes non-equilibrium steady state dynamics: application to edge of chaos},
  author={Qiu, Junbin and Huang, Haiping},
  journal={arXiv preprint arXiv:2401.10009},
  year={2024}
}

@article{stubenrauch2025fixed,
  title={Fixed point geometry in chaotic neural networks},
  author={Stubenrauch, Jakob and Keup, Christian and Kurth, Anno C and Helias, Moritz and van Meegen, Alexander},
  journal={Physical Review Research},
  volume={7},
  number={2},
  pages={023203},
  year={2025},
  publisher={APS}
}

@article{wainrib2013topological,
  title={Topological and dynamical complexity of random neural networks},
  author={Wainrib, Gilles and Touboul, Jonathan},
  journal={Physical review letters},
  volume={110},
  number={11},
  pages={118101},
  year={2013},
  publisher={APS}
}

@article{ringel2025applications,
  title={Applications of Statistical Field Theory in Deep Learning},
  author={Ringel, Zohar and Rubin, Noa and Mor, Edo and Helias, Moritz and Seroussi, Inbar},
  journal={arXiv preprint arXiv:2502.18553},
  year={2025}
}

@article{engelken2023lyapunov,
  title={Lyapunov spectra of chaotic recurrent neural networks},
  author={Engelken, Rainer and Wolf, Fred and Abbott, Larry F},
  journal={Physical Review Research},
  volume={5},
  number={4},
  pages={043044},
  year={2023},
  publisher={APS}
}

@book{helias2020statistical,
  title={Statistical field theory for neural networks},
  author={Helias, Moritz and Dahmen, David},
  volume={970},
  year={2020},
  publisher={Springer}
}

@article{hakim2025theory,
  title={Theory of Temporal Pattern Learning in Echo State Networks},
  author={Hakim, Vincent and Karma, Alain},
  journal={bioRxiv},
  pages={2025--06},
  year={2025},
  publisher={Cold Spring Harbor Laboratory}
}

@article{pereira2023forgetting,
  title={Forgetting leads to chaos in attractor networks},
  author={Pereira-Obilinovic, Ulises and Aljadeff, Johnatan and Brunel, Nicolas},
  journal={Physical Review X},
  volume={13},
  number={1},
  pages={011009},
  year={2023},
  publisher={APS}
}

@article{mastrogiuseppe2018linking,
  title={Linking connectivity, dynamics, and computations in low-rank recurrent neural networks},
  author={Mastrogiuseppe, Francesca and Ostojic, Srdjan},
  journal={Neuron},
  volume={99},
  number={3},
  pages={609--623},
  year={2018},
  publisher={Elsevier}
}

@article{clark2023dimension,
  title={Dimension of activity in random neural networks},
  author={Clark, David G and Abbott, LF and Litwin-Kumar, Ashok},
  journal={Physical Review Letters},
  volume={131},
  number={11},
  pages={118401},
  year={2023},
  publisher={APS}
}

@article{clark2024theory,
  title={Theory of coupled neuronal-synaptic dynamics},
  author={Clark, David G and Abbott, LF},
  journal={Physical Review X},
  volume={14},
  number={2},
  pages={021001},
  year={2024},
  publisher={APS}
}

@article{fournier2025generative,
  title={Generative modeling through internal high-dimensional chaotic activity},
  author={Fournier, Samantha J and Urbani, Pierfrancesco},
  journal={Physical Review E},
  volume={111},
  number={4},
  pages={045304},
  year={2025},
  publisher={APS}
}

@article{cugliandolo2023recent,
  title={Recent applications of dynamical mean-field methods},
  author={Cugliandolo, Leticia F},
  journal={Annual Review of Condensed Matter Physics},
  volume={15},
  year={2023},
  publisher={Annual Reviews}
}

@article{sussillo2009generating,
  title={Generating coherent patterns of activity from chaotic neural networks},
  author={Sussillo, David and Abbott, Larry F},
  journal={Neuron},
  volume={63},
  number={4},
  pages={544--557},
  year={2009},
  publisher={Elsevier}
}

@article{kamali2023dynamical,
  title={Dynamical mean field theory for models of confluent tissues and beyond},
  author={Kamali, Persia Jana and Urbani, Pierfrancesco},
  journal={SciPost Physics},
  volume={15},
  number={5},
  pages={219},
  year={2023}
}

@article{parisi1986asymmetric,
  title={Asymmetric neural networks and the process of learning},
  author={Parisi, Giorgio},
  journal={Journal of Physics A: Mathematical and General},
  volume={19},
  number={11},
  pages={L675},
  year={1986},
  publisher={IOP Publishing}
}

@article{sompolinsky1988chaos,
  title={Chaos in random neural networks},
  author={Sompolinsky, Haim and Crisanti, Andrea and Sommers, Hans-Jurgen},
  journal={Physical review letters},
  volume={61},
  number={3},
  pages={259},
  year={1988},
  publisher={APS}
}

@article{BennetinWolf1,
  title = {Ergodic theory of chaos and strange attractors},
  author = {Eckmann, J. -P. and Ruelle, D.},
  journal = {Rev. Mod. Phys.},
  year = {1985},
  publisher = {American Physical Society},
}

@book{BennetinWolf2,
    title={Lyapunov Exponents: A Tool to Explore Complex Dynamics},
    publisher={Cambridge University Press},
    author={Pikovsky, Arkady and Politi, Antonio},
    year={2016}
}

@article{stariolo2025zero,
  title={Zero-temperature dynamics of the spherical model with non-reciprocal interactions},
  author={Stariolo, Daniel A and Metz, Fernando L},
  journal={arXiv preprint arXiv:2511.16836},
  year={2025}
}

\end{document}